\documentclass[aps,prd,twocolumn,nofootinbib,superscriptaddress]{revtex4-2}

\usepackage[utf8]{inputenc}
\usepackage[T1]{fontenc}
\usepackage{amsmath,amssymb,amsfonts,bm}
\usepackage{graphicx}
\usepackage{booktabs}
\usepackage{hyperref}
\usepackage{xcolor}
\usepackage{verbatim}

\hypersetup{
    colorlinks=true,
    linkcolor=blue,
    citecolor=blue,
    urlcolor=blue
}

\newcommand{\dd}{\mathrm{d}}
\newcommand{\e}{\mathrm{e}}

\newcommand{\Msun}{M_\odot}
\newcommand{\phib}{\phi}
\newcommand{\varphidef}{\varphi}
\newcommand{\RF}{R_{\rm F}}
\newcommand{\RB}{R^{99}_{\rm B}}

\begin{document}

\title{Fermion-Boson Stars in Scalar-Tensor Gravity}

\author{Jorge Castelo Mourelle}
\email{jorge.castelo@correo.nucleares.unam.mx}
\affiliation{
Instituto de Ciencias Nucleares, Universidad Nacional Aut\'onoma de M\'exico, Circuito Exterior C.U., A.P. 70-543, M\'exico D.F. 04510, M\'exico.}

\author{Saeed Fakhry} 
\email{saeed.fakhry@uv.es}
\affiliation{Departamento de Astronomía y Astrofísica, Universitat de València, Avinguda Vicent 
Andrés Estellés 19, 46100, Burjassot (València), Spain}
\date{\today}

\author{Néstor Ortiz}
\email{nestor.ortiz@nucleares.unam.mx}
\affiliation{
Instituto de Ciencias Nucleares, Universidad Nacional Aut\'onoma de M\'exico, Circuito Exterior C.U., A.P. 70-543, M\'exico D.F. 04510, M\'exico.}

\begin{abstract}
In the massless Damour-Esposito-Farèse scalar-tensor theory of gravity, we numerically construct static, spherically symmetric star models composed of a mixture of fermionic and bosonic matter, interacting purely gravitationally. The fermionic sector is modeled as a perfect fluid, while the bosonic sector is described by a massive, free, complex scalar field. We demonstrate that mixed fermion–boson stars can undergo spontaneous scalarization, and we systematically investigate the role played by the bosonic field across different regimes of bosonic mass, central fermionic pressure, and central bosonic amplitude. Our analysis reveals a complex interplay among the parameters, giving rise to a variety of scalarized configurations, ranging from fermionic matter embedded within extended bosonic halos to bosonic cores confined within fermionic matter. We examine sequences of solutions, in particular mass–radius surfaces, and we determine the critical coupling parameter for the onset of scalarization.
\end{abstract}

\maketitle

%%%%%%%%%%%%%%%%%%%%%%%%%%%%%%%%%%%%%%%%%%%%%%%%%%%%%%%%%%%%
%%%%%%%%%%%%%%%%%%%%%%%%%%%%%%%%%%%%%%%%%%%%%%%%%%%%%%%%%%%%
%%%%%%%%%%%%%%%%%%%%%%%%%%%%%%%%%%%%%%%%%%%%%%%%%%%%%%%%%%%%
\section{Introduction}
%%%%%%%%%%%%%%%%%%%%%%%%%%%%%%%%%%%%%%%%%%%%%%%%%%%%%%%%%%%%
%%%%%%%%%%%%%%%%%%%%%%%%%%%%%%%%%%%%%%%%%%%%%%%%%%%%%%%%%%%%
%%%%%%%%%%%%%%%%%%%%%%%%%%%%%%%%%%%%%%%%%%%%%%%%%%%%%%%%%%%%
Compact stars provide one of the cleanest strong-field laboratories for testing gravity and the possible existence of dark-sector fields. Neutron stars (NSs) probe matter at supranuclear densities~\cite{Chatziioannou:2024jsr,Baiotti:2016qnr}, while boson stars (BSs) represent self-gravitating condensates of complex scalar fields~\cite{Jetzer:1991jr,Liebling:2012fv}, which have been widely used as models of compact dark objects, black-hole (BH) mimickers, and galactic halo-like structures; see~\cite{Visinelli:2021uve,Hui:2021tkt} for reviews. A natural generalization is obtained by allowing both sectors to coexist in equilibrium, producing mixed fermion-boson stars (FBSs) in which a fermionic fluid and a bosonic condensate are bound by gravity. Such solutions of the Einstein equations have been obtained and studied in different regimes and frameworks. The first works considered static, spherically symmetric spacetimes, polytropic equations of state (EOS) for the fluid, and non-self-interacting scalar fields \cite{Henriques:1989ar,Henriques:1989ez,Jetzer:1990xa,Pisano:1995yk,deSousa:1995ye}. More recent studies consider realistic NSs coupled to self-interacting fields in different rotation regimes~\cite{Mourelle:2024qgo,Konstantinou:2024ynd,Valdez-Alvarado:2024eze,Cipriani:2025wem,Zhou:2025kku}. Not only have static configurations been analyzed, but also dynamical ones, including the formation process~\cite{Valdez-Alvarado:2012rct,DiGiovanni:2020frc,Nyhan:2022pda}. 
In addition, FBS solutions have been found in theories of gravity beyond GR, such as Horndeski~\cite{Roque:2021lvr,daRocha:2021sqd}, Einstein-Scalar-Gauss-Bonnet~\cite{Q:2025ycf}, and $f(R)$-theories~\cite{Fakhry:2026cmi}, increasing the number of models put in place during the last decade. These models open a window that may explain some mass regimes of compact objects, which are higher in mass than pure NSs and lower than intermediate BHs.

While alternative scenarios involving the merger of exotic compact objects have been proposed \cite{Bustillo:2020syj,bustillo2023searching}, the vast majority of compact-binary gravitational-wave (GW) observations to date are consistently interpreted as mergers involving BHs and/or NSs in well-established mass regimes~\cite{GWTC1_PRX,LIGOScientific:2020ibl,LIGOScientific:2021usb,KAGRA:2021vkt,GWTC4}, However, some observations challenge this picture. A notable example is GW190814 \cite{LIGOScientific:2020zkf}, whose secondary component has a mass of approximately $2.6M_\odot$, placing it close to the transition between the heaviest NSs and the lightest stellar-mass BHs. More recently, GW230529 \cite{GW230529} provided further evidence for compact objects populating the so-called ``lower-mass black-hole gap'', with a primary component in the approximate range $2.5-4.5M_\odot$. These observations indicate that this region of the compact-object mass spectrum, traditionally expected to be sparsely populated by standard stellar evolution scenarios, may not be completely empty. While scenarios involving primordial BHs \cite{Vattis:2020iuz,Clesse:2020ghq} and dark matter \cite{lee2021could,Das:2021yny} have been proposed to explain compact objects in this mass range, FBSs can also naturally populate it, providing an alternative interpretation~\cite{DiGiovanni:2021ejn,Mourelle:2024qgo}.

Although fully relativistic simulations of binary mergers involving FBSs remain largely unexplored, there are at least two potential avenues through which the presence of a bosonic component could be identified or constrained. First, GWs emitted during the late inspiral encode information about the tidal response of the binary components. The tidal deformability of FBSs can differ significantly from that of ordinary NSs, depending on the bosonic fraction and on whether the bosonic component forms a compact core or a more extended cloud \cite{Diedrichs:2023trk}. Precise measurements of tidal effects could therefore provide a way of distinguishing mixed stars from their purely fermionic counterparts. Second, the bosonic component may significantly modify the merger dynamics and the remnant's nature, potentially affecting the threshold for BH formation, the post-merger matter distribution, and the corresponding post-merger GW emission. Determining these effects will ultimately require numerical relativity simulations of FBS binaries. Detecting such signatures, particularly those associated with the high-frequency post-merger regime, will greatly benefit from next-generation observatories such as Einstein Telescope \cite{Punturo:2010zza,Hild:2010id}, Cosmic Explorer \cite{Reitze:2019iox,LIGOScientific:2016wof}, or the Neutron Star Extreme Matter Observatory (NEMO) \cite{Ackley:2020atn}.

Scalar-tensor theories (STTs) of gravity add a further ingredient to this picture. In these theories, gravity is mediated not only by the metric tensor but also by a nonminimally coupled, real-valued scalar field; see ~\cite{Fujii_Maeda_2003} for a review.
In a particular class of massless STTs, Damour and Esposito-Far\`ese (DEF) discovered that sufficiently compact NS models admit a non-trivial scalar field profile~\cite{Damour:1992we,Damour:1993hw}. This phenomenon, known as \textit{spontaneous scalarization}, corresponds to the endpoint of a tachyonic instability~\cite{Harada:1997mr,Sotani:2005qx,Sotani:2014tua} that can evolve nonlinearly toward stable scalarized configurations~\cite{Novak:1998rk,Palenzuela:2015ima,Mendes:2016fby}.
In addition to NSs, other types of compact objects have been found to be susceptible to scalarization in various contexts and theories. For a review, see~\cite{Doneva:2022ewd} and references therein. Similarly to the NS case, scalarized BSs feature a rich phenomenology, and can also display larger masses and radii than their non-scalarized counterparts~\cite{Whinnett:1999ma,Whinnett:1999sc,Alcubierre:2010ea,Ruiz:2012jt,Brihaye:2019puo,Brihaye:2020klz,Evstafyeva:2023kfg}. 
Most of the interest in the phenomenology of scalarized compact objects lies in the possibility of probing STTs in the strong-gravity regime through astrophysical observations~\cite{Damour:1993hw,Berti:2015itd}. However, the lack of detection of scalarized objects has put stringent bounds on the parameters of massless STTs~\cite{Bertotti:2003rm,Freire:2012mg,Antoniadis:2013pzd,Shao:2017gwu,Niu:2021nic,Tuna:2022qqr}.
In this regard, STTs involving a massive scalar field have gained attention, mainly because the effects of spontaneous scalarization of compact objects can be screened, so that current bounds on the parameter space could be avoided~\cite{PhysRevD.92.124016,Ramazanoglu:2016kul,Yazadjiev:2016pcb,Doneva:2016xmf,Sperhake:2017itk,Staykov:2018hhc,Rosca-Mead:2019seq,Rosca-Mead:2020bzt,Huang:2021tpu,Kuan:2023hrh,Asakawa:2023obq,Degollado:2024oyo,Lam:2024wpq}.

In this work and its Companion~\cite{CMjOnFs2026a}, we report the construction of static, spherical models of scalarized FBS in the context of a \emph{massless} DEF scalar-tensor theory. In this scenario, the gravitational scalar field is sourced by the combined effective trace of the fermionic and bosonic sectors, and the equilibrium structure is controlled by both a hydrostatic surface and a bosonic localization scale. For a set of sequential solutions, we explore the conditions for the onset of scalarization, as well as the effects of scalarization in the mass-radius relation.

This paper is organized as follows. In Sec.~\ref{sec:theory} we define the theory, the matter sectors, and the radial structure equations. In Sec.~\ref{sec:numerics} we describe the numerical construction of equilibrium stellar models. In Sec.~\ref{sec:representative_solutions} we discuss representative examples of scalarized pure NSs, pure BSs, and fully mixed FBSs. In Sec.~\ref{sec:results}, we analyze the features of scalarized FBSs, mass-radius surfaces, and the onset of scalarization. We conclude in Sec.~\ref{sec:conclusions} with a discussion of results and perspectives for future work. In the Appendix, we validate numerical tools in limiting cases (\ref{app:validation}); we calculate global quantities (\ref{app:magnitudes}); and we comment on the maximum value of the DEF scalar field (\ref{app:varphimax}). We use geometrized units such that \(G=c=1\), unless explicitly stated.

%%%%%%%%%%%%%%%%%%%%%%%%%%%%%%%%%%%%%%%%%%%%%%%%%%%%%%%%%%%%
%%%%%%%%%%%%%%%%%%%%%%%%%%%%%%%%%%%%%%%%%%%%%%%%%%%%%%%%%%%%
%%%%%%%%%%%%%%%%%%%%%%%%%%%%%%%%%%%%%%%%%%%%%%%%%%%%%%%%%%%%
\section{Theoretical setup}
\label{sec:theory}
%%%%%%%%%%%%%%%%%%%%%%%%%%%%%%%%%%%%%%%%%%%%%%%%%%%%%%%%%%%%
%%%%%%%%%%%%%%%%%%%%%%%%%%%%%%%%%%%%%%%%%%%%%%%%%%%%%%%%%%%%
%%%%%%%%%%%%%%%%%%%%%%%%%%%%%%%%%%%%%%%%%%%%%%%%%%%%%%%%%%%%
We consider a massless scalar-tensor theory of gravity with two matter sources: a perfect fluid describing fermionic matter and a complex, massive scalar field describing bosonic matter. In the Einstein-frame formulation, the class of massless DEF theories is given by the action
\begin{equation}
\begin{aligned}
S
={}&
\frac{1}{16\pi}
\int \dd^4x\,\sqrt{-\tilde g}
\left[
\tilde R
-2
\tilde \nabla_\mu\varphidef
\tilde \nabla^\mu\varphidef
\right]
\\
&+
S_{\rm F}\!\left[g_{\mu\nu},\Theta_{\rm F}\right]
+
S_{\rm B}\!\left[g_{\mu\nu},\Phi\right],
\end{aligned}
\label{eq:total_action}
\end{equation}
where \(\tilde R\) is the Ricci scalar associated with the Einstein-frame metric \(\tilde{g}_{\mu\nu}\), and \(\varphidef\) is the real scalar gravitational field associated with the DEF theory. \(\Theta_{\rm F}\) collectively denotes the fluid variables, and \(\Phi\) is the complex scalar field describing bosonic matter. 
The gravitational sector takes the canonical Einstein-Hilbert form in this frame, while both matter sectors couple universally to the Jordan-frame metric
\begin{equation}
    g_{\mu\nu}
    =
    a^2(\varphidef)\tilde g_{\mu\nu},
    \label{eq:jordan_metric}
\end{equation}
where $a(\varphi)$ is the coupling function that defines a particular theory. Consequently, freely falling matter follows the geometry determined by \(g_{\mu\nu}\), whereas the field equations are most conveniently formulated in terms of \(\tilde{g}_{\mu\nu}\).

For the bosonic sector, the corresponding matter action is
\begin{equation}
S_{\rm B}
=
-\frac{1}{2}
\int \dd^4x\,\sqrt{-g}
\left[
\nabla_\mu\Phi^*
\nabla^\mu\Phi
+
V\!\left(|\Phi|^2\right)
\right].
\label{eq:bosonic_action}
\end{equation}

The fermionic action is not specified since the fluid is introduced directly through its Jordan-frame energy-momentum tensor and conservation equations.

%%%%%%%%%%%%%%%%%%%%%%%%%%%%%%%%%%%%%%%%%%%%%%%%%%%%%%%%%%%%
%%%%%%%%%%%%%%%%%%%%%%%%%%%%%%%%%%%%%%%%%%%%%%%%%%%%%%%%%%%%
\subsection{Field equations}
%%%%%%%%%%%%%%%%%%%%%%%%%%%%%%%%%%%%%%%%%%%%%%%%%%%%%%%%%%%%
%%%%%%%%%%%%%%%%%%%%%%%%%%%%%%%%%%%%%%%%%%%%%%%%%%%%%%%%%%%%
Variation of the action~\eqref{eq:total_action} with respect to the Einstein-frame metric gives the field equations
\begin{equation}
\tilde G_{\mu\nu}
=
8\pi \tilde T_{\mu\nu}
+
2\tilde\nabla_{\mu}\varphidef\tilde\nabla_{\nu}\varphidef
-
\tilde g_{\mu\nu}
\tilde\nabla_{\sigma}\varphidef\tilde\nabla^{\sigma}\varphidef,
\label{eq:einstein_equations}
\end{equation}
where
\begin{equation}
\tilde T_{\mu\nu}
=
\tilde T_{\mu\nu}^{\mathrm{F}}
+
\tilde T_{\mu\nu}^{\mathrm{B}}
\label{eq:total_matter_tensor}
\end{equation}
is the total Einstein-frame energy-momentum tensor from the fermionic $(\tilde T_{\mu\nu}^{\mathrm{F}})$ and bosonic $(\tilde T_{\mu\nu}^{\mathrm{B}})$ contributions. For matter variables defined in the Jordan frame (denoted without tilde), the energy-momentum tensor relations are given by
\begin{equation}
\tilde T_{\mu\nu}
=
a^{2}(\varphidef)T_{\mu\nu},
\qquad
\tilde T
=
a^{4}(\varphidef)T,
\label{eq:stress_tensor_transformation}
\end{equation}
where \(\tilde T=\tilde g^{\mu\nu}\tilde T_{\mu\nu}\) and
\(T=g^{\mu\nu}T_{\mu\nu}\).

Variation of the action~\eqref{eq:total_action} with respect to $\varphi$ gives the equation of motion
\begin{equation}
\tilde\Box\varphidef
=
-4\pi
\alpha(\varphidef)\tilde T,
\label{eq:covariant_DEF_equation}
\end{equation}
where $\alpha(\varphidef) \equiv \dd\ln a(\varphidef)/\dd\varphidef$. It shows that the DEF-field dynamics is sourced by the trace of the total matter energy-momentum tensor,
\begin{equation}
\tilde T
=
\tilde T_{\mathrm{F}}
+
\tilde T_{\mathrm{B}}.
\label{eq:total_trace}
\end{equation}

%%%%%%%%%%%%%%%%%%%%%%%%%%%%%%%%%%%%%%%%%%%%%%%%%%%%%%%%%%%%
%%%%%%%%%%%%%%%%%%%%%%%%%%%%%%%%%%%%%%%%%%%%%%%%%%%%%%%%%%%%
\subsection{Fermionic sector}
\label{sec:Fermionic sector}
%%%%%%%%%%%%%%%%%%%%%%%%%%%%%%%%%%%%%%%%%%%%%%%%%%%%%%%%%%%%
%%%%%%%%%%%%%%%%%%%%%%%%%%%%%%%%%%%%%%%%%%%%%%%%%%%%%%%%%%%%
The fermionic component is modeled as a cold perfect fluid in the Jordan frame, with pressure $p$, total energy density $\varepsilon$, and rest-mass density $\rho_0$. The Jordan-frame stress-energy tensor of the fluid,
\begin{equation}
    T^{\mu\nu}_{\rm F}
    = (\varepsilon+p)u^\mu u^\nu+p g^{\mu\nu} ,
    \label{eq:Tmunu_fluid}
\end{equation}
is indeed covariantly conserved,
\begin{equation}
\nabla_{\mu}T_{\rm F}^{\mu}{}_{\nu}
= 0 .
\label{eq:Jordan_frame_conservation}
\end{equation}
For the sake of comparison with previous results on the scalarization of pure NSs, we follow Ref.~\cite{Mendes:2016fby} on using the polytropic equation of state
\begin{equation}
    p=K \rho_\text{ref} \left( \frac{\rho_0}{\rho_\text{ref}} \right)^\Gamma, \qquad
    \varepsilon=\rho_0+\frac{p}{\Gamma-1}.
    \label{eq:eos}
\end{equation}
with $\Gamma = 3$, $K=0.005$, and the reference density $\rho_\text{ref} = 1.66\times 10^{14}$ g/cm$^{3}$.\footnote{Internally, our code uses variables in geometrized units, for which the equation of state reads $p=K \rho_0^\Gamma$, with $\Gamma = 3$ and $K=6.923299\times10^{4}$.} The choice in Eq.~\eqref{eq:eos} is certainly not astrophysically realistic. Nevertheless, it suffices for our purpose of demonstrating spontaneous scalarization in a simplified scenario. In the Companion Paper~\cite{CMjOnFs2026a}, however, we employ a more realistic matter model, replacing the polytrope with a tabulated nuclear EOS that satisfies current observational constraints.

Finally, we define the Jordan-frame fermionic radius, $\RF$, as the radial position at which the fluid pressure vanishes. See Appendix~\ref{app:radii} for details.

%%%%%%%%%%%%%%%%%%%%%%%%%%%%%%%%%%%%%%%%%%%%%%%%%%%%%%%%%%%%
%%%%%%%%%%%%%%%%%%%%%%%%%%%%%%%%%%%%%%%%%%%%%%%%%%%%%%%%%%%%
\subsection{Bosonic sector}
\label{sec:Bosonic sector}
%%%%%%%%%%%%%%%%%%%%%%%%%%%%%%%%%%%%%%%%%%%%%%%%%%%%%%%%%%%%
%%%%%%%%%%%%%%%%%%%%%%%%%%%%%%%%%%%%%%%%%%%%%%%%%%%%%%%%%%%%
Variation of the action~\eqref{eq:bosonic_action} with respect to \(\Phi^{*}\) yields the Klein-Gordon equation, which we write in the Einstein frame consistently with~\eqref{eq:einstein_equations} and~\eqref{eq:covariant_DEF_equation},
\begin{equation}
\tilde\nabla_{\mu}
\left[
a^{2}(\varphidef)\tilde\nabla^{\mu}\Phi
\right]
=
a^{4}(\varphidef)
\frac{\dd V}{\dd |\Phi|^{2}}
\Phi.
\label{eq:covariant_KG_equation}
\end{equation}

The Einstein-frame energy-momentum tensor obtained from Eq.~\eqref{eq:bosonic_action} is
\begin{align}
\tilde T_{\mu\nu}^{\mathrm{B}}
={}&
a^{2}(\varphidef)
\tilde\nabla_{(\mu}\Phi^{*}
\tilde\nabla_{\nu)}\Phi
\nonumber\\
-
&\frac{1}{2}\tilde g_{\mu\nu}
\left[
a^{2}(\varphidef)
\tilde\nabla_{\alpha}\Phi^{*}
\tilde\nabla^{\alpha}\Phi
+
a^{4}(\varphidef)
V\left(|\Phi|^{2}\right)
\right].
\label{eq:bosonic_energy_momentum}
\end{align}

As usual, we assume a stationary harmonic time dependence for the bosonic field,
\begin{equation}
    \Phi(t,r)=\phib(r)\e^{-i\omega t},
    \label{eq:boson_ansatz}
\end{equation}
where $\omega$ is the field frequency and defines its central value through an eigenvalue problem. For simplicity, we consider a free, massive bosonic field governed by the potential
\begin{equation}
    V(\phib)=\mu^2\phib^2,
    \label{eq:free_potential}
\end{equation}
where $\mu$ denotes the boson mass.

Although self-interaction terms can be included as long as they preserve the $U(1)$ symmetry, we restrict the potential to the mass term alone to avoid introducing additional degrees of freedom and parameters. However, in the Companion Paper~\cite{CMjOnFs2026a}, we do consider a self-interaction term, which impacts the maximum supported mass, compactness, and spatial extension of the bosonic component.

In our numerical implementation, the radial coordinate is normalized
by the geometrized solar-mass length,
$L_\odot=GM_\odot/c^2=1.47667\,{\rm km}$. Consequently, the physical
boson rest energy associated with the dimensionless mass parameter
$\mu$ is
\begin{equation}
m_{\rm B} c^2
=
\mu\,\frac{\hbar c}{L_\odot}.
\end{equation}
Thus, $\mu=1$ corresponds to
$m_{\rm B} c^2\simeq1.336\times10^{-10}\,{\rm eV}$. The values
$\mu^2=\{0.5, 1.0, 1.5\}$ considered below therefore correspond,
respectively, to
$m_{\rm B} c^2\simeq\{9.45\times10^{-11},
1.34\times10^{-10},1.64\times10^{-10}\}\,{\rm eV}$.
Hence, all the configurations studied here involve ultralight bosons with masses of the order $10^{-10}\,{\rm eV}/c^2$.

We define the Jordan-frame bosonic radius, $\RB$, as the radius enclosing $99\%$ of the total bosonic matter; the precise definition can be found in Appendix~\ref{app:radii}. This definition conveniently characterizes the spatial extent of the bosonic component while accounting for its extended but rapidly decaying tail.

%%%%%%%%%%%%%%%%%%%%%%%%%%%%%%%%%%%%%%%%%%%%%%%%%%%%%%%%%%%%
%%%%%%%%%%%%%%%%%%%%%%%%%%%%%%%%%%%%%%%%%%%%%%%%%%%%%%%%%%%%
\subsection{Coupling function and gravitational scalar field}
\label{sec:coupling_function}
%%%%%%%%%%%%%%%%%%%%%%%%%%%%%%%%%%%%%%%%%%%%%%%%%%%%%%%%%%%%
%%%%%%%%%%%%%%%%%%%%%%%%%%%%%%%%%%%%%%%%%%%%%%%%%%%%%%%%%%%%
We adopt the DEF coupling function
\begin{equation}
    a(\varphidef)
    =
    \exp\left(
        \alpha_0 \varphidef
        +\frac{1}{2}\beta\varphidef^2
    \right),
    \label{eq:def_factor}
\end{equation}
whose logarithmic derivative is
\begin{equation}
    \alpha(\varphidef)
    =
    \frac{\dd\ln a(\varphidef)}{\dd\varphidef}
    =
    \alpha_0 + \beta\varphidef.
    \label{eq:def_alpha}
\end{equation}
The parameter \(\alpha_0\) controls the weak-field matter-scalar coupling, and it has been constrained by Solar System experiments to \(|\alpha_0| \lesssim 3.4 \times 10^{-3}\)~\cite{Bertotti:2003rm}, so in this work we set \(\alpha_0 \equiv 0 \) for simplicity. In turn, the parameter \(\beta\) governs the leading nonlinear dependence of the coupling on the scalar field. Neutron stars can spontaneously scalarize if \(\beta \lesssim -4.35\)~\cite{Damour:1993hw,Harada:1998ge}. However, we take \(\beta = -7\) as a representative value which, as we shall see in Sec.~\ref{sec:onset_of_scalarization}, is deep inside the scalarization window for most configurations studied here.

We also assume that the gravitational scalar field vanishes asymptotically. In massless DEF theory, the falloff corresponding to a scalarized solution is
\begin{equation}
    \varphidef(\tilde r)=\frac{Q_\varphi}{\tilde r}+\mathcal{O}(\tilde r^{-2}),
    \label{eq:def_tail}
\end{equation}
where $Q_\varphi$ is a scalar charge that characterizes the object; see Appendix~\ref{app:scalar_charge}.

%%%%%%%%%%%%%%%%%%%%%%%%%%%%%%%%%%%%%%%%%%%%%%%%%%%%%%%%%%%%
%%%%%%%%%%%%%%%%%%%%%%%%%%%%%%%%%%%%%%%%%%%%%%%%%%%%%%%%%%%%
\subsection{Structure equations}
%%%%%%%%%%%%%%%%%%%%%%%%%%%%%%%%%%%%%%%%%%%%%%%%%%%%%%%%%%%%
%%%%%%%%%%%%%%%%%%%%%%%%%%%%%%%%%%%%%%%%%%%%%%%%%%%%%%%%%%%%
In order to construct static and spherically symmetric models of FBSs, we write the Einstein-frame line element of spacetime in the radial gauge and polar slicing,
\begin{equation}
    \dd \tilde s^2
    =
    -\e^{\nu(\tilde r)}\dd \tilde t^2
    +\e^{\lambda(\tilde r)}\dd \tilde r^2
    +\tilde r^2\dd\Omega^2,
    \label{eq:metric}
\end{equation}
where \(\tilde r\) is the Einstein-frame areal radius. For convenience, we introduce the auxiliary functions
\begin{equation}
    A(\tilde r)\equiv\e^{\nu(\tilde r)},
    \qquad
    B(\tilde r)\equiv\e^{\lambda(\tilde r)},
    \qquad
    \vartheta(\tilde r)\equiv\varphidef'(\tilde r).
    \label{eq:metric_functions}
\end{equation}
Substituting the metric ansatz~\eqref{eq:metric}, the perfect-fluid energy-momentum tensor~\eqref{eq:Tmunu_fluid}, and the harmonic ansatz of the bosonic field~\eqref{eq:boson_ansatz} into equations~\eqref{eq:einstein_equations},~\eqref{eq:covariant_DEF_equation},~\eqref{eq:Jordan_frame_conservation}, and~\eqref{eq:covariant_KG_equation}, the equilibrium problem reduces to the following system of ordinary differential equations. The independent field equations are
\begin{align}
\nu'
={}&
8\pi \tilde r B
\left[
a^{4}(\varphidef)
\left(
p-\frac{V}{2}
\right)
%\right.
%\nonumber
%&\left.
%\qquad
+
\frac{a^{2}(\varphidef)}{2}
\left(
\frac{\omega^{2}\phib^{2}}{A}
+
\frac{\phib'^{\,2}}{B}
\right)
\right]
\nonumber\\
&+
\tilde r\vartheta^{2}
+
\frac{B-1}{\tilde r},
\label{eq:nu_eq}
\\
\lambda'
={}&
8\pi \tilde r B
\left[
a^{4}(\varphidef)
\left(
\varepsilon+\frac{V}{2}
\right)
%\right.
%\nonumber\\
%&\left.
%\qquad
+
\frac{a^{2}(\varphidef)}{2}
\left(
\frac{\omega^{2}\phib^{2}}{A}
+
\frac{\phib'^{\,2}}{B}
\right)
\right]
\nonumber\\
&+
\tilde r\vartheta^{2}
-
\frac{B-1}{\tilde r}.
\label{eq:lambda_eq}
\end{align}

The radial component of the fluid conservation equation~\eqref{eq:Jordan_frame_conservation} gives the modified hydrostatic equilibrium condition
\begin{equation}
p'
=
-(\varepsilon+p)
\left[
\frac{\nu'}{2}
+
\alpha(\varphidef)\vartheta
\right].
\label{eq:pressure_eq}
\end{equation}

The radial Klein-Gordon equation for the bosonic amplitude is
\begin{align}
\phib''
={}&
\phib B
\left[
a^{2}(\varphidef)
\frac{\dd V}{\dd(\phib^{2})}
-
\frac{\omega^{2}}{A}
\right]
\nonumber\\
&+
\frac{\phib'}{\tilde r}
\left[
4\pi B \tilde r^{2}a^{4}(\varphidef)
\left(
\varepsilon-p+V
\right)
\right.
\nonumber\\
&\left.
\qquad
-(1+B)
-2\tilde r\alpha(\varphidef)\vartheta
\right].
\label{eq:boson_eq}
\end{align}

Finally, the radial equation for the DEF scalar field can be written as
\begin{align}
\varphidef'
={}&
\vartheta,
\label{eq:def_first_order}
\\
\vartheta'
={}&
-\vartheta
\left[
\frac{2}{\tilde r}
+
\frac{1}{2}
\left(
\nu'-\lambda'
\right)
\right]
\nonumber\\
&+
4\pi B\,\alpha(\varphidef)
\left[
a^{4}(\varphidef)
\left(
\varepsilon-3p+2V
\right)
\right.
\nonumber\\
&\left.
\qquad
+
a^{2}(\varphidef)
\left(
\frac{\phib'^{\,2}}{B}
-
\frac{\omega^{2}\phib^{2}}{A}
\right)
\right].
\label{eq:def_scalar_eq}
\end{align}

%%%%%%%%%%%%%%%%%%%%%%%%%%%%%%%%%%%%%%%%%%%%%%%%%%%%%%%%%%%%
%%%%%%%%%%%%%%%%%%%%%%%%%%%%%%%%%%%%%%%%%%%%%%%%%%%%%%%%%%%%
%%%%%%%%%%%%%%%%%%%%%%%%%%%%%%%%%%%%%%%%%%%%%%%%%%%%%%%%%%%%
\section{Numerical construction}
\label{sec:numerics}
%%%%%%%%%%%%%%%%%%%%%%%%%%%%%%%%%%%%%%%%%%%%%%%%%%%%%%%%%%%%
%%%%%%%%%%%%%%%%%%%%%%%%%%%%%%%%%%%%%%%%%%%%%%%%%%%%%%%%%%%%
%%%%%%%%%%%%%%%%%%%%%%%%%%%%%%%%%%%%%%%%%%%%%%%%%%%%%%%%%%%%

All numerical calculations are performed with respect to the Einstein-frame radial coordinate $\tilde r$. However, results are reported in terms of the Jordan-frame areal coordinate $r$, which is given by
\begin{equation}
r
=
a[\varphi(\tilde r)]\,\tilde r.
\label{eq:jordan_radius}
\end{equation}

For fixed central pressure, $p_c:=p(0)$, and central bosonic amplitude, $\phib_c:=\phib(0)$, equilibrium configurations are obtained by solving Eqs.~\eqref{eq:nu_eq}-\eqref{eq:def_scalar_eq} as a nonlinear boundary-value problem (BVP) \cite{press1992numerical}. In contrast with a direct initial-value integration, the BVP solver determines the complete radial profiles together with the unknown bosonic eigenfrequency $\omega$. The central DEF scalar field, $\varphidef_c:=\varphidef(0)$, is not prescribed independently, but is determined by the boundary conditions imposed simultaneously at the origin and at the outer boundary.\footnote{As an independent cross-check, we have also solved the same equilibrium system using a nested double-shooting method, in which $\omega$ and $\varphidef_c$ are adjusted iteratively through bisection \cite{press1992numerical}. Within their respective numerical
tolerances, the BVP and double-shooting implementations recover the
same physical branches and observables.}

The conditions at the origin, including regularity, are
\begin{equation}
\begin{split}
&B(0)=1,
\qquad
\phib(0)=\phib_c,
\qquad
\phib'(0)=0,
\\
&
p(0)=p_c,
\qquad
\vartheta(0)=0.
\end{split}
\label{eq:bvp_origin_conditions}
\end{equation}
In practice, integration starts at a small but finite radius
$\tilde r_{\min}=10^{-4}$ in code units, where the regular central
expansions of the fields are imposed.

We select physical bosonic configurations by requiring the bosonic field to be localized and nodeless. The fundamental branch satisfies
\begin{equation}
\phib(\tilde r)\rightarrow0
\quad
(\tilde r\rightarrow\infty),
\label{eq:boson_asymptotic_condition}
\end{equation}
with no radial nodes in $\phib$. This localization condition fixes $\omega$ for each choice of $(p_c,\phib_c)$.

Since the numerical domain is finite, the asymptotic conditions are implemented at $\tilde r=\tilde r_{\max}$ using the leading large-radius behavior of the fields. For the massless DEF scalar field we impose the Robin condition from Eq.~\eqref{eq:def_tail},
\begin{equation}
\mathcal{R}_{\varphi}
:=
\varphidef'(\tilde r_{\max})
+
\frac{\varphidef(\tilde r_{\max})}{\tilde r_{\max}}
=
0,
\label{eq:def_residual}
\end{equation}
which follows from the asymptotic behavior $\varphidef\simeq Q_\varphi/\tilde r$. For the bosonic field, the corresponding condition is
\begin{equation}
\phib'(\tilde r_{\max})
+
\left[
k+\frac{1+\epsilon}{\tilde r_{\max}}
\right]
\phib(\tilde r_{\max})
=
0,
\label{eq:boson_robin}
\end{equation}
where
\begin{equation}
k=\sqrt{\mu^2-\omega^2},
\qquad
\epsilon
=
\frac{M_{\rm ADM}\left(2k^2-\mu^2\right)}{k},
\label{eq:boson_asymptotic_parameters}
\end{equation}
where $M_{\rm ADM}$ is the Arnowitt-Deser-Misner (ADM) mass, which is the total gravitational mass in our framework. The Robin condition incorporates the leading exponentially decaying tail of the massive bosonic field and is more accurate at finite $\tilde r_{\max}$ than imposing $\phib(\tilde r_{\max})=0$.

The mass aspect function entering the outer boundary conditions is evaluated from the Einstein-frame radial metric function as
\begin{equation}
M(\tilde r)
=
\frac{\tilde r}{2}
\left[
1-\frac{1}{B(\tilde r)}
\right].
\label{eq:einstein_mass_function}
\end{equation}
Asymptotic normalization of the temporal metric is imposed directly
as part of the BVP through
\begin{equation}
\ln\sqrt{A(\tilde r_{\max})}
+
\frac{M(\tilde r_{\max})}{\tilde r_{\max}}
=
0.
\label{eq:bvp_lapse_boundary}
\end{equation}
This is the leading finite-radius form of the condition
\begin{equation}
A(\tilde r\rightarrow\infty)
B(\tilde r\rightarrow\infty)
=
1.
\label{eq:asymptotic_metric_normalization}
\end{equation}
Consequently, the BVP normalizes the time coordinate, so no independent post-integration rescaling of $A$ is required.

More generally, if $\omega_{\rm num}$ were obtained using an unnormalized numerical time coordinate, the frequency measured by an asymptotic observer would be \cite{Lai:2004fw}
\begin{equation}
\omega_{\infty}
=
\frac{\omega_{\rm num}}
{\sqrt{
A(\tilde r_{\max})B(\tilde r_{\max})
}}.
\label{eq:physical_frequency}
\end{equation}
The eigenfrequency returned by the solver can therefore be identified with $\omega_{\infty}$ up to the residual finite-boundary correction.

The nonlinear BVP is solved using an adaptive collocation method. The initial radial mesh contains $900$ points, but is automatically refined according to the local collocation residual; hence, unlike in the Runge-Kutta shooting implementation, there is no fixed radial step $\Delta\tilde r$. For production calculations, we use a relative collocation tolerance of $10^{-5}$ and allow up to $5\times10^4$ mesh nodes. A solution is accepted only when the BVP solver converges, all fields remain finite, the Jordan-frame radial coordinate is monotonic, and the bosonic profile is nodeless.

The production domain extends to $\tilde r_{\max}=200$ in code units. Since one radial code unit corresponds to $GM_\odot/c^2=1.47667\,{\rm km}$, this is equivalent to
\begin{equation}
\tilde r_{\max}\simeq295.334\,{\rm km}.
\end{equation}
We explicitly checked the dependence of the solutions on the location of the outer boundary. Increasing $\tilde r_{\max}$ beyond the production value changes the ADM mass of the most demanding configurations by approximately $0.1\%$ or less. In the extended long-domain comparison between $\tilde r_{\max}=480$ and $640$, the relative changes in $M_{\rm ADM}$ and $\varphidef_c$ were approximately $0.034\%$ and $0.021\%$, respectively. The finite-radius estimate of the scalar charge converges more slowly and is therefore monitored separately.

Continuation is used to obtain sequences of solutions across parameter space. Once a convergent solution is obtained, it is interpolated onto the initial mesh of a neighboring point and used as the starting profile for the next BVP solve. This procedure is essential for tracking the nontrivial scalarized branch without allowing the nonlinear solver to collapse spuriously onto the coexisting solution with $\varphidef=0$.

For each model, we extract the fermionic and bosonic radii, $\RF$ and $\RB$, respectively, the total gravitational mass $M_{\rm ADM}$, the asymptotically normalized bosonic frequency, the baryonic and bosonic particle numbers, the scalar charge, and the boundary and collocation residuals. Masses, radii, particle numbers, and other quantities are defined and discussed in Appendix~\ref{app:magnitudes}.

%%%%%%%%%%%%%%%%%%%%%%%%%%%%%%%%%%%%%%%%%%%%%%%%%%%%%%%%%%%%
%%%%%%%%%%%%%%%%%%%%%%%%%%%%%%%%%%%%%%%%%%%%%%%%%%%%%%%%%%%%
%%%%%%%%%%%%%%%%%%%%%%%%%%%%%%%%%%%%%%%%%%%%%%%%%%%%%%%%%%%%
\section{Representative solutions}
\label{sec:representative_solutions}
%%%%%%%%%%%%%%%%%%%%%%%%%%%%%%%%%%%%%%%%%%%%%%%%%%%%%%%%%%%%
%%%%%%%%%%%%%%%%%%%%%%%%%%%%%%%%%%%%%%%%%%%%%%%%%%%%%%%%%%%%
%%%%%%%%%%%%%%%%%%%%%%%%%%%%%%%%%%%%%%%%%%%%%%%%%%%%%%%%%%%%
Here we discuss three representative solutions. We emphasize that all results are presented in terms of the Jordan-frame coordinate $r$, defined above in Sec.~\ref{sec:numerics}. Our aim in this section is to examine separately the two limiting cases, namely, the fermionic limit, where a NS scalarizes, and the bosonic limit, where a BS scalarizes, and then a fully mixed case, in which a FBS scalarizes. For each case, we analyze the gravitational scalar field and matter profiles together with the metric functions, and discuss how these quantities determine the global observables introduced in the previous section. This comparison illustrates the non-trivial interplay between the DEF scalar field and the relative contribution of the different matter components.

%%%%%%%%%%%%%%%%%%%%%%%%%%%%%%%%%%%%%%%%%%%%%%%%%%%%%%%%%%%%
%%%%%%%%%%%%%%%%%%%%%%%%%%%%%%%%%%%%%%%%%%%%%%%%%%%%%%%%%%%%
\subsection{Fermionic limit: Scalarized neutron stars}
\label{limit_NS}
%%%%%%%%%%%%%%%%%%%%%%%%%%%%%%%%%%%%%%%%%%%%%%%%%%%%%%%%%%%%
%%%%%%%%%%%%%%%%%%%%%%%%%%%%%%%%%%%%%%%%%%%%%%%%%%%%%%%%%%%%
In this case, we prescribe a non-vanishing central fermionic pressure while setting the bosonic scalar field to zero at the origin, i.e., $p_c\neq0$ and $\phi_c=0$. Then the bosonic sector is consistently switched off, and the system reduces to that of a scalarized NS in the massless DEF theory for the coupling function parameters specified in Sec.~\ref{sec:coupling_function}. This limiting case provides an important consistency check of our numerical implementation. Further checks can be found in Appendix~\ref{app:validation}. For the parameters adopted for our polytropic EOS and a central fermionic pressure $p_c=7\times10^{-4}\,M_\odot\,{\rm km}^{-3}$, we obtain a NS of radius $\RF\simeq 9.858\,{\rm km}$ and a total gravitational mass $M_{\rm ADM}\simeq2.094\,\Msun$.

Let us analyze in more detail the radial profiles shown in the upper part of Fig.~\ref{fig:all_def} (for a qualitative comparison see, for instance, Fig. 1 of Ref.~\cite{Novak:1997hw}). Starting with the top row, the left panel confirms that the bosonic scalar field vanishes identically throughout the entire radial domain, as expected for a pure NS. The middle panel shows the energy density profile, which follows the standard behavior expected for a compact fluid configuration: it reaches its maximum at the center and decreases monotonically outward until it vanishes at the stellar surface. Finally, the right panel displays the corresponding DEF scalar field profile, which starts from a central value $\varphi_c=0.0592$, increases smoothly to a maximum $\varphi\sim0.0820$, and then decreases monotonically. This profile is shell-like in that there is a layer away from the origin where it has a greater amplitude than in the center. Remarkably, the field shows an inflection point very close to the fermionic radius.

Now consider the bottom row of the upper part of Fig.~\ref{fig:all_def}. The left panel shows the metric function $A(r)$, which exhibits the expected regular behavior throughout the stellar interior and smoothly approaches its asymptotically flat value at large radii. The fermionic radius also lies where $A(r)$ has an inflection point.
The middle panel shows the radial metric function, $B(r)$. Here, the stellar radius lies just after the metric function's maximum, as expected.
The last subplot shows the derivative of the DEF field. As expected, it vanishes at the radial position where $\varphi$ reaches its maximum, subsequently becoming negative as the scalar field decreases toward its asymptotic value. Its most negative value occurs where the scalar field profile changes most rapidly, after which the derivative progressively tends to zero at spatial infinity, consistently with the asymptotic flattening of $\varphi$.

%%%%%%%%%%%%%%%%%%%%%%%%%%%%%%%%%%%%%%%%%%%%%%%%%%%%%%%%%%%%
%%%%%%%%%%%%%%%%%%%%%%%%%%%%%%%%%%%%%%%%%%%%%%%%%%%%%%%%%%%%
\subsection{Bosonic limit: Scalarized boson stars}
%%%%%%%%%%%%%%%%%%%%%%%%%%%%%%%%%%%%%%%%%%%%%%%%%%%%%%%%%%%%
%%%%%%%%%%%%%%%%%%%%%%%%%%%%%%%%%%%%%%%%%%%%%%%%%%%%%%%%%%%%

\begin{figure*}[htbp]
    \centering
    \includegraphics[width=0.8\textwidth]{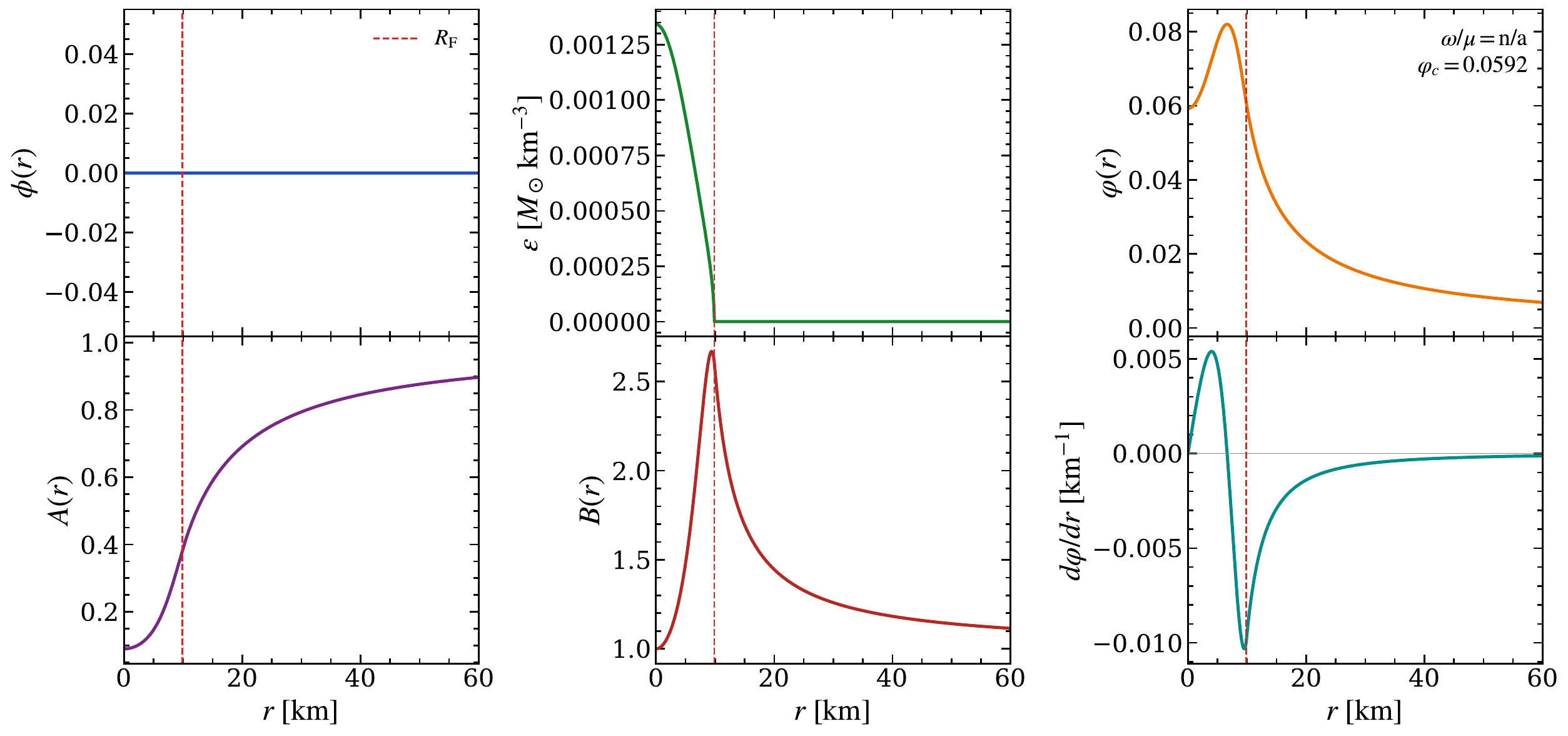}
    \includegraphics[width=0.8\textwidth]{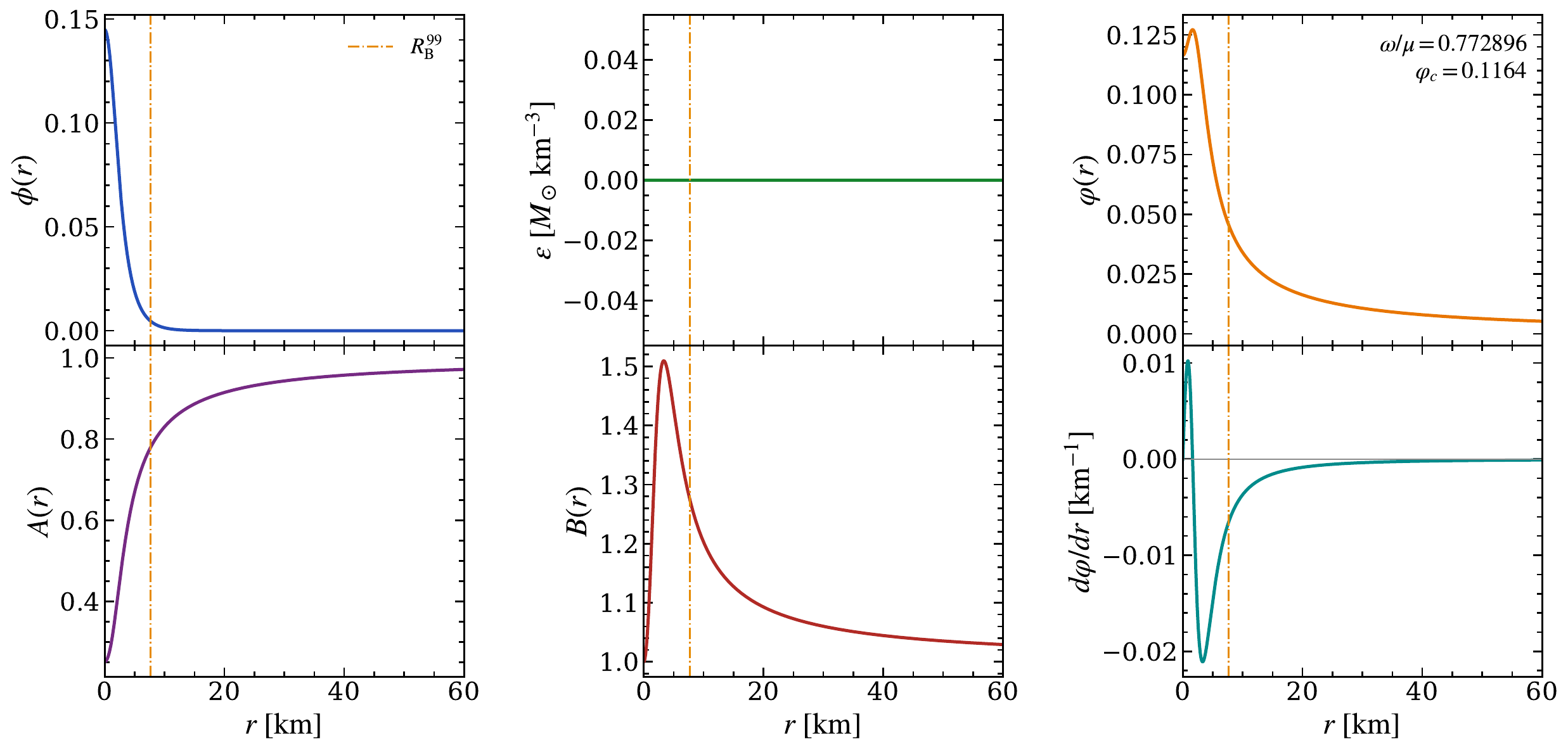}
    \includegraphics[width=0.8\textwidth]{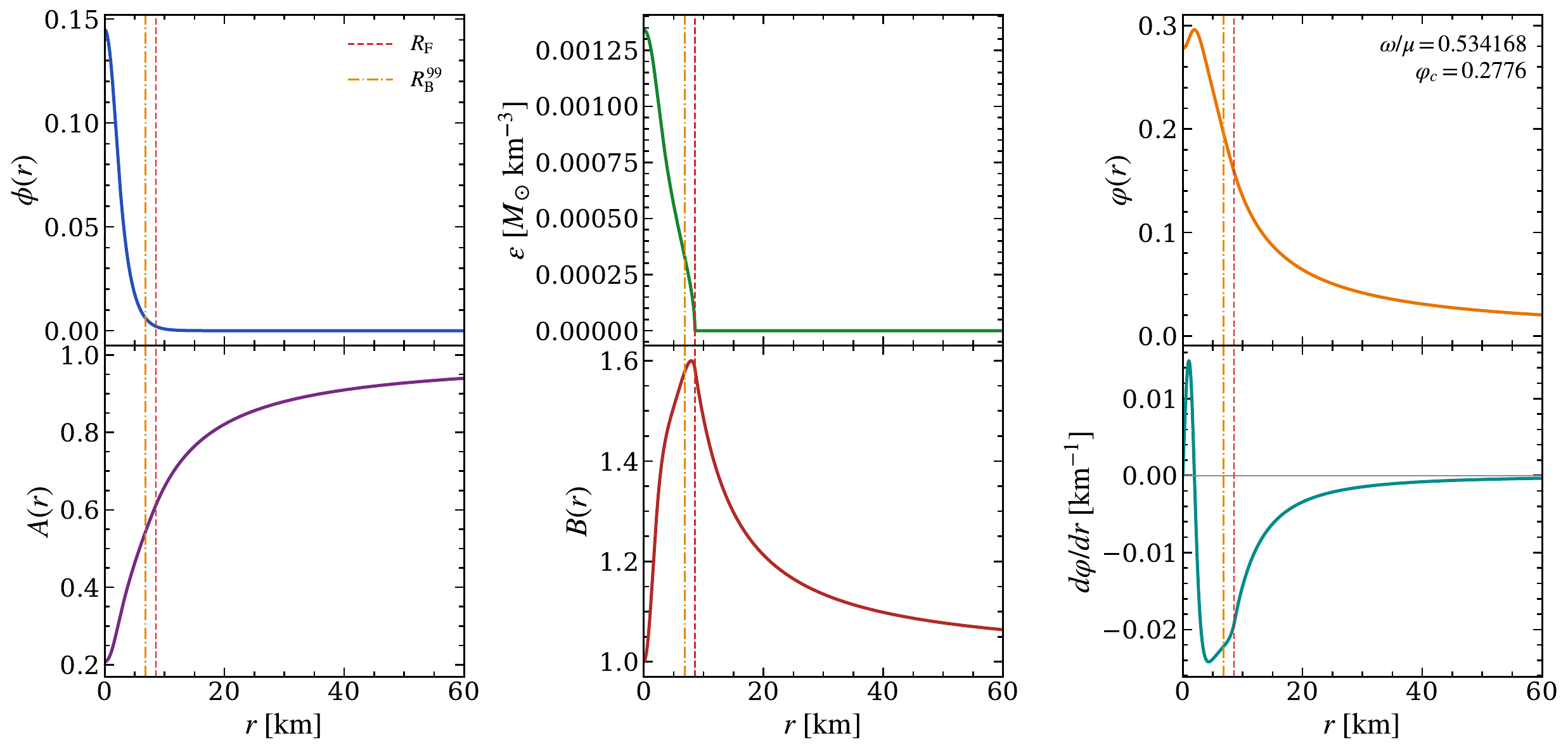}
    \caption{Representative scalarized configurations in massless DEF scalar-tensor gravity for $\beta=-7$: pure fermionic limit (upper part); pure bosonic limit (middle); and a mixed FBS (lower). In each case, the six panels show, as functions of the Jordan-frame radial coordinate $r$, the bosonic field $\phi(r)$, the fermionic energy density $\varepsilon(r)$, the DEF scalar field $\varphi(r)$, the metric functions $A(r)$ and $B(r)$, and the radial derivative $\varphi'(r)$. The Einstein-frame metric functions are displayed parametrically against the Jordan-frame radius to facilitate comparison with the matter profiles. The dashed and dash-dotted vertical lines mark the fermionic and bosonic radii $\RF$ and $\RB$, respectively. Table~\ref{tab:representative_models} shows the corresponding parameters and magnitudes.}
    \label{fig:all_def}
\end{figure*}

Here, we set $p_c=0$ while keeping $\phi_c\neq0$, switching off the fermionic sector and reducing the system to a scalarized BS. As in the previous case, we tested this limit against results in the literature, particularly those reported in Ref.~\cite{Evstafyeva:2023kfg} (see Appendix~\ref{app:validation} for details). This case is technically more involved than the pure fermionic limit. Both the bosonic frequency and the central value of the DEF scalar are treated as unknown parameters of the BVP and are determined by imposing the required asymptotic behavior of the bosonic and DEF scalar fields. For the central value $\phi_c=0.144883$, the resulting total mass is $M_{\rm ADM}\simeq0.577\,\Msun$ and the bosonic radius is $\RB\simeq7.684\,{\rm km}$.

Results are displayed in the middle part of Fig.~\ref{fig:all_def} (for comparison, see the top-left panel of Fig. 6 of Ref.~\cite{Evstafyeva:2023kfg}). The left panel of the first row shows the bosonic field $\phi(r)$, which exhibits the characteristic profile of a boson star: it reaches its maximum value at the origin and subsequently decreases monotonically, displaying the expected exponential decay toward zero as $r\rightarrow\infty$. Most of the bosonic field is concentrated within a characteristic ``bulk'' region extending up to approximately $8~{\rm km}$, while a non-vanishing but strongly suppressed tail persists beyond this scale.

The middle panel confirms that the fermionic energy density vanishes identically throughout the domain, as expected in the pure bosonic limit.
Finally, the right panel displays the corresponding DEF scalar field profile $\varphi(r)$, whose maximum is slightly shifted away from $r=0$. Although the DEF field approaches its prescribed value at spatial infinity, its decay is considerably slower than that of the bosonic field. As a consequence, the DEF scalar extends over substantially larger radial scales and can therefore contribute to the total mass at radii that are significantly larger than those associated with the bosonic matter distribution alone.

Let us now analyze the second row of the middle part of Fig.~\ref{fig:all_def}. The left panel displays the metric function $A(r)$, which exhibits the regular behavior expected for a typical BS configuration. We notice that the bosonic radius does not match the neighborhood of the inflection point, as in the NS case. The smooth approach to the asymptotically flat value is present, as expected.
The middle panel, showing the radial metric function $B(r)$, again behaves as expected, with a maximum within the star.
Finally, the right panel shows the radial derivative of the DEF scalar field, which is qualitatively similar to that of the pure fermionic case, except that its minimum is not reached as close to the object's radius.

\begin{table*}[t]
\centering
\caption{
Main parameters and global observables of the representative
fermionic, bosonic, and mixed configurations shown in
Fig.~\ref{fig:all_def}. All solutions have $\beta=-7$ and $\mu^2=1$.
The outer compactness is defined as
$C_{\rm out}=GM_{\rm ADM}/(c^2R_{\rm out})$, where
$R_{\rm out}=\max\!\left(\RF,\RB\right)$, and the binding energy
is $E_{\rm bind}=M_{\rm ADM}-M_{\rm 0,F}-M_{{\rm 0,B}}$.
The scalar charge is evaluated asymptotically as
$Q_\varphi=-r^2\varphi'(r)$. See definitions in Appendix~\ref{app:magnitudes}.
}
\label{tab:representative_models}
\footnotesize
\renewcommand{\arraystretch}{1.15}
\setlength{\tabcolsep}{3.2pt}
\resizebox{\textwidth}{!}{%
\begin{tabular}{
c
c
c
c
c
c
c
c
c
c
c
c
c
c
c
}
\hline\hline
Configuration &
$p_c$ &
$\phi_c$ &
$\mu^2$ &
$\varphi_c$ &
$\lvert\varphi_{\max}\rvert$ &
$\omega/\mu$ &
$\RF$ &
$\RB$ &
$M_{\rm ADM}$ &
$M_{\rm 0,F}$ &
$M_{{\rm 0,B}}$ &
$E_{\rm bind}$ &
$C_{\rm out}$ &
$Q_\varphi$
\\
&
$\left[M_\odot\,{\rm km}^{-3}\right]$ &
&
&
&
&
&
$[{\rm km}]$ &
$[{\rm km}]$ &
$[M_\odot]$ &
$[M_\odot]$ &
$[M_\odot]$ &
$[M_\odot]$ &
&
\\
\hline
Fermionic
&
$7.0\times10^{-4}$
&
$0$
&
$1$
&
$0.0592$
&
$0.0820$
&
$0$
&
$9.858$
&
$0$
&
$2.094$
&
$2.566$
&
$0$
&
$-0.472$
&
$0.3136$
&
$0.2717$
\\

Bosonic
&
$0$
&
$0.144883$
&
$1$
&
$0.1164$
&
$0.1272$
&
$0.7729$
&
$0$
&
$7.684$
&
$0.577$
&
$0$
&
$1.169$
&
$-0.592$
&
$0.1109$
&
$0.2127$
\\
Mixed
&
$7.0\times10^{-4}$
&
$0.144883$
&
$1$
&
$0.2776$
&
$0.2960$
&
$0.5342$
&
$8.545$
&
$6.823$
&
$1.229$
&
$0.928$
&
$1.102$
&
$-0.801$
&
$0.2124$
&
$0.8106$
\\
\hline\hline
\end{tabular}
}
\end{table*}

%%%%%%%%%%%%%%%%%%%%%%%%%%%%%%%%%%%%%%%%%%%%%%%%%%%%%%%%%%%%
%%%%%%%%%%%%%%%%%%%%%%%%%%%%%%%%%%%%%%%%%%%%%%%%%%%%%%%%%%%%
\subsection{Scalarized fermion-boson stars}
%%%%%%%%%%%%%%%%%%%%%%%%%%%%%%%%%%%%%%%%%%%%%%%%%%%%%%%%%%%%
%%%%%%%%%%%%%%%%%%%%%%%%%%%%%%%%%%%%%%%%%%%%%%%%%%%%%%%%%%%%
Let us now examine the mixed case, in which a FBS configuration is constructed within the framework of DEF gravity. Here, both the fermionic and bosonic matter sectors are present and interact only through gravity.

We solve the full DEF-Fluid-Klein-Gordon system \eqref{eq:nu_eq}-\eqref{eq:def_scalar_eq} by prescribing non-vanishing central values for both the bosonic field and the fermionic pressure. For the representative solution considered here, we choose $\phi_c=0.144883$ and $p_c=7\times10^{-4}\,M_\odot\,{\rm km}^{-3}$, for the sake of comparison with the previous cases. With $\beta=-7$, these values lead to a total gravitational mass $M_{\rm ADM}\simeq1.229\,\Msun$, with radii $\RF=8.545\,{\rm km}$ and $\RB=6.823\,{\rm km}$. The presence of two distinct matter components makes the metric functions notably richer.

Let us begin the analysis of the configuration shown in the lower part of Fig.~\ref{fig:all_def} by focusing on the top row. The left panel displays the radial profile of the bosonic field. Its qualitative behavior resembles that of the pure BS, showing the same monotonic decrease from its central value and the expected asymptotic decay. Nevertheless, despite having the same central field value, $\phi(r)$ decreases more rapidly, leading to a smaller bosonic radius $\RB\simeq6.823\,{\rm km}$, compared with $\RB\simeq7.684\,{\rm km}$ in the pure BS case shown in the middle part of Fig.~\ref{fig:all_def}.

The middle panel displays the fermionic energy density profile $\varepsilon(r)$. Although the central value is the same as in the pure NS configuration, the presence of the bosonic component modifies its radial distribution. In particular, the energy density initially decreases with a certain slope in the region populated by the bosonic field, but once it becomes negligible, the $\varepsilon(r)$ profile transitions to a different decay rate, producing an effective two-region profile associated with the spatial extent of the bosonic field. Interestingly, although it has the same central pressure as in the pure NS case, the fermionic matter extends to a smaller radius in the mixed configuration. Clearly, the bosonic component modifies the gravitational field and, consequently, the fermionic hydrostatic equilibrium, redistributing the fluid toward smaller radii.

The right panel shows the corresponding DEF scalar field profile $\varphi(r)$, which increases from $\varphi_c$, reaches a global maximum, and then decreases monotonically with a decay rate that changes as the bosonic field becomes negligible. This behavior reflects the fact that the DEF scalar field responds to the combined matter distribution and therefore retains information about the different spatial scales associated with the fermionic and bosonic components.

Another remarkable feature is the enhanced central value $\varphi_c=0.2776$, which is considerably larger with respect to either of the two limiting cases discussed above. The coexistence of the fermionic and bosonic components therefore not only modifies the radial structure of the scalar field but also increases its central amplitude.

As expected, the internal structure of the mixed object reflects the spatial distribution of the different matter components, as well as the global properties of the solution. For the representative model considered here, the gravitational mass decreases from approximately $2.094M_\odot$ in the pure NS case to about $1.229M_\odot$ in the mixed configuration, while the radius goes from roughly $9.858$ to $8.545$~km, making the FBS \emph{less} compact. This behavior illustrates the nonlinear interplay between the fermionic, bosonic, and DEF scalar sectors, and shows that mixed solutions cannot be understood as a simple superposition of the corresponding pure limits.

We now turn to the bottom row of the lower part of Fig.~\ref{fig:all_def}. The left panel shows the metric function $A(r)$, which increases monotonically with radius, but with slightly different slopes across regions populated by fermionic and bosonic matter fields. Accordingly, the radial metric function $B(r)$, shown in the middle panel, features a slight slope change before reaching its global maximum. Evidently, the simultaneous presence of the fermionic and bosonic matter components modifies the internal gravitational field in a nontrivial way, providing a geometrical signature of the multi-component nature of the mixed configuration.

Finally, the right panel displays the radial derivative of the DEF scalar field. Because the maximum $\varphi$ is not at the origin, its derivative increases, reaches a global maximum, and then decreases until it reaches a global minimum within the bulk matter. Again, the radial behavior changes across regions where the relative contributions of the bosonic and fermionic components differ, reflecting the multi-scale nature of the matter distribution. At larger radii, the derivative remains negative and progressively approaches zero, consistent with the asymptotic behavior required for the DEF scalar field.

The three representative configurations discussed above show that our implementation consistently covers physical regimes that are usually studied separately. The mixed configurations combine both ingredients, fermionic and bosonic, within a single equilibrium system, but their behavior is not simply obtained by superposing the two limiting solutions. The coexistence of fermionic and bosonic components introduces distinct radial scales, modifies the scalarization strength, and leaves a nontrivial imprint on the matter profiles, spacetime geometry, and global observables.

%%%%%%%%%%%%%%%%%%%%%%%%%%%%%%%%%%%%%%%%%%%%%%%%%%%%%%%%%%%%
%%%%%%%%%%%%%%%%%%%%%%%%%%%%%%%%%%%%%%%%%%%%%%%%%%%%%%%%%%%%
%%%%%%%%%%%%%%%%%%%%%%%%%%%%%%%%%%%%%%%%%%%%%%%%%%%%%%%%%%%%
\section{Analysis of scalarized fermion-boson stars}
\label{sec:results}
%%%%%%%%%%%%%%%%%%%%%%%%%%%%%%%%%%%%%%%%%%%%%%%%%%%%%%%%%%%%
%%%%%%%%%%%%%%%%%%%%%%%%%%%%%%%%%%%%%%%%%%%%%%%%%%%%%%%%%%%%
%%%%%%%%%%%%%%%%%%%%%%%%%%%%%%%%%%%%%%%%%%%%%%%%%%%%%%%%%%%%
In the following subsections, we analyze how the bosonic field modifies the DEF scalar field, how it affects the existence and development of scalarized branches of solutions, and how this impacts global mass-radius relations. We aim to provide a picture of both the internal structure and macroscopic properties of mixed FBS configurations.

%%%%%%%%%%%%%%%%%%%%%%%%%%%%%%%%%%%%%%%%%%%%%%%%%%%%%%%%%%%%
%%%%%%%%%%%%%%%%%%%%%%%%%%%%%%%%%%%%%%%%%%%%%%%%%%%%%%%%%%%%
\subsection{Role of the bosonic field}
\label{sec:role_of_the_bosonic_field}
%%%%%%%%%%%%%%%%%%%%%%%%%%%%%%%%%%%%%%%%%%%%%%%%%%%%%%%%%%%%
%%%%%%%%%%%%%%%%%%%%%%%%%%%%%%%%%%%%%%%%%%%%%%%%%%%%%%%%%%%%
Here, we investigate how the bosonic field modifies the system's properties using the purely scalarized NS solution as a reference. We consider representative fermionic configurations, characterized by four different values of the central pressure,
\begin{equation}
p_c \in
\{1,~2,~5,~8\}\times10^{-4}
~M_\odot{\rm km}^{-3}.
\end{equation}
These configurations span a wide range of fermionic models and let us assess whether the bosonic sector's effect depends on the underlying fermionic structure. Each of the four panels of Fig.~\ref{fig:mu_varphi_set1} corresponds to a fixed value of $p_c$. For each value of $p_c$, we progressively introduce a bosonic component and analyze its impact on the overall stellar configuration. Namely, we vary the two parameters that control the bosonic sector: the central field amplitude $\phi_c$, and the boson mass $\mu$. For each of the four fermionic configurations, we consider three values of the central bosonic field,
\begin{equation}
\phi_c \in \{0.05,~0.10,~0.15\},
\end{equation}
which are displayed, respectively, in the top, middle, and bottom subplots of each panel of Fig.~\ref{fig:mu_varphi_set1}. Each subplot contains four equilibrium solutions: the pure NS configuration as a reference, together with three FBSs for which
\begin{equation}
\mu^2 \in
\{0.5,~1.0,~1.5\}.
\end{equation}
For a given pair $(p_c,\phi_c)$, all remaining parameters are kept fixed, so differences among mixed profiles can be attributed to changes in $\mu$.

In Fig.~\ref{fig:mu_varphi_set1}, the solid and dashed curves represent the $\varphi(r)$ and $\phi(r)$ profiles, respectively. The vertical lines indicate fermionic radii. Dots and squares correspond to the values of $\varphi$ and $\phi$ at the fermionic and bosonic radii, respectively. By comparing the x-positions of dots and squares, we see whether the bosonic component is confined inside the fluid or vice versa. In the pure NS case, the blue vertical line necessarily intersects the blue $\varphi(r)$ profile at the same position in all three panels, since neither the fermionic configuration nor the DEF scalar field changes. In contrast, in mixed solutions, both the DEF scalar field profile and the fermionic surface location depend on $\mu$ and $\phi_c$, reflecting the bosonic component's backreaction on the equilibrium configuration. For each subplot of fixed $\phi_c$, the mass of the boson $\mu^2$ induces changes in the bosonic radii. 

In the following, we analyze trends that should not be taken as universal because they depend on the specific regime under consideration.

%%%%%%%%%%%%%%%%%%%%%%%%%%%%%%%%%%%%%%%%%%%%%%%%%%%%%%%%%%%%
\subsubsection{Effects of the central bosonic amplitude}
\label{sec:effects_of_phi_c}
\begin{figure*}[htbp]
    \centering

    \includegraphics[width=0.45\textwidth]{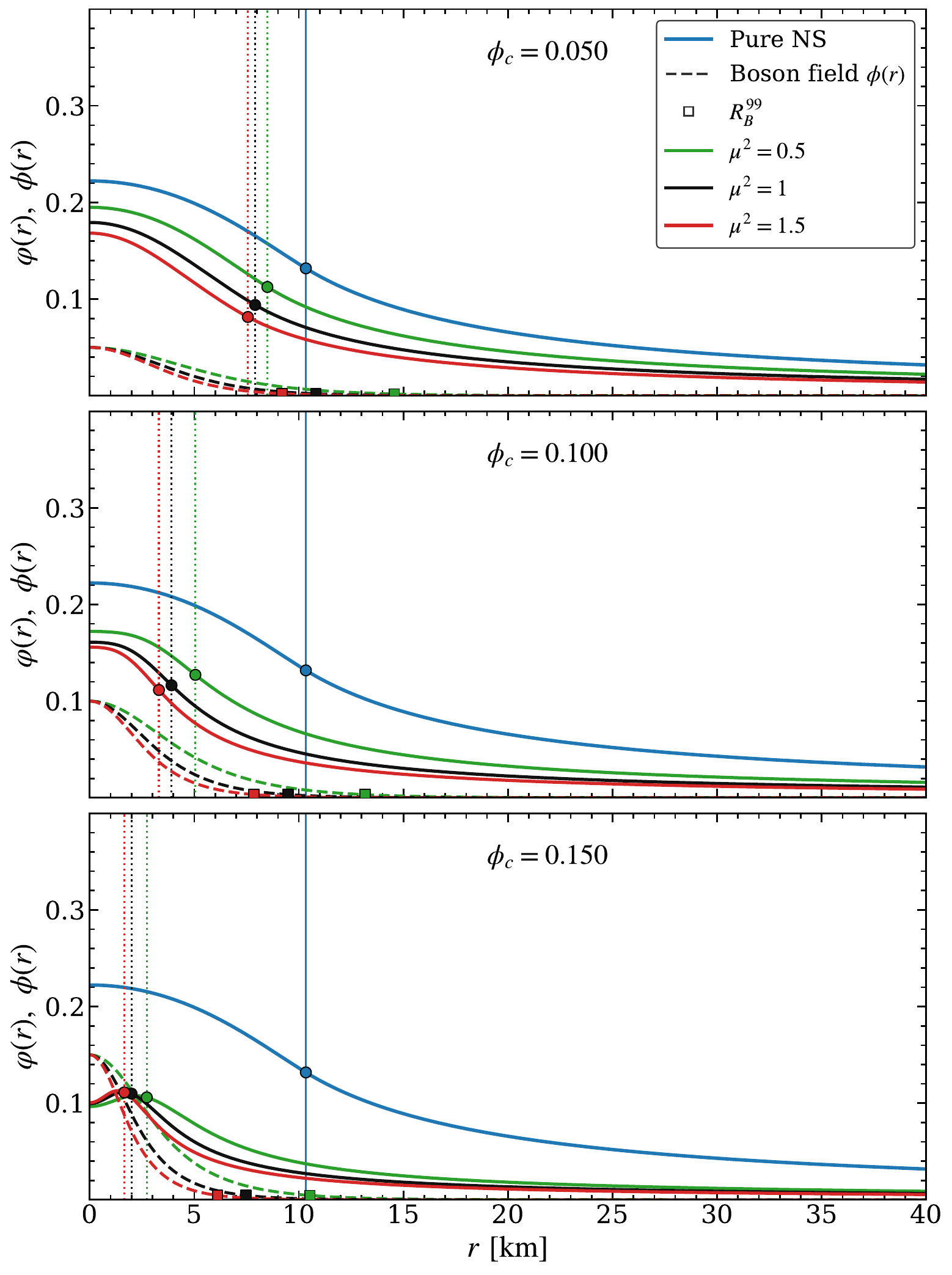}
    \hfill
    \includegraphics[width=0.45\textwidth]{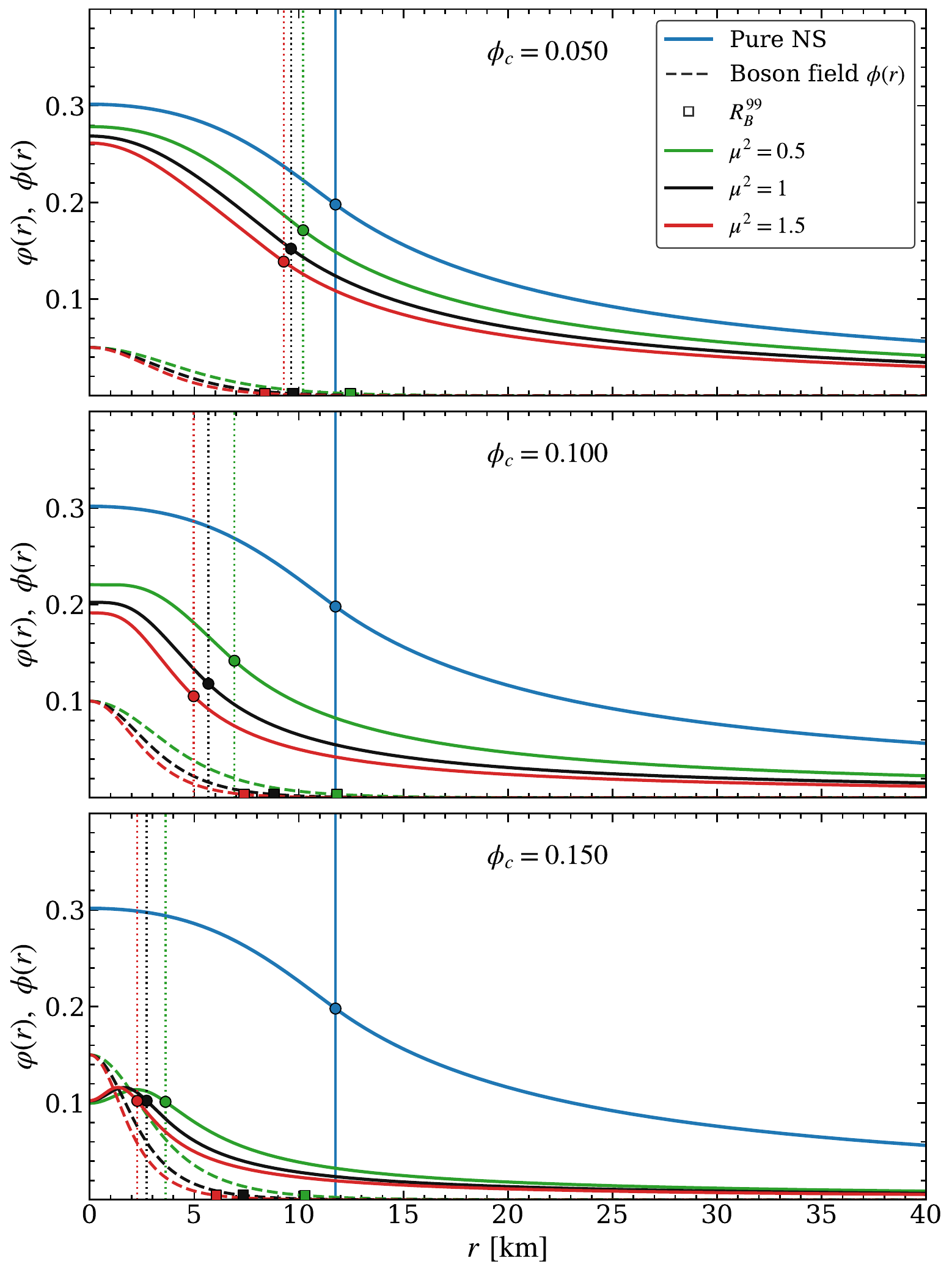}

    \vspace{0.05cm}

    \includegraphics[width=0.45\textwidth]{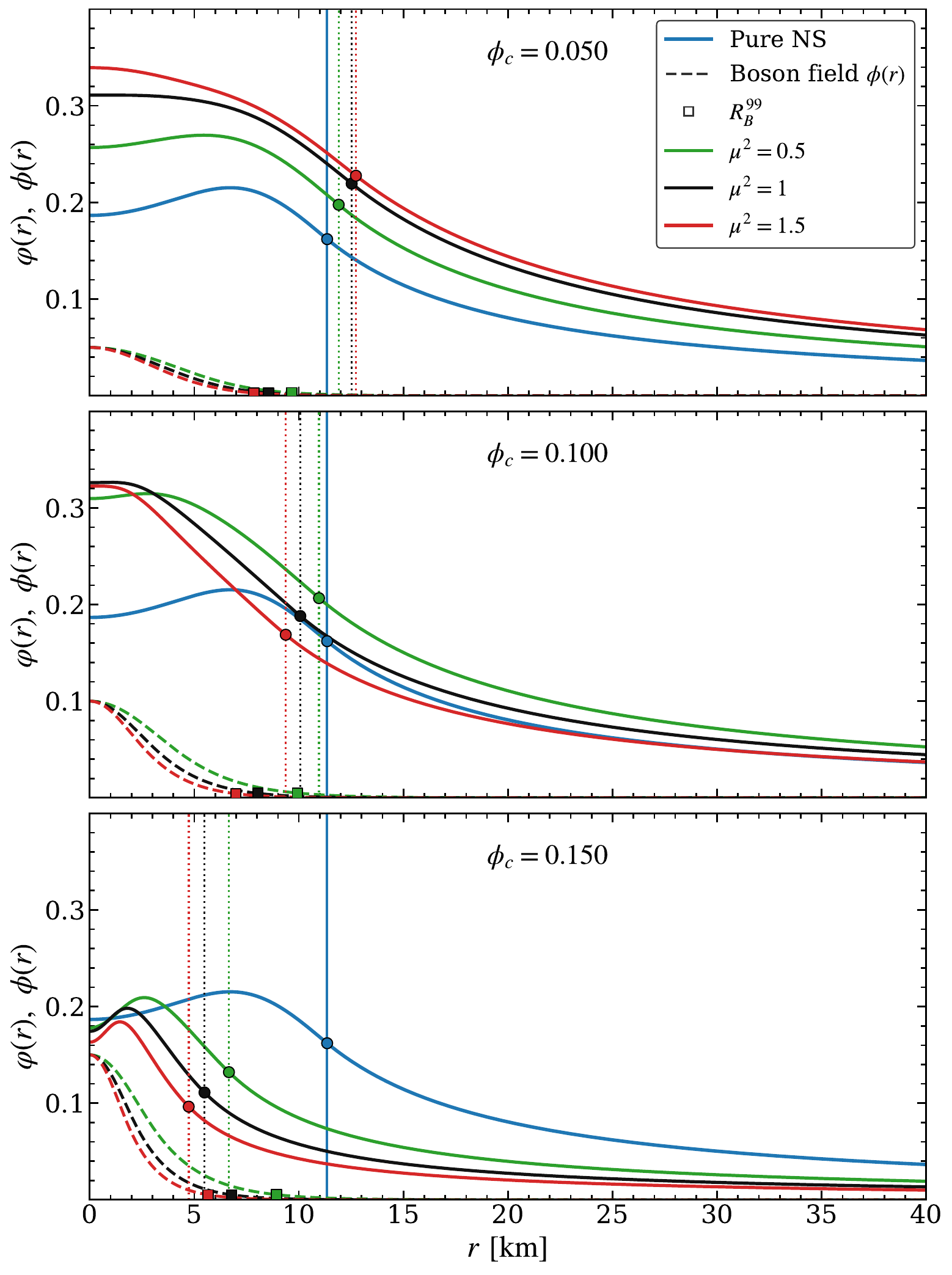}
    \hfill
    \includegraphics[width=0.45\textwidth]{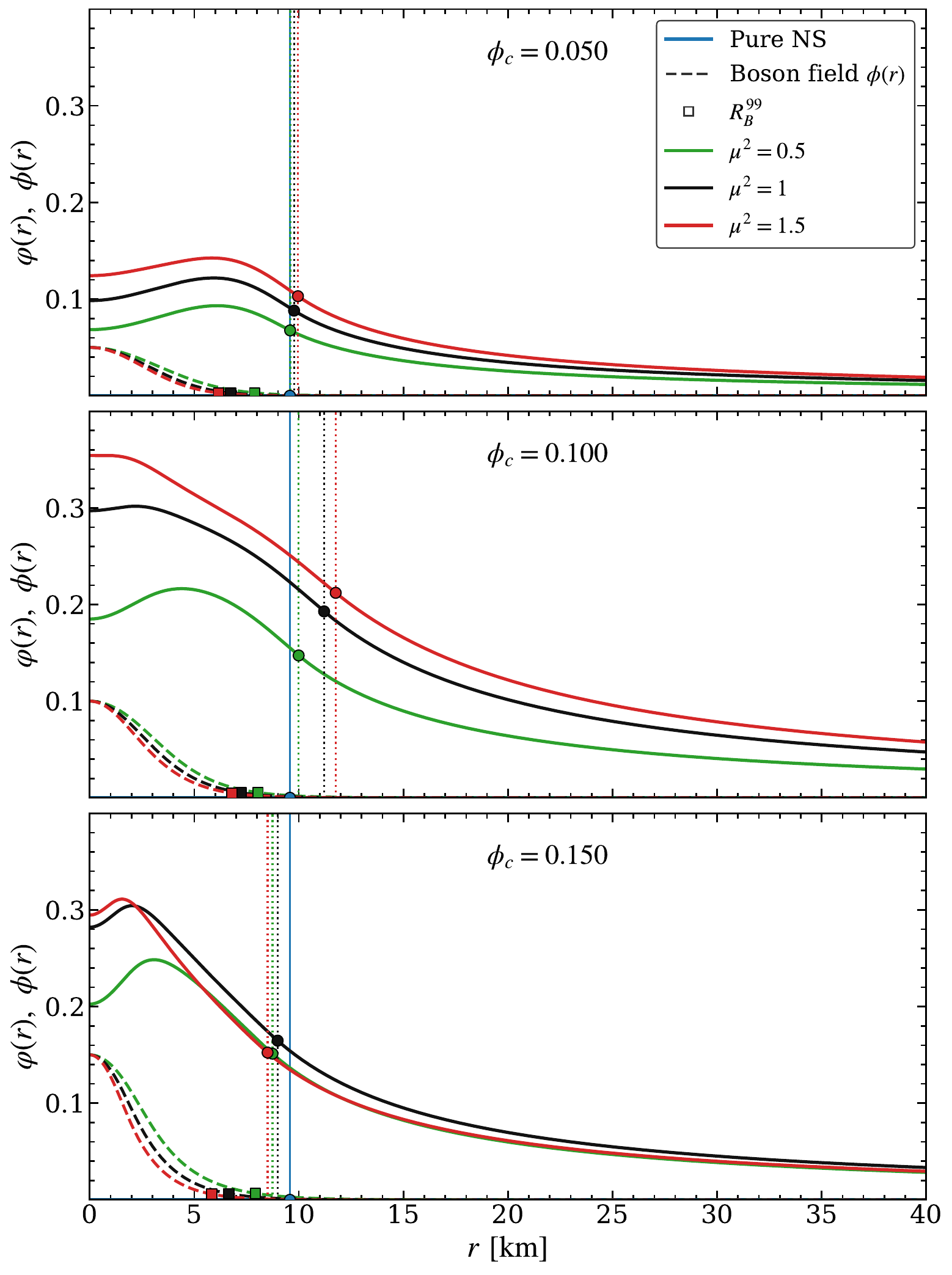}

    \caption{Radial profiles of the DEF scalar field $\varphi(r)$ (solid lines) and the bosonic field $\phi(r)$ (dashed lines) for four fermionic-pressure families, $p_c\in\{1,2,5,8\}\times10^{-4}~\mathrm{M_\odot/km^3}$, arranged from left to right and top to bottom. Within each family, the three subplots correspond to fixed values of $\phi_c\in\{0.05,0.1,0.15\}$, while colors distinguish the boson mass $\mu^2\in\{0.5,1.0,1.5\}$. The solid blue curve indicates the reference pure NS. Dots (squares) indicate fermionic (bosonic) radii, also indicated by vertical lines.}
    \label{fig:mu_varphi_set1}
\end{figure*}

%%%%%%%%%%%%%%%%%%%%%%%%%%%%%%%%%%%%%%%%%%%%%%%%%%%%%%%%%%%%
For sufficiently low central pressures, $p_c$, increasing the central bosonic amplitude $\phi_c$ reduces the central value of the DEF scalar field, $\varphi_c$, and induces a faster decay of $\varphi(r)$ inside the object. This behavior is visible in Fig.~\ref{fig:mu_varphi_set1} for the cases $p_c\in\{1,2,5\}\times 10^{-4}~\mathrm{\Msun/km^{3}}$, and shows that the bosonic component can contribute to suppress scalarization. Once the fermionic pressure becomes sufficiently large, the trend becomes less straightforward. For the representative case $p_c=8\times10^{-4}~M_\odot/{\rm km}^3$, the smallest $\phi_c$ corresponds to the smallest $\varphi_c$. Increasing the bosonic amplitude initially increases $\varphi_c$, whereas further increases cause it to decrease slightly again. Even in this latter case, however, $\varphi_c$ remains larger than in the configuration with the smallest bosonic amplitude. 
An additional layer of complexity is introduced by the boson mass. For smaller $\mu^2$, variations in $\varphi_c$ induced by changes in $\phi_c$ become less pronounced.

%%%%%%%%%%%%%%%%%%%%%%%%%%%%%%%%%%%%%%%%%%%%%%%%%%%%%%%%%%%%
\subsubsection{Gravitational scalar field profile}
\label{sec:shape_of_varphi}
%%%%%%%%%%%%%%%%%%%%%%%%%%%%%%%%%%%%%%%%%%%%%%%%%%%%%%%%%%%%
Another important regime-dependent effect concerns the shape of the gravitational scalar-field profiles, $\varphi(r)$. Increasing the central bosonic amplitude, $\phi_c$, tends to favor configurations in which the maximum of $\varphi(r)$ is displaced away from the origin. However, this behavior is not universal and depends on the pressure regime. In particular, increasing $p_c$ makes off-center extrema possible even for comparatively small values of $\phi_c$. The location and shape of the DEF-field maximum therefore depend on a non-trivial interplay between the fermionic and bosonic sectors rather than on $\phi_c$ alone.
For the high-pressure case, $p_c=8\times10^{-4}~M_\odot/{\rm km}^3$, the behavior becomes even more complex. In the intermediate panel, some DEF-field profiles develop a non-monotonic radial structure, with more than one decay rate away from the center.

% Force all pending figures to appear before the appendix
\clearpage

%%%%%%%%%%%%%%%%%%%%%%%%%%%%%%%%%%%%%%%%%%%%%%%%%%%%%%%%%%%%
\subsubsection{Effects of the boson mass}
\label{sec:effects_of_mu}
%%%%%%%%%%%%%%%%%%%%%%%%%%%%%%%%%%%%%%%%%%%%%%%%%%%%%%%%%%%%
In the low-pressure regime, as $\mu^2$ increases, the DEF field becomes progressively more confined and its central value typically decreases. For the largest values of $\phi_c$, this trend is partially modified: the absolute maximum of the DEF field can occur for the largest $\mu^2$, although the smallest $\mu^2$ still produces the most spatially extended configuration. The high-pressure regime displays the opposite tendency. In this case, larger $\mu^2$ leads to less confined DEF scalar field profiles with enhanced central values.
However, as $\phi_c$ increases, the trend becomes progressively less systematic, and there is no simple monotonic dependence on $\mu^2$.
The intermediate-pressure configurations interpolate between these two limiting behaviors in a highly non-trivial way. The case of $p_c=2\times10^{-4}~M_\odot/{\rm km}^3$ remains qualitatively close to the low-pressure regime, while $p_c=5\times10^{-4}~M_\odot/{\rm km}^3$ shows a much richer dependence on $\phi_c$. For small $\phi_c$, the behavior resembles the high-pressure limit; for intermediate $\phi_c$, the DEF profiles show a more intricate dependence on $\mu^2$; and for the largest $\phi_c$, the system again approaches the qualitative behavior of the low-pressure regime. This indicates that the transition between the two limiting cases occurs mainly in the intermediate-pressure sector, where the effect of $\mu^2$ becomes strongly coupled to the central bosonic amplitude.

%%%%%%%%%%%%%%%%%%%%%%%%%%%%%%%%%%%%%%%%%%%%%%%%%%%%%%%%%%%%
\subsubsection{Extent of matter components}
\label{sec:extent}
%%%%%%%%%%%%%%%%%%%%%%%%%%%%%%%%%%%%%%%%%%%%%%%%%%%%%%%%%%%%
Regarding the fermionic radius in the low-$p_c$ regime and fixed $\mu^2$, increasing $\phi_c$ decreases $\RF$. The bosonic matter therefore affects not only the amplitude and radial distribution of the DEF scalar field, but also the size of the fluid component. At sufficiently large $p_c$, however, this behavior becomes non-monotonic. For example, in the case of $p_c=8\times10^{-4}~M_\odot/{\rm km}^3$, we find
\begin{equation}
    \RF|_{\phi_c=0.15}<\RF|_{\phi_c=0.05}<\RF|_{\phi_c=0.1}.
\end{equation}
Thus, although the largest value of $\phi_c$ still produces the smallest fermionic radius, the intermediate bosonic amplitude yields the largest one. This non-monotonic behavior indicates that, in the high-$p_c$ regime, the fluid's response to the bosonic field cannot be described by a simple one-to-one dependence on $\phi_c$.
Similarly, at low pressures and fixed $\mu^2$, increasing $\phi_c$ reduces $\RB$, but as pressure increases this trend is no longer monotonic.
In contrast, at fixed $\phi_c$, increasing $\mu^2$ seems to generically reduce $\RB$, regardless of the pressure regime.

%%%%%%%%%%%%%%%%%%%%%%%%%%%%%%%%%%%%%%%%%%%%%%%%%%%%%%%%%%%%
\subsubsection{Bosonic core vs. Bosonic halo}
\label{sec:core_vs_halo}
%%%%%%%%%%%%%%%%%%%%%%%%%%%%%%%%%%%%%%%%%%%%%%%%%%%%%%%%%%%%
Interesting features emerge when comparing bosonic and fermionic radii across different pressure regimes. Again, no universal trend is found. We start with the lowest-pressure configuration, $p_c=1\times10^{-4}~M_\odot/{\rm km}^3$, and use it as a reference for the remaining cases. As shown in the upper-left panel of Fig.~\ref{fig:mu_varphi_set1} for FBSs, bosonic radii are always larger than fermionic ones,
\begin{equation}
    \RB > \RF.
\end{equation}
For fixed $\phi_c$, the two radii decrease as $\mu^2$ increases, but they tend to approach each other. Increasing $\phi_c$ does not eliminate this hierarchy. The physical picture is therefore clear in this pressure regime: \emph{the fermionic fluid is contained within the bosonic matter}, thus making a halo-like configuration.

Turning now to the upper-right panel of Fig.~\ref{fig:mu_varphi_set1}, we find a qualitatively different interplay between the fermionic and bosonic radii. In particular, for the first subcase with $\phi_c=0.05$, the relative order of the two radii changes markedly with $\mu^2$. For $\mu^2=0.5$, the configuration remains halo-like. At $\mu^2=1$, the two radii become nearly identical,
\begin{equation}
    \RB \simeq \RF,
\end{equation}
indicating that both matter components occupy approximately the same spatial region. By contrast, for $\mu^2=1.5$, the hierarchy is reversed, and the system develops a bosonic-core configuration in which the bosonic matter is contained inside the fermionic fluid.
This sequence is particularly interesting because it shows that, at fixed $\phi_c$, varying the boson mass can produce three qualitatively distinct spatial arrangements: a halo-like configuration, an approximately coextensive one, and a bosonic-core configuration. However, this behavior is not preserved as $\phi_c$ increases. For $\phi_c>0.05$, the system returns to the qualitative hierarchy found in the lower-pressure case $p_c=1\times10^{-4}~M_\odot/{\rm km}^3$, with the bosonic distribution again extending beyond the fermionic surface. This again illustrates that the relative spatial organization of the two matter sectors is controlled by a non-trivial interplay between $\phi_c$, $\mu^2$, and the underlying fermionic configuration.

The lower-left panel, corresponding to $p_c=5\times10^{-4}~M_\odot/{\rm km}^3$, again exhibits a highly non-trivial spatial distribution of the two matter components. For the smallest central bosonic amplitude, the bosonic matter is confined within the fermionic fluid. In the intermediate subplot, the ordering of the radii for different values of $\mu^2$ is qualitatively preserved, although the bosonic radii move progressively closer to the fermionic surface. This indicates that the two matter components become increasingly coextensive as the central bosonic amplitude grows. For the largest value of $\phi_c$, the situation changes. The fermionic component contracts and the bosonic radius becomes larger than the fermionic one. The configuration therefore transitions from a bosonic core-like to a halo-like structure. This is another example of the non-monotonic interplay among $p_c$, $\phi_c$, and $\mu^2$, showing that the same central pressure can support qualitatively different matter distributions depending on the bosonic parameters.

Finally, the lower-right panel reveals a behavior opposite to the one in the upper-left panel. In this high-pressure regime, the bosonic component remains contained within the fermionic matter, so that
\begin{equation}
    \RB < \RF
\end{equation}
for all the considered values of $\phi_c$. \emph{The bosonic matter is therefore entirely contained within the fermionic fluid}, so the system consistently exhibits a bosonic-core structure in this pressure regime.

Together, these results suggest a broader qualitative picture. Low-pressure fermionic configurations are generally more prone to develop extended bosonic halos, whereas sufficiently high central pressures favor solutions in which the bosonic matter remains confined inside the fermionic fluid. Intermediate-pressure regimes are considerably more intricate, since the relative ordering of the fermionic and bosonic radii can change depending on both $\phi_c$ and $\mu^2$. The transition between halo-like and core-like configurations should therefore not be associated with a single parameter alone.

%%%%%%%%%%%%%%%%%%%%%%%%%%%%%%%%%%%%%%%%%%%%%%%%%%%%%%%%%%%%
%%%%%%%%%%%%%%%%%%%%%%%%%%%%%%%%%%%%%%%%%%%%%%%%%%%%%%%%%%%%
\subsection{Sequences of solutions}
\label{sec:sequences_of_solutions}
%%%%%%%%%%%%%%%%%%%%%%%%%%%%%%%%%%%%%%%%%%%%%%%%%%%%%%%%%%%%
%%%%%%%%%%%%%%%%%%%%%%%%%%%%%%%%%%%%%%%%%%%%%%%%%%%%%%%%%%%%
We now investigate how scalarization modifies sequences of FBS solutions. As before, $\mu^2\in\{0.5,~1.0,~1.5\}$. For each chosen value of $\mu^2$, we obtain the corresponding mass-radius diagram by varying two central quantities: the fermionic central pressure, $p_c$, or equivalently the central rest-mass density $\rho_{0c}$, and the central bosonic-field amplitude, $\phi_c$.\footnote{Some studies parametrize scalarized solutions with the maximum value of the DEF field, $\varphi_{\rm max}$, as it does not always coincides with the central value $\varphi_c$ (see examples in Fig.~\ref{fig:mu_varphi_set1}). Both choices are physically meaningful, but they emphasize slightly different aspects of the solution. We use $\varphi_c$, since it is the parameter that defines the equilibrium configuration. Nevertheless, for selected subsets of data, in Appendix~\ref{app:varphimax} we compare behaviors of $\varphi_{\rm max}$ and $\varphi_c$, and we find that they follow similar qualitative trends.}
More explicitly, we consider a set of central pressures
\begin{equation}
    \mathbf{P}_c
    =
    \left[
    p_{c,0},
    p_{c,1},
    \ldots,
    p_{c,n}
    \right],
\end{equation}
together with a set of central bosonic amplitudes
\begin{equation}
    \mathbf{\Phi}_c
    =
    \left[
    \phi_{c,0},
    \phi_{c,1},
    \ldots,
    \phi_{c,m}
    \right],
\end{equation}
where $m$ and $n$ need not be equal. The full set of initial conditions is therefore given by the Cartesian product
\begin{equation}
   \mathbf{M}= \mathbf{P}_c\times\mathbf{\Phi}_c,
\end{equation}
so that, for each value of $p_c$, we obtain a sequence of solutions by scanning the selected values of $\phi_c$. We employ a grid of  $n=1~250$ and $m=250$, which leads to $\dim\mathbf{M}=312~500$.

For each fixed boson mass $\mu^2$, we generate a two-parameter family of equilibrium configurations spanning an extended region of the solution space. This differs fundamentally from the familiar pure NS or pure BS cases, where varying a single central parameter produces a one-dimensional mass-radius curve. Mixed FBS configurations instead depend simultaneously on $p_c$ and $\phi_c$, producing a mass-radius surface. As before, our coupling parameter is fixed to $\beta=-7$.

%%%%%%%%%%%%%%%%%%%%%%%%%%%%%%%%%%%%%%%%%%%%%%%%%%%%%%%%%%%%
\subsubsection{Radius of a Fermion-Boson Star}
\label{sec:radius_definition}
%%%%%%%%%%%%%%%%%%%%%%%%%%%%%%%%%%%%%%%%%%%%%%%%%%%%%%%%%%%%
We define the radius of a FBS as
\begin{equation}
\label{eq:R-definition}
R=\max\left(\RF,\RB\right),
\end{equation}
with the fermionic and bosonic radii, $\RF$ and $\RB$, defined previously in sections~\ref{sec:Fermionic sector} and~\ref{sec:Bosonic sector}, respectively. The definition~\eqref{eq:R-definition} introduces a natural branch change in the mass-radius diagram whenever the dominant spatial extent switches from the fermionic to the bosonic component. As long as $\RF>\RB$, the plotted radius is set by the fermionic surface. However, once $\RB$ exceeds $\RF$, the effective radius is instead determined by the bosonic distribution. From that point onward, the sequence can move toward larger radii even though the fermionic radius continues to decrease. The resulting sequences are thus continuous but not smooth, with the breaking point signaling the transition from configurations of larger fermionic radii to larger bosonic radii. The choice~\eqref{eq:R-definition} differs from prescriptions that use the fermionic radius throughout the sequence, or that define a single $99\%$ radius for the total matter distribution. Instead, our definition follows the physical structure of the mixed configurations. When the fermionic component is the most extended one and the bosonic field is confined within it, the physical size of the object is naturally set by $\RF$. In this regime, the outer bosonic tail contributes negligibly to the overall extent, and imposing a $99\%$ criterion would artificially shift the radius associated with the clearly defined fermionic surface. Conversely, when the bosonic component extends beyond the fermionic matter, using $\RF$ alone would underestimate the size of the configuration, especially in halo-like solutions where the fermionic component may constitute only a compact core embedded within a much more extended bosonic cloud. In that case, $\RB$ provides a more representative measure of the total spatial extent. The prescription in Eq.~\eqref{eq:R-definition} therefore adapts the radius to the component that actually determines the outer size of the mixed object. We have verified that this choice does not introduce significant quantitative differences relative to a total-$99\%$-radius prescription: for the observables considered here, deviations are generally below $1\%$.

%%%%%%%%%%%%%%%%%%%%%%%%%%%%%%%%%%%%%%%%%%%%%%%%%%%%%%%%%%%%
\subsubsection{Mass-radius surfaces}
%%%%%%%%%%%%%%%%%%%%%%%%%%%%%%%%%%%%%%%%%%%%%%%%%%%%%%%%%%%%

\begin{figure}[htbp]
    \centering
    \includegraphics[width=\columnwidth]{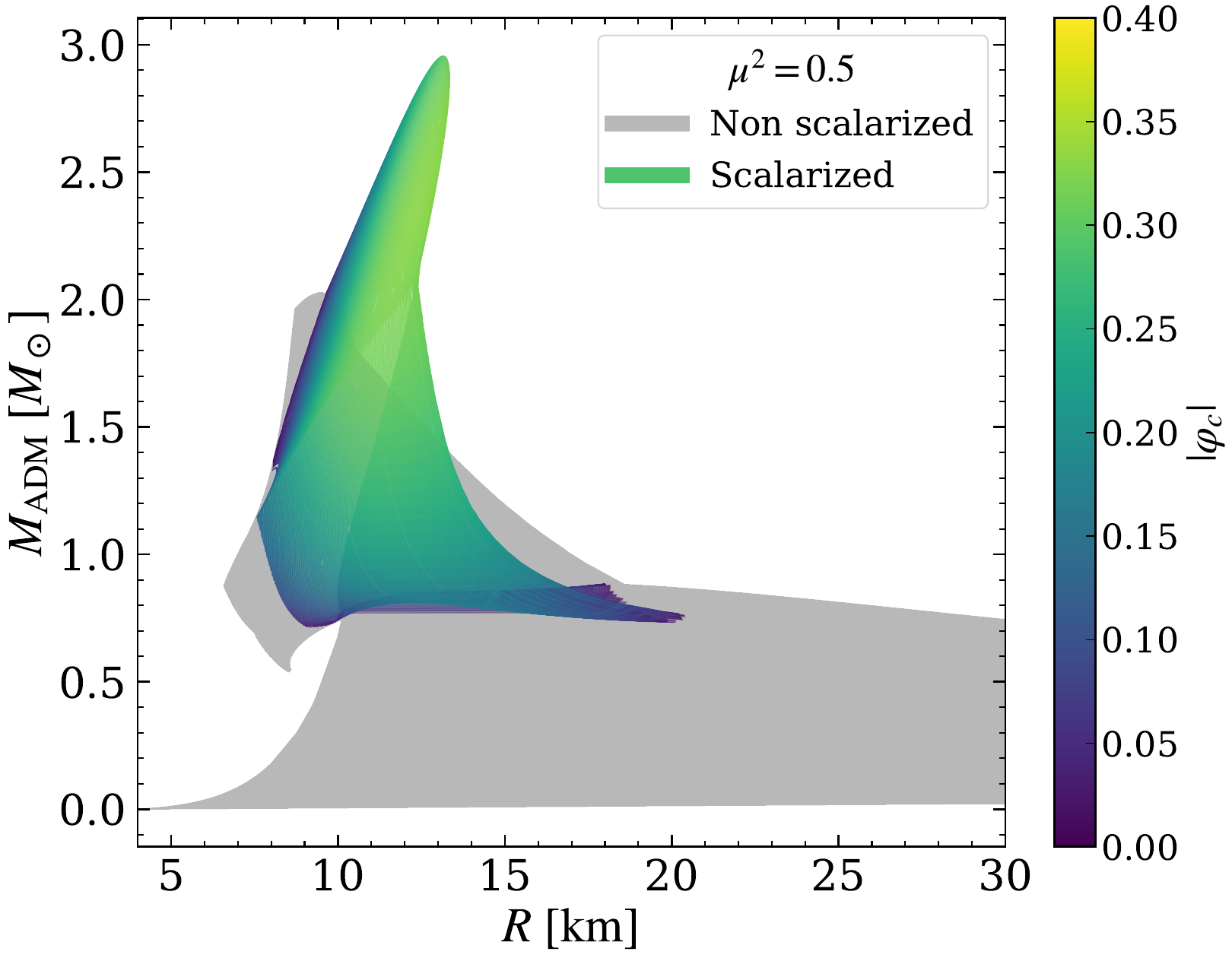}
     \includegraphics[width=\columnwidth]{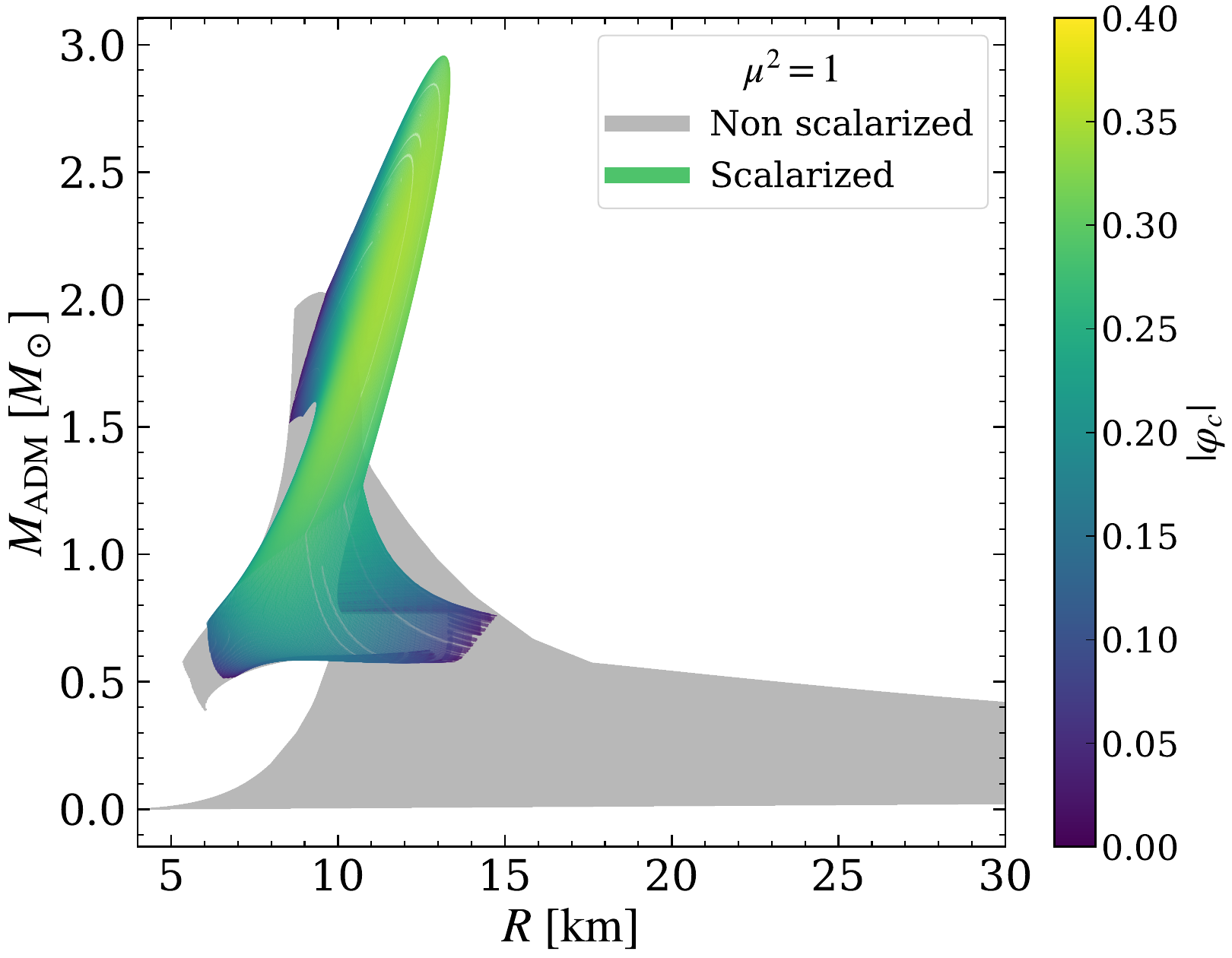}
      \includegraphics[width=\columnwidth]{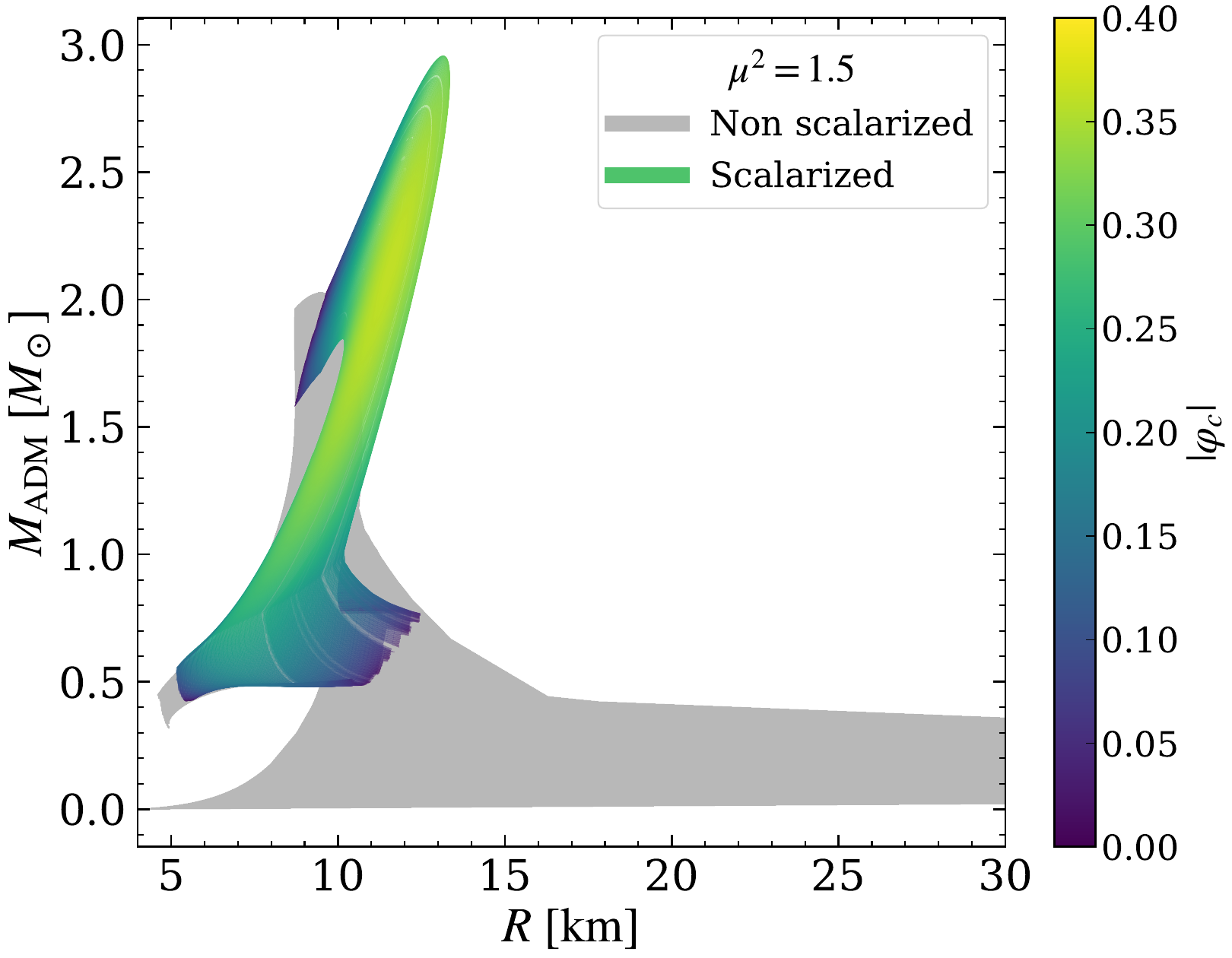}
       
    \caption{Mass-radius surfaces obtained from the initial parameter matrix $\mathbf{M}$. The upper, middle, and lower panels correspond to $\mu^2=0.5$, $1.0$, and $1.5$, respectively. Scalarized solutions are shown with colors parametrized by $|\varphi_c|$, while non-scalarized solutions are shown in gray.}
    \label{fig_MR_SEQ}
\end{figure}

\begin{figure}[htbp]
    \centering
    \includegraphics[width=\columnwidth]{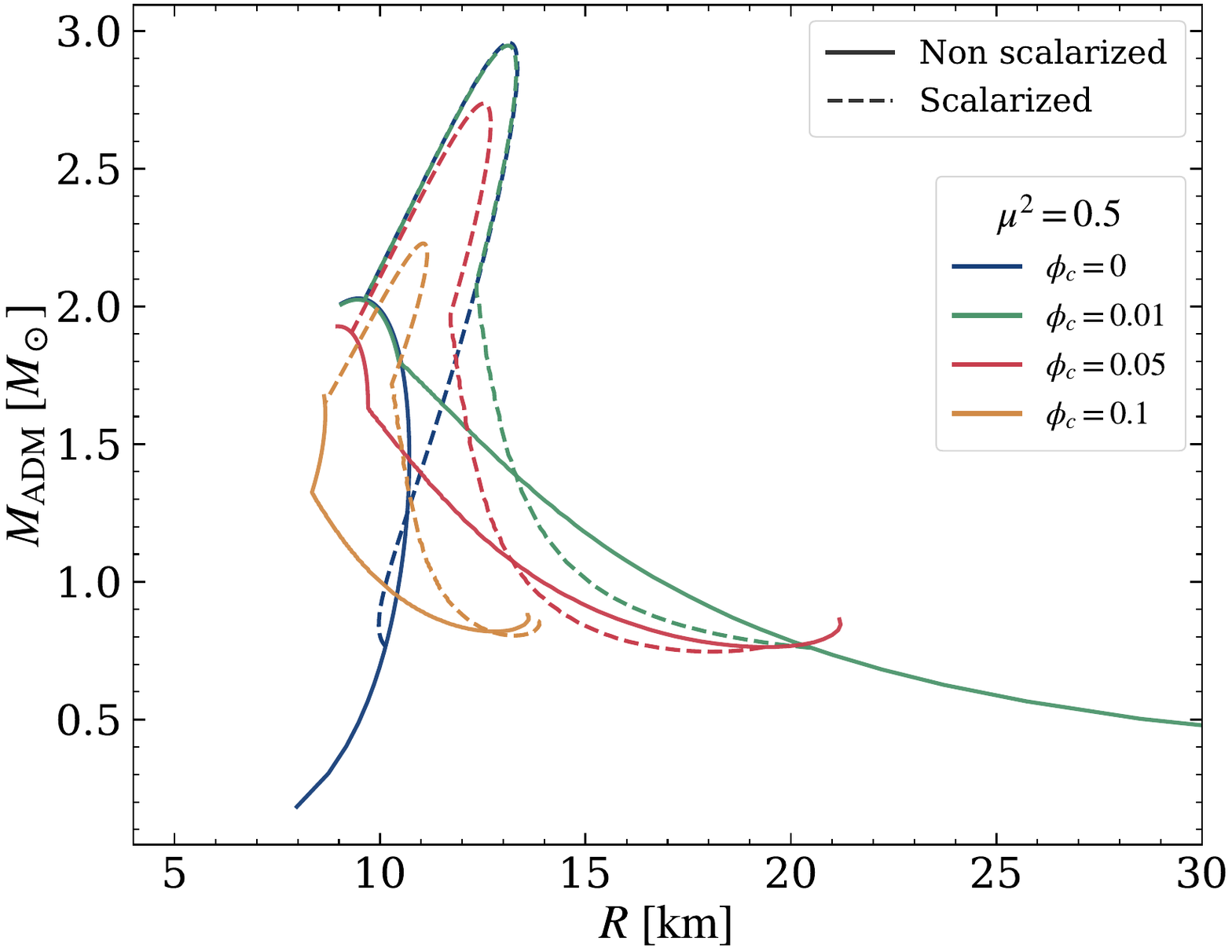}
     \includegraphics[width=\columnwidth]{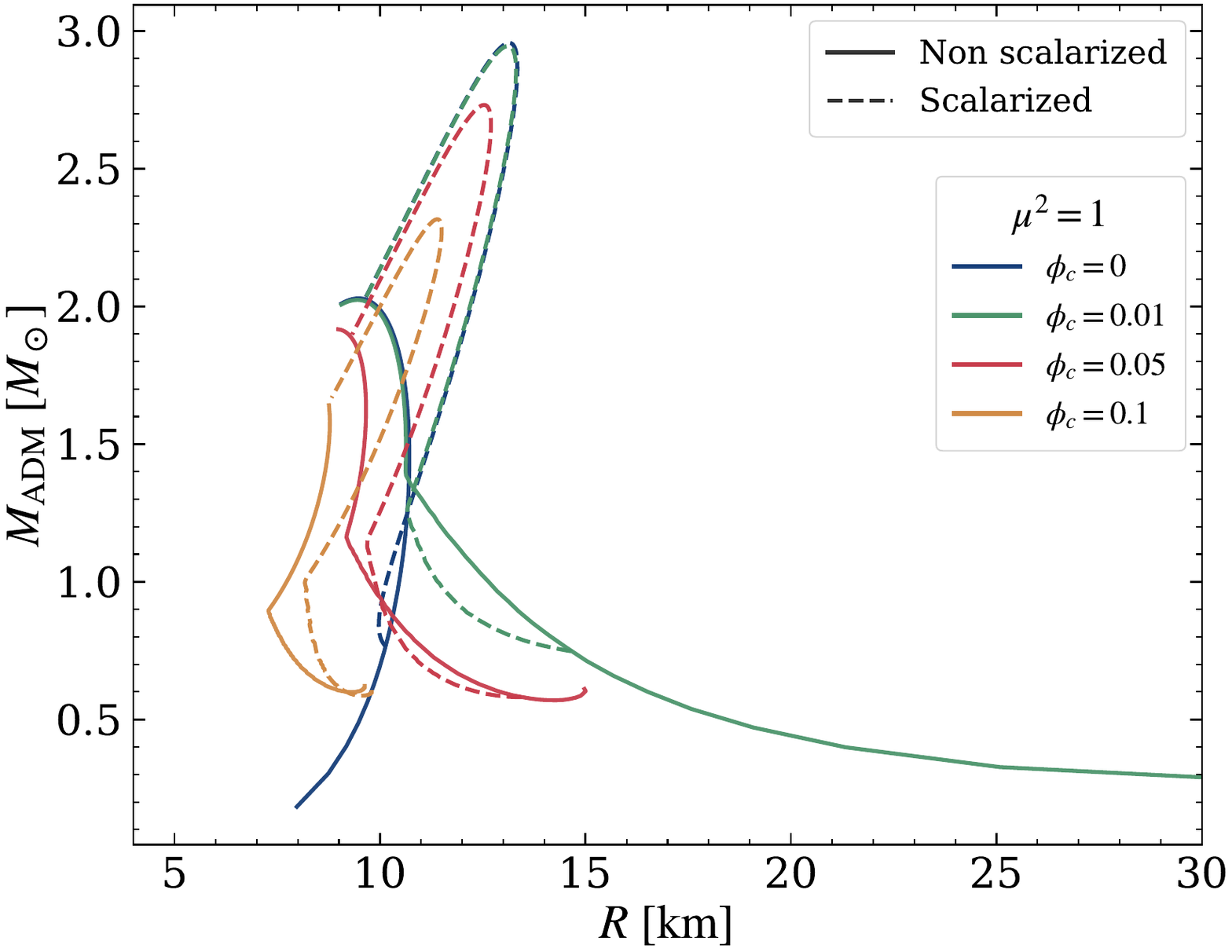}
      \includegraphics[width=\columnwidth]{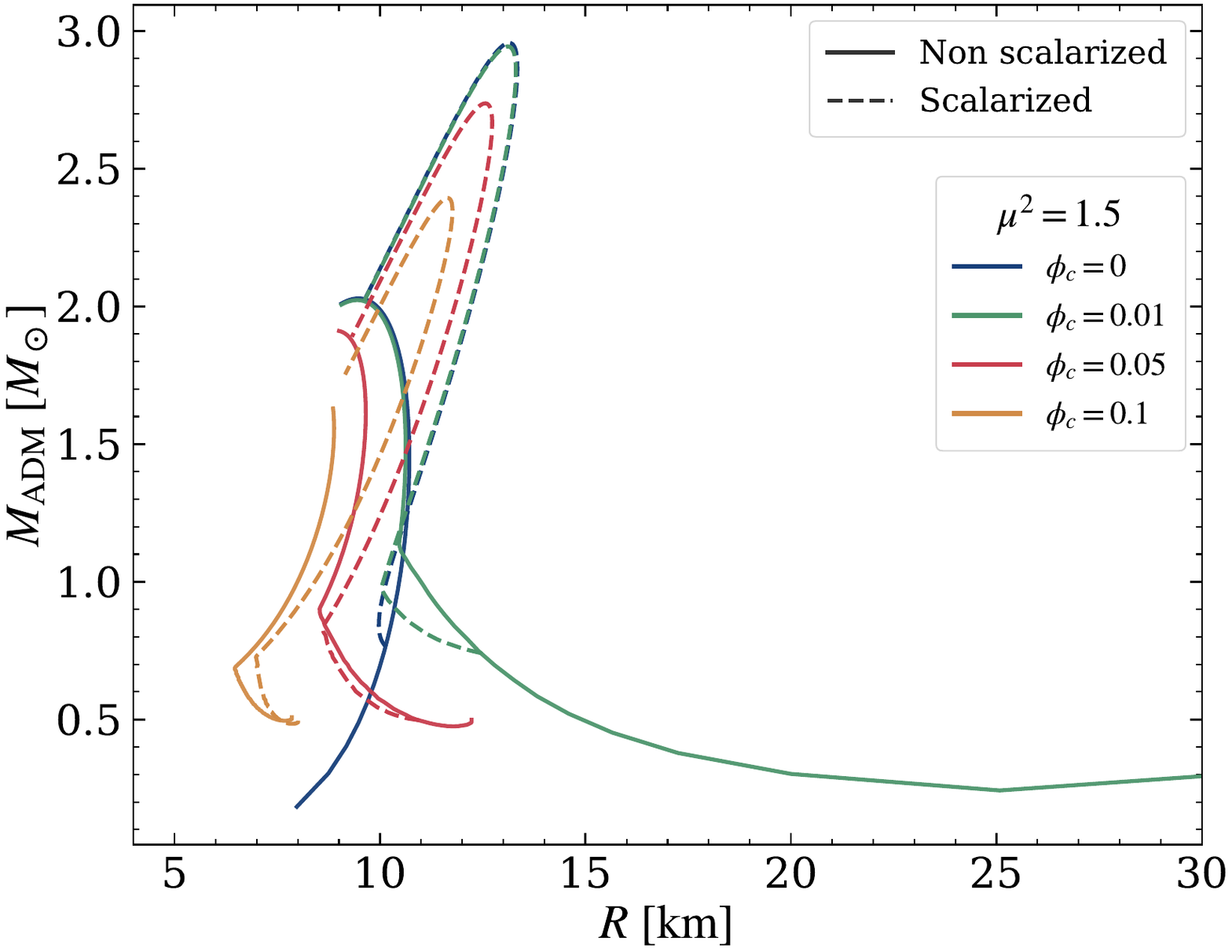}
       
    \caption{Subset of mass-radius curves obtained by varying the full parameter set $\mathbf{P}_c$ for four fixed values of $\phi_c$. The upper, middle, and lower panels correspond to $\mu^2=0.5$, $1.0$, and $1.5$, respectively. Solid (dashed) lines indicate non-scalarized (scalarized) solutions, and each color identifies a fixed $\phi_c$.}
    \label{fig_MR_lineas}
\end{figure}

Figure~\ref{fig_MR_SEQ} shows mass-radius surfaces, each panel corresponding to a different value of the boson mass $\mu^2$. Gray regions represent non-scalarized configurations, whereas colored regions represent scalarized solutions for which the color scale encodes the central value of the gravitational scalar field, $\varphi_c$. The diagrams thus show both the onset and strength of scalarization. Depending on the region of parameter space, some scalarized configurations may have smaller masses than their non-scalarized counterparts, while some extend to substantially higher masses. This qualitative behavior is similar to the one previously identified in scalarized pure BSs (see, e.g., Ref.~\cite{Evstafyeva:2023kfg}).

Starting from the non-scalarized region, the central DEF field grows and solutions progressively move toward the yellow part of the colormap, producing a characteristic ``wing'' in which the departure from non-scalarized solutions is most pronounced. Scalarization does not increase indefinitely: the central DEF field eventually decreases until it vanishes and the scalarized family reconnects with the non-scalarized surface. This behavior is analogous to the case of pure NS and BS sequences when the coupling parameter $\alpha_0=0$.

We now examine how the boson mass modifies mass-radius surfaces by comparing the three panels of Fig.~\ref{fig_MR_SEQ}. Decreasing $\mu^2$ shifts both scalarized and non-scalarized configurations toward larger radii, reflecting the more extended nature of the bosonic component. This trend is already expected from the bosonic sector itself and should not be interpreted as a consequence of scalarization. The color scale, however, reveals that configurations associated with larger values of $\mu^2$ tend to be more strongly scalarized in the sense that the maximum of $\varphi_c$ increases with $\mu^2$. The boson mass also affects the region of the mass-radius plane where scalarization occurs. For higher values of $\mu^2$, the scalarized surface spans a narrower range of both masses and radii.

%%%%%%%%%%%%%%%%%%%%%%%%%%%%%%%%%%%%%%%%%%%%%%%%%%%%%%%%%%%%
\subsubsection{Mass-radius curves}
%%%%%%%%%%%%%%%%%%%%%%%%%%%%%%%%%%%%%%%%%%%%%%%%%%%%%%%%%%%%
In order to examine the sequences of FBS solutions more thoroughly, we now isolate some of the mass-radius curves that produce the surfaces of Fig.~\ref{fig_MR_SEQ}. We retain the three values of $\mu^2$ and continue to vary the fermionic central-pressure set $\mathbf{P}_c$, while fixing $\phi_c$. For each chosen value of $\phi_c$, this procedure generates a one-parameter sequence of mixed solutions containing both scalarized and non-scalarized branches.
We select four representative values of $\phi_c\in\{0,\,0.01,\,0.05,\,0.1\}$ and plot the corresponding mass-radius curves in Fig.~\ref{fig_MR_lineas}. This strategy allows us to follow how mass-radius curves deform as the bosonic field is introduced and strengthened.

The case of $\phi_c=0$ corresponds to the pure NS, which exhibits the well known behavior for a polytropic EoS. The cases of $\phi_c > 0$ progressively increase the bosonic content in mixed configurations. As discussed in the previous subsection, the radial interval over which scalarization develops strongly depends on $\phi_c$.

It should be noted that, for every fixed $\phi_c$, the maximum mass is reached in the branch of scalarized solutions; the largest maximum mass corresponds to the pure NS case, $\phi_c=0$, and it decreases as $\phi_c$ increases.

Next, we notice that mass-radius curves for $\phi_c \ne 0$ are not smooth. This is a subtle consequence of our definition of FBS radius, Eq.~\eqref{eq:R-definition}; see the discussion in Sec.~\ref{sec:radius_definition}.

Yet another feature is that, at large radii, after the sequences reach a minimum mass, they slightly increase as the radius continues to grow. The same qualitative feature is present along the scalarized branches. This growth of the mass with radius starts at smaller radii as $\phi_c$ and $\mu^2$ increase.

It is also noticeable that, for fixed $\phi_c$, increasing $\mu^2$ tends to shorten the radial regime where scalarized solutions exist, while the maximum mass increases. The combined dependence on $\phi_c$ and $\mu^2$ therefore determines not only where scalarization appears along a given mass-radius sequence, but also how strongly the scalarized configurations depart from their non-scalarized counterparts and the range of masses and radii over which this departure persists.

%%%%%%%%%%%%%%%%%%%%%%%%%%%%%%%%%%%%%%%%%%%%%%%%%%%%%%%%%%%%
\subsubsection{Central value of gravitational scalar field}
%%%%%%%%%%%%%%%%%%%%%%%%%%%%%%%%%%%%%%%%%%%%%%%%%%%%%%%%%%%%
We now analyze how the (absolute value of the) central DEF scalar field, $|\varphi_c|$, varies with the central bosonic amplitude $\phi_c$. As discussed previously, one could alternatively analyze $|\varphi_{\rm max}|$. In Appendix~\ref{app:varphimax} we verify that the same qualitative conclusions hold for $|\varphi_{\rm max}|$.

\begin{figure}[h!]
    \centering
    \includegraphics[width=\columnwidth]{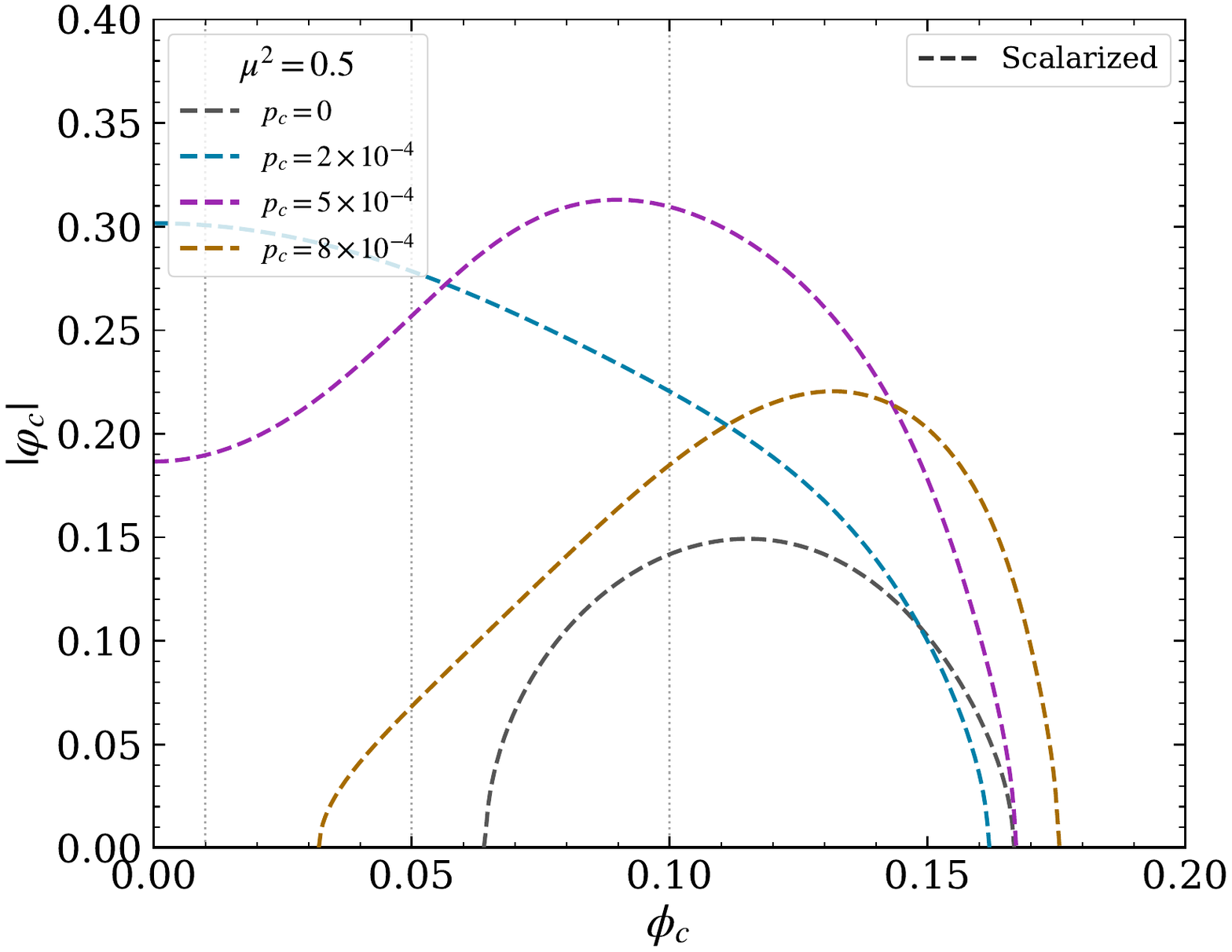}
     \includegraphics[width=\columnwidth]{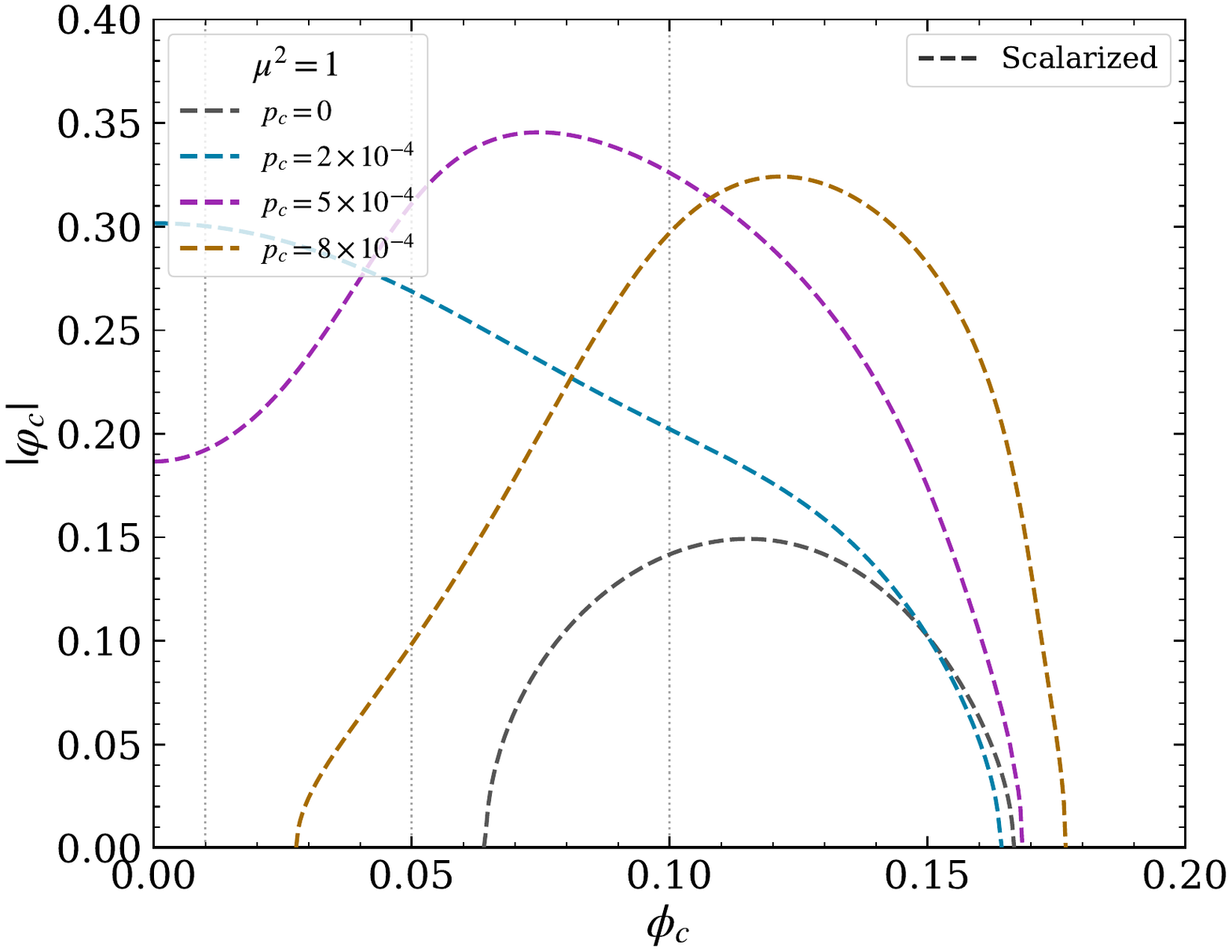}
      \includegraphics[width=\columnwidth]{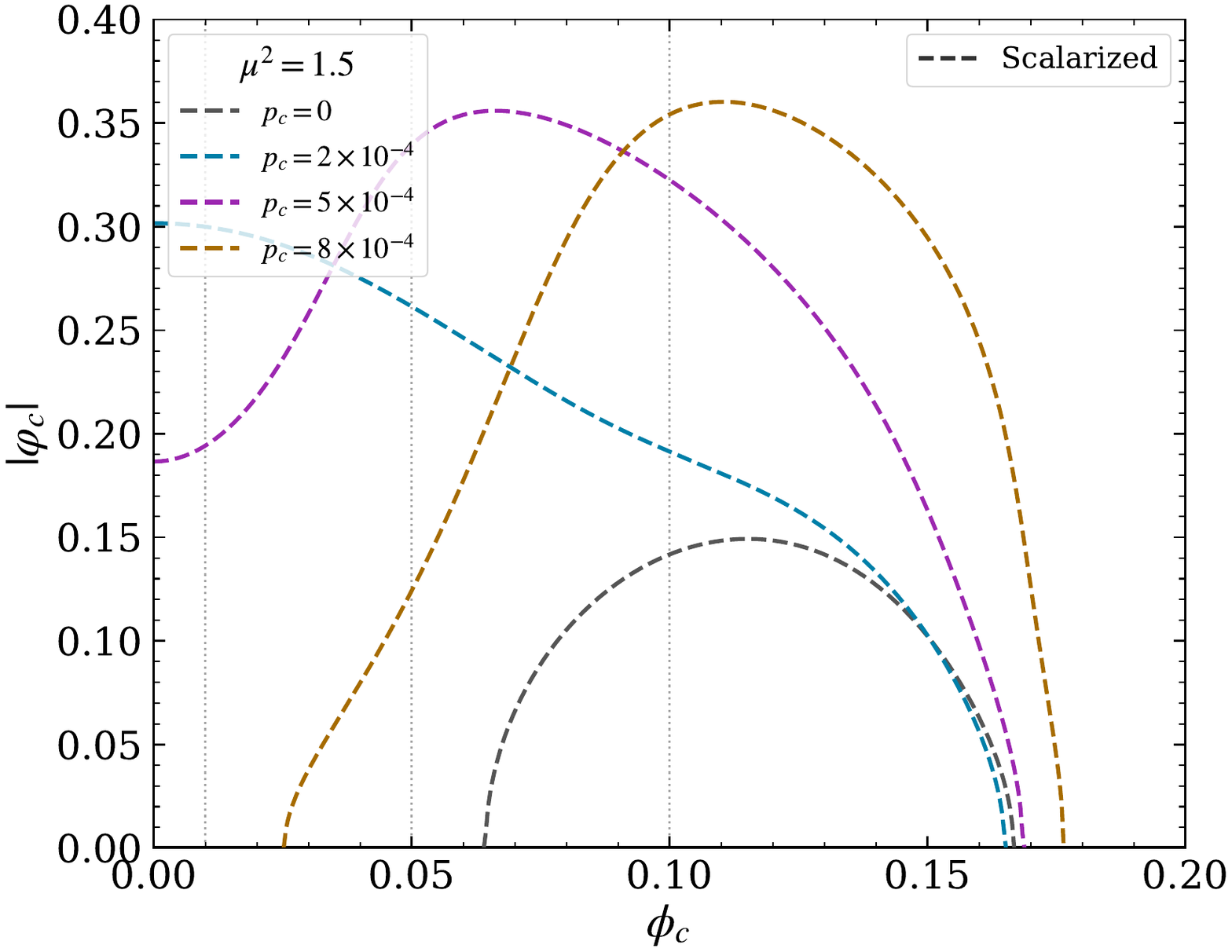}
       
    \caption{Central value of the DEF scalar field, $|\varphi_c|$, as a function of central bosonic field, $\phi_c$, for four central pressures, $p_c$. The upper, middle, and lower panels correspond to $\mu^2=0.5$, $1.0$, and $1.5$, respectively. Dashed colored lines identify scalarized solutions for different fixed-$p_c$ families. Vertical gray lines indicate $\phi_c = 0.01$, 0.05, and 0.1. } 
    \label{fig_varphi_phi}
\end{figure}

For our purpose, we scan the array $\mathbf{\Phi_c}$ for several fixed values of the fermionic central pressure. For each of the three representative values of $\mu^2$, we consider $p_c\in\{0,\,2,\,5,\,8\}\times10^{-4}~M_\odot/{\rm km}^3$. Figure~\ref{fig_varphi_phi} shows how the fermionic pressure modifies the relation between $|\varphi_c|$ and $\phi_c$, for each fixed value of $\mu^2$. Clearly, pure BSs ($p_c=0$) scalarize only over a narrow interval of $\phi_c$ compared to the mixed cases ($p_c>0$). For non-zero values of $p_c$, the domain of scalarization extends and the functional form of $|\varphi_c(\phi_c)|$ is no longer approximately semicircular as in the pure BS case. Instead, different monotonic and non-monotonic patterns appear, with an enhanced maximum of $|\varphi_c|$ occurring either at $\phi_c=0$ or $\phi_c \ne 0$, depending on the fermionic pressure.

The transition from the pure BS limit to increasingly fermionic pressures shows that the fluid component not only changes the strength of scalarization, but also reshapes the domain of $\phi_c$ over which scalarized solutions exist. For high enough $p_c$, for instance, the highest-pressure considered here, there is a minimum and a maximum value of $\phi_c$ for which scalarized solutions exist. 

\begin{figure}[h!]
    \centering
    \includegraphics[width=\columnwidth]{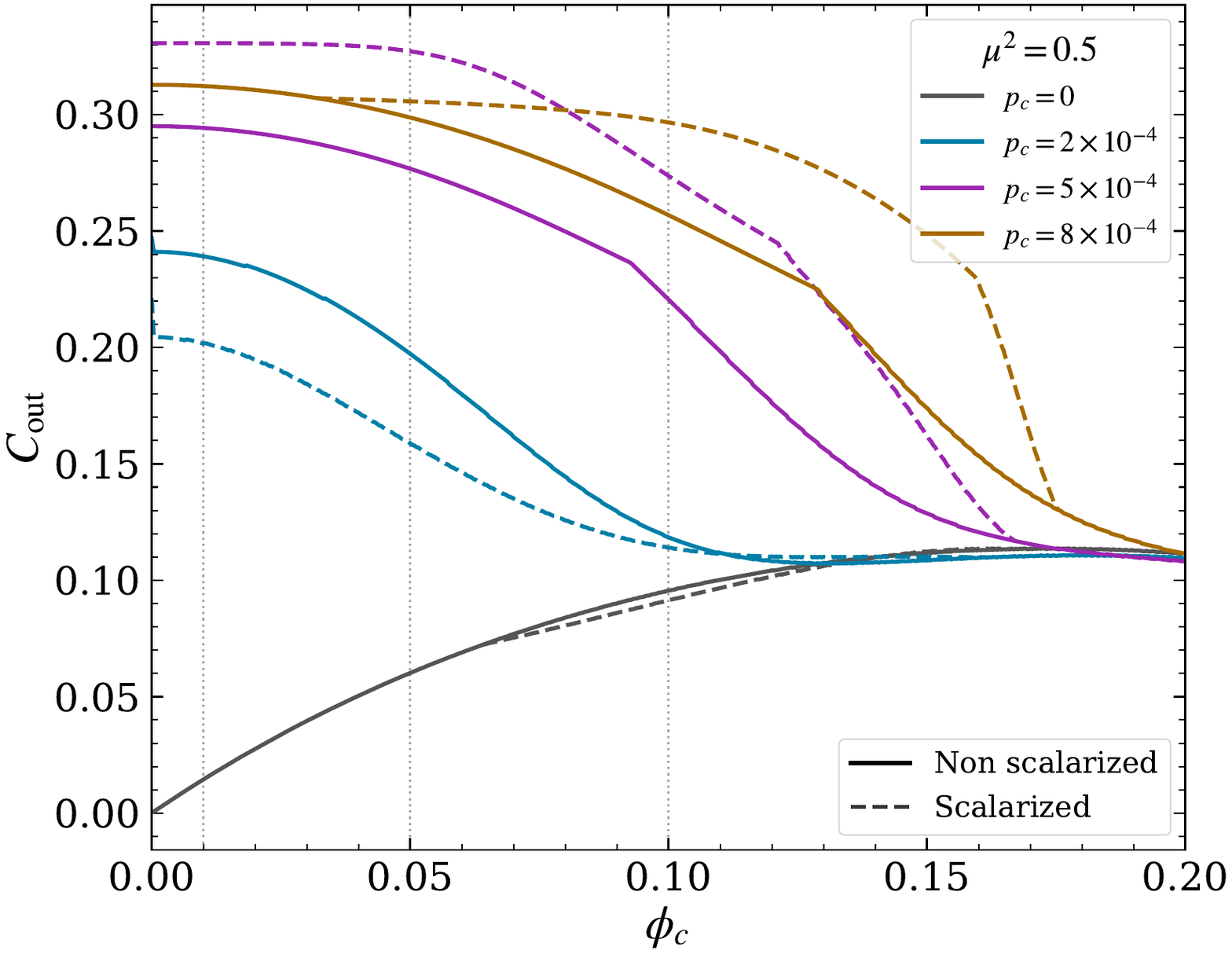}
     \includegraphics[width=\columnwidth]{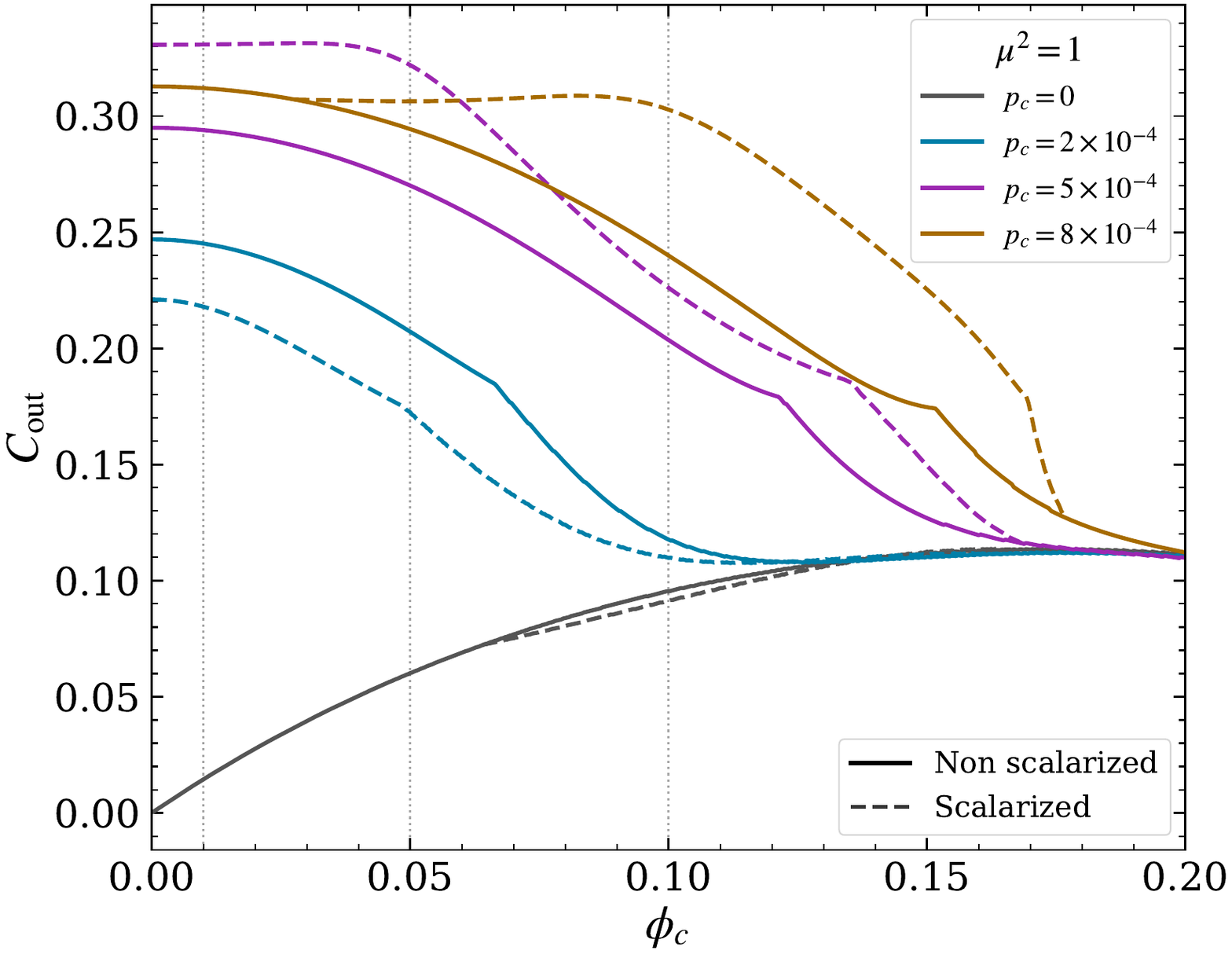}
      \includegraphics[width=\columnwidth]{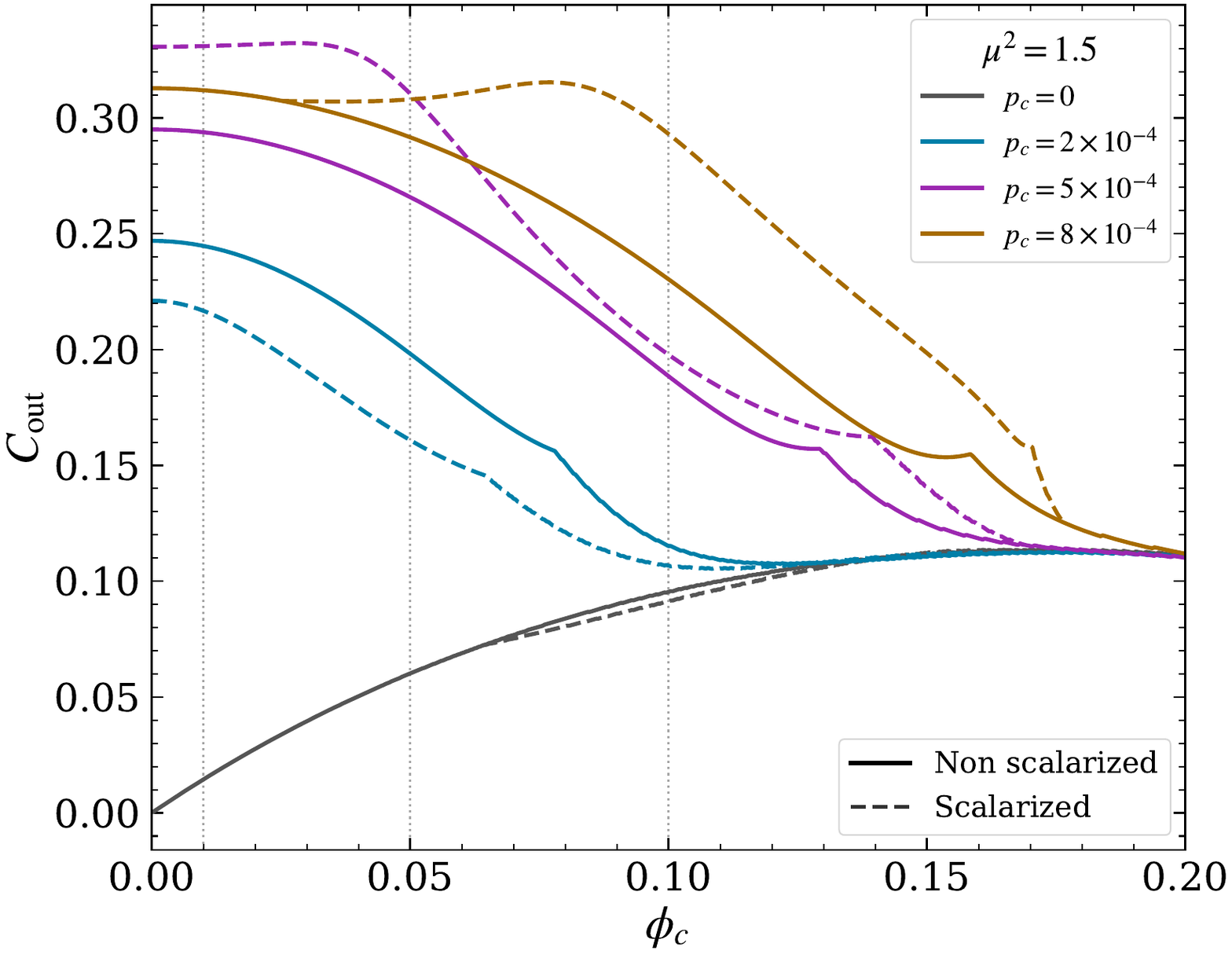}
       
    \caption{Compactness as a function of $\phi_c$. The upper, middle, and lower panels correspond to $\mu^2=0.5$, $1.0$, and $1.5$, respectively. Different colors identify fixed-$p_c$ families, while solid and dashed lines denote non-scalarized and scalarized solutions, respectively. Vertical gray lines correspond to $\phi_c = 0.01$, 0.05, and 0.1.}
    \label{fig_compact}
\end{figure}

Regarding the boundaries of the scalarization domain in $\phi_c$, we observe that the minimum value of $\phi_c$ for which scalarization occurs can vary substantially across families, while the corresponding maximum values remain comparatively close. Although this upper limit depends on $p_c$, among all families considered, the high-pressure case ultimately supports the largest $\phi_c$ for which scalarization persists.

Finally, increasing the boson mass generally shifts the maximum allowed $|\varphi_c|$ toward larger values, but this effect depends on the fermionic pressure. For the lower-$p_c$ families, the displacement is comparatively small, whereas it becomes increasingly pronounced as $p_c$ grows. Also, for every fixed $p_c$, the range of scalarized solutions slightly increases with $\mu^2$.

%%%%%%%%%%%%%%%%%%%%%%%%%%%%%%%%%%%%%%%%%%%%%%%%%%%%%%%%%%%%
\subsubsection{Compactness}
%%%%%%%%%%%%%%%%%%%%%%%%%%%%%%%%%%%%%%%%%%%%%%%%%%%%%%%%%%%%
We define compactness as
\begin{equation}
C_{\rm out}=
\frac{G M_{\rm ADM}}{c^2 R},
\label{compactness}
\end{equation}
where, as before, $R=\max\left(\RF,\RB\right)$.
Figure~\ref{fig_compact} shows $C_{\rm out}$ as a function of $\phi_c$ for several fixed-pressure families, with the three panels corresponding to the same reference values of $\mu^2$ as before. A number of qualitative features are common to all three cases. First, pure BSs ($p_c=0$) systematically feature the lowest compactness. Adding a fermionic component ($p_c>0$) generally increases the compactness, although the dependence on $p_c$ and $\phi_c$ is highly non-trivial. At sufficiently large $\phi_c$, the different pressure families tend to approach the pure BS compactness.

The pure BS family provides the simplest behavior. The scalarized branch initially has lower compactness and eventually becomes slightly more compact before reconnecting with the non-scalarized sequence. The maximum compactness, $\simeq 0.115$, is reached by a non-scalarized solution at about $\phi_c \simeq 0.175$.

The mixed families display a richer structure. As in the mass-radius diagrams, the curves are not smooth because the effective radius changes from $\RF$ to $\RB$. For $p_c=2\times10^{-4}~M_\odot/{\rm km}^3$, the compactness at fixed $\phi_c$ is lower than in higher-pressure cases. At $\phi_c=0$, the non-scalarized solution is more compact than the scalarized one, and as $\phi_c$ increases, both branches decrease and eventually intersect. Beyond this crossing, the scalarized configurations become slightly more compact, although the difference remains small, and the two branches eventually merge again at larger $\phi_c$.

The family with $p_c=5\times10^{-4}~M_\odot/{\rm km}^3$ exhibits the largest compactness, $\simeq 0.33$, among the cases considered. Interestingly, the ordering between scalarized and non-scalarized solutions is now reversed with respect to the lower-pressure family. For small and intermediate values of $\phi_c$, the scalarized branch is systematically more compact than the corresponding non-scalarized branch. As $\phi_c$ increases further, the two branches progressively approach each other and eventually become indistinguishable over the high-$\phi_c$ region.

Finally, the $p_c=8\times10^{-4}~M_\odot/{\rm km}^3$ family exhibits an even more intricate interplay between scalarization and compactness, confirming that, once the fermionic pressure becomes sufficiently large, the effect of scalarization on compactness can no longer be described by the simple patterns observed in the lower-pressure families. As discussed previously, in the case of $p_c=8\times10^{-4}~M_\odot/{\rm km}^3$, scalarization is absent at low values of $\phi_c$, so the sequence initially follows only the non-scalarized branch, starting from a relatively large value of $C_{\rm out}$. 
When the two branches separate, non-scalarized configurations remain systematically less compact, while the scalarized configurations keep high compactness over a broader interval of $\phi_c$. Only at sufficiently large $\phi_c$ does the scalarized branch drop and reconnect with the non-scalarized sequence.

Varying $\mu^2$ while keeping the remaining parameters fixed introduces additional differences among the various families of solutions. For $p_c=0$ and $p_c=2\times10^{-4}~M_\odot/{\rm km}^3$, these differences are comparatively small. This may seem surprising, since changing $\mu$ modifies the characteristic length scale of the bosonic field and, consequently, its spatial distribution. However, for a pure BS with the quadratic potential considered here, mass and radius satisfy approximate scaling relations of the form
\begin{equation}
M_{\rm ADM}\propto\mu^{-1},
\qquad
\RB\propto\mu^{-1}.
\label{eq:scaling_with_mu}
\end{equation}
The leading dependence on $\mu$ therefore cancels in the compactness \eqref{compactness}, making the pure-BS compactness nearly invariant under changes in the boson mass. At $p_c=2\times10^{-4}~M_\odot/{\rm km}^3$, the configurations remain predominantly bosonic and lie mainly in the halo regime, $\RB>\RF$; hence, their total mass and outer radius are still largely controlled by the bosonic sector, and the comparatively small fermionic core only weakly breaks the pure-BS scaling. The small residual differences visible in the numerical sequences may also reflect the finite integration domain and the numerical extraction of the radius containing $99\%$ of the bosonic Noether charge.

The scaling symmetry~\eqref{eq:scaling_with_mu} is broken in FBSs because the fermionic sector introduces an additional physical scale through the EOS. In particular, $\RF$ does not rescale with $\mu$ in the same way as $\RB$. Consequently, the dependence of $C_{\rm out}$ on $\mu$ becomes increasingly apparent in configurations with a substantial fermionic component.

%%%%%%%%%%%%%%%%%%%%%%%%%%%%%%%%%%%%%%%%%%%%%%%%%%%%%%%%%%%%
%%%%%%%%%%%%%%%%%%%%%%%%%%%%%%%%%%%%%%%%%%%%%%%%%%%%%%%%%%%%
\subsection{Onset of scalarization}
\label{sec:onset_of_scalarization}
%%%%%%%%%%%%%%%%%%%%%%%%%%%%%%%%%%%%%%%%%%%%%%%%%%%%%%%%%%%%
%%%%%%%%%%%%%%%%%%%%%%%%%%%%%%%%%%%%%%%%%%%%%%%%%%%%%%%%%%%%
\begin{figure}[]
    \centering
    \includegraphics[width=\columnwidth]{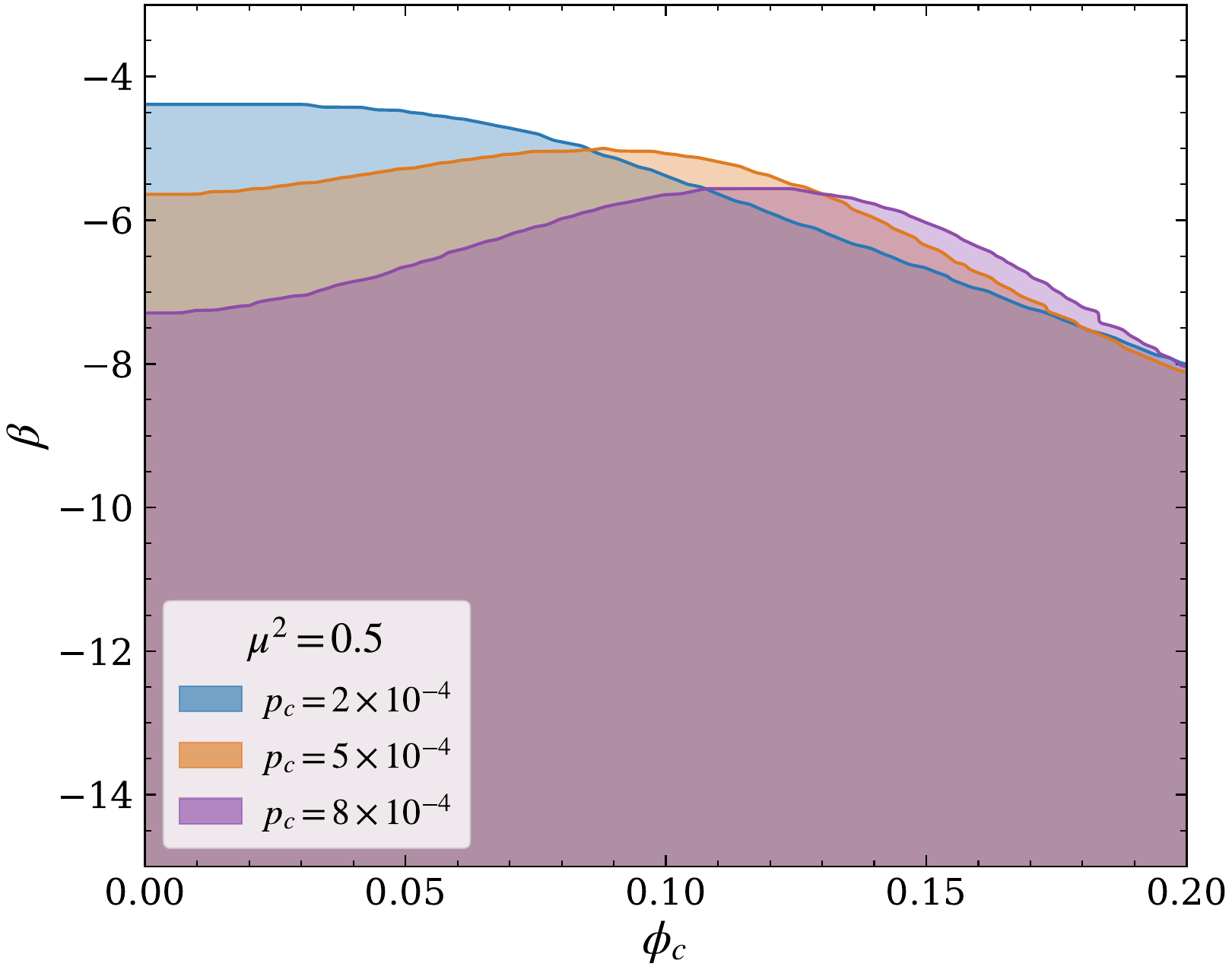}
     \includegraphics[width=\columnwidth]{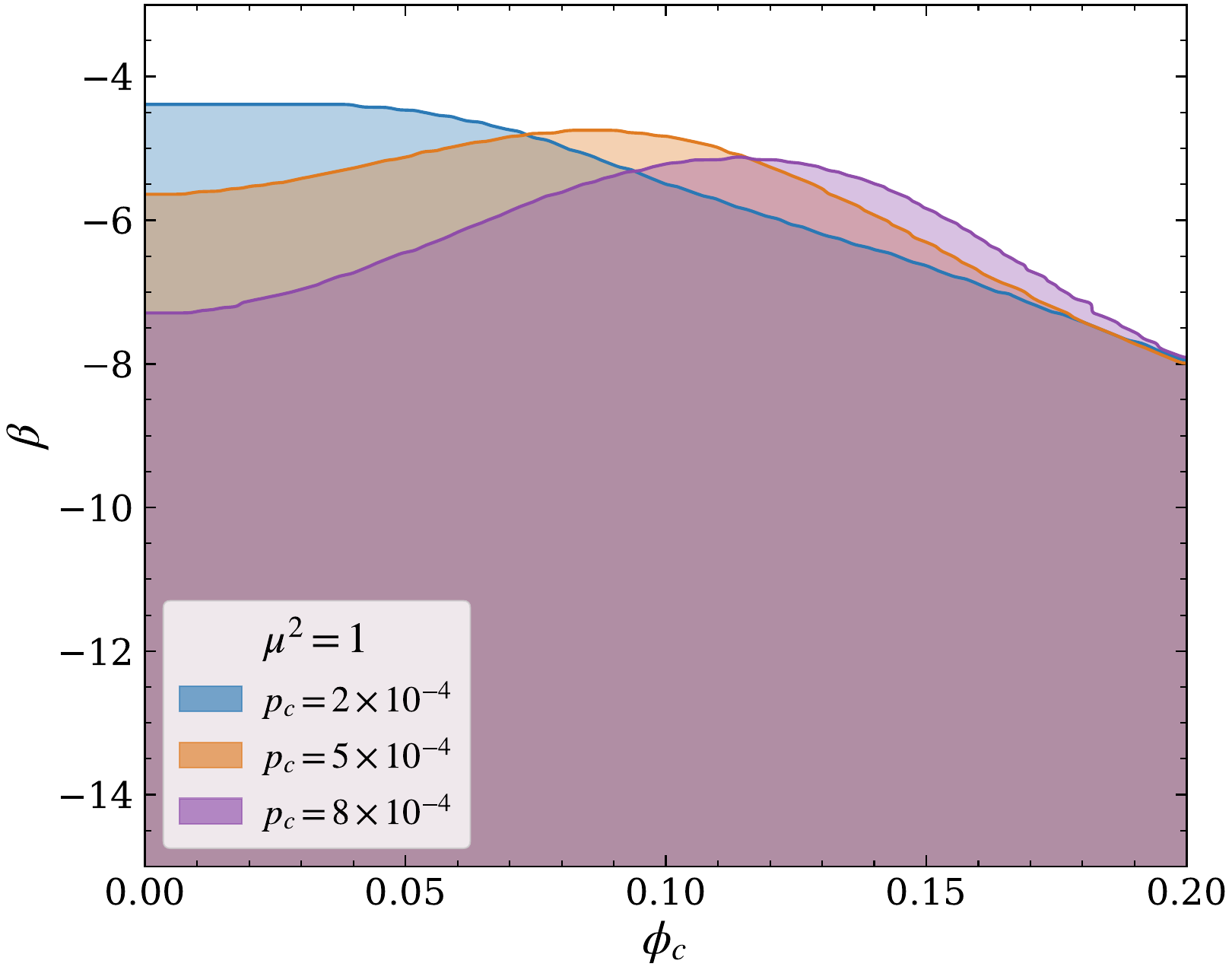}
      \includegraphics[width=\columnwidth]{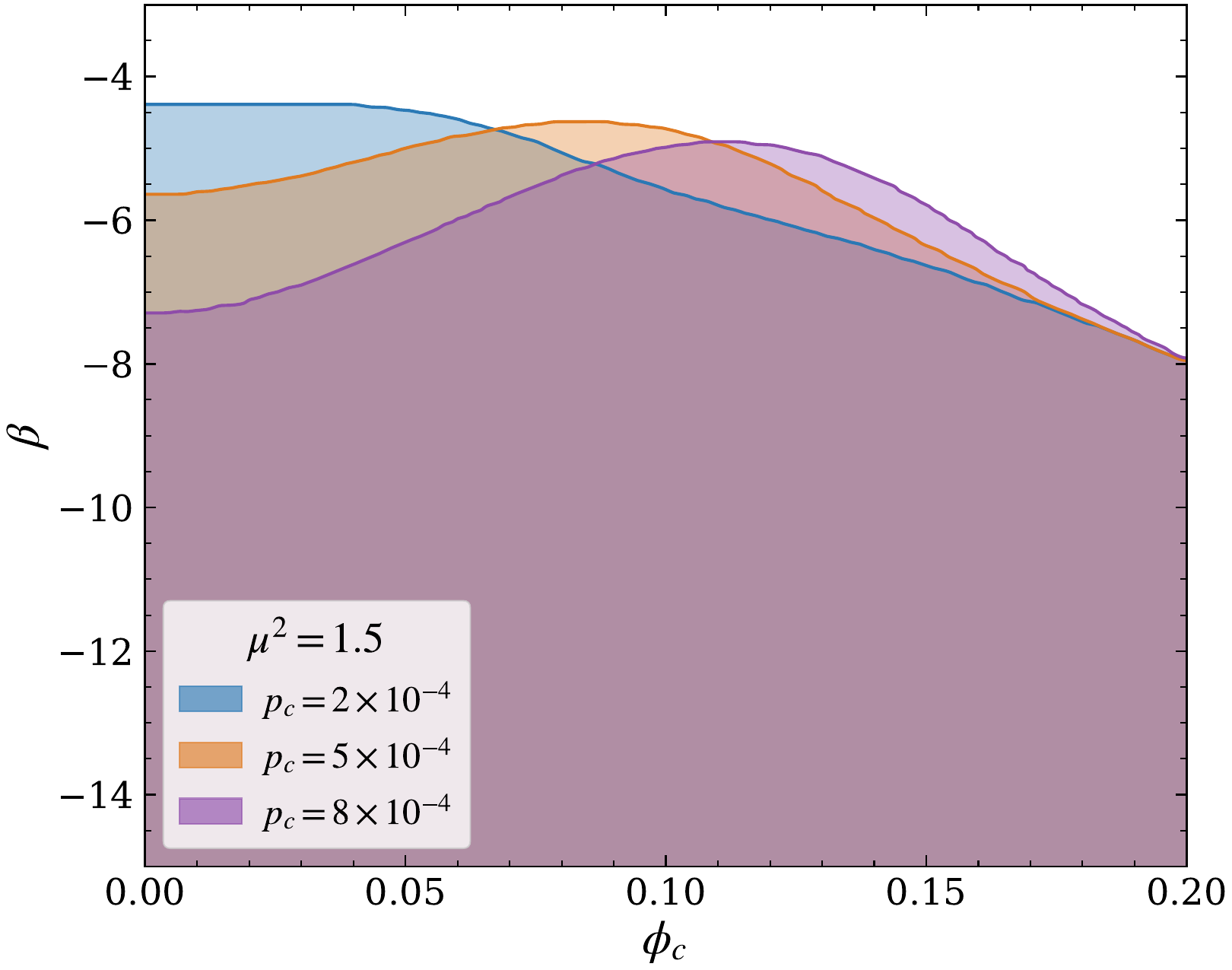}
       
    \caption{Parameter space $\beta \times \phi_c$. Scalarized solutions are found in colored regions corresponding to fixed fermionic central pressures $p_c\in\{2,5,8\}\times10^{-4}~\mathrm{\Msun/km^{3}}$. Their boundaries define critical values for the onset of scalarization, $\beta_{\rm crit}(\phi_c)$. Each subplot corresponds to a fixed boson mass, $\mu^2\in\{0.5,1.0,1.5\}$.}
    \label{fig:onset}
\end{figure}

In this subsection, we investigate the conditions under which FBSs undergo spontaneous scalarization and how the onset of this phenomenon depends on the properties of both the fermionic and bosonic components.
Up to this point, we have examined the roles of $p_c$, $\phi_c$, and $\mu$ in shaping already scalarized configurations, always for a fixed coupling parameter $\beta=-7$. We now extend the analysis by varying $\beta$ to determine how the scalarization threshold depends on $p_c$, $\phi_c$, and $\mu$. We aim to identify which combinations of parameters favor or suppress the development of a nontrivial DEF scalar field. 

To sample different fermionic regimes, we consider three representative values of the central pressure, $p_c \in \{2,5,8\}\times10^{-4}~\mathrm{\Msun/km^{3}}$. For the bosonic sector, we likewise consider three representative values of $\mu^2 \in \{0.5,1.0,1.5\}$, in code units. These combinations define nine distinct frameworks in which we explore the scalarization domain. For each of these nine cases, we scan the two-dimensional parameter space spanned by the DEF coupling and the central bosonic-field amplitude,
\begin{equation}
\beta \in [-15,0],
\qquad
\phi_c \in [0,0.2],
\end{equation}
with a grid of $150\times150$ for the $\beta\times\phi_c$ space.
The interval $\beta \in [-15,0]$ is an arbitrary choice corresponding to a typical regime for the onset of scalarization, whereas the interval $\phi_c \in [0,0.2]$ is sufficiently large to cover most of the existence domain of equilibrium solutions.

Figure~\ref{fig:onset} shows the parameter space $\beta\times\phi_c$, where colored surfaces represent scalarized solutions for fixed values of $p_c$; their boundary defines a critical curve for scalarization, $\beta_{\rm crit}(\phi_c)$. Each panel of Fig.~\ref{fig:onset} corresponds to a different value of $\mu^2$. We have verified that in the pure NS case ($\phi_c=0$), our results coincide with those in Fig.~2 of Ref.~\cite{Mendes:2016fby}.

For the lowest-pressure case, $p_c=2\times10^{-4}~\mathrm{\Msun/km^3}$, the fraction of scalarized solutions is almost independent of $\mu^2$. The percentages of scalarized solutions are $55.30\%$, $55.28\%$, and $55.20\%$ for $\mu^2=0.5$, $1.0$, and $1.5$, respectively. The total relative variation across this range is only about $0.18\%$, which is small enough to be consistent with the scan's numerical resolution. Thus, in this low-pressure regime, changing the boson mass has a negligible effect on the overall size of the scalarized region.

The situation changes for the intermediate-pressure family. For $p_c=5\times10^{-4}~\mathrm{\Msun/km^3}$, the fractions of scalarized solutions increase from $50.39\%$ to $53.29\%$ and $54.11\%$ as $\mu^2$ varies from $0.5$ to $1.0$ and $1.5$. This corresponds to a relative change of approximately $7.38\%$ between the lowest- and highest-$\mu^2$ cases.

This trend is even more pronounced in the high-pressure regime. For $p_c=8\times10^{-4}~\mathrm{\Msun/km^3}$, the scalarized fractions are $39.84\%$, $43.42\%$, and $46.06\%$ for $\mu^2=0.5$, $1.0$, and $1.5$, respectively. The corresponding relative increase is about $15.6\%$, showing that the scalarization domain becomes substantially more sensitive to the boson mass as the fermionic central pressure grows.

A complementary trend emerges when comparing scalarization fractions as $p_c$ varies for a fixed $\mu^2$. From the above percentages, we see that, independent of $\mu^2$, increasing $p_c$ reduces the fraction of scalarized configurations within the parameter range under consideration. 

While increasing the boson mass tends to enlarge the scalarized domain at fixed intermediate or high $p_c$, increasing the fermionic pressure produces the opposite effect. The scalarization threshold is therefore determined by a competition between the fermionic and bosonic sectors rather than by either parameter independently.

We now examine the different threshold curves $\beta_{\rm crit}(\phi_c)$. For the low-pressure case, $\beta_{\rm crit}\simeq-4.409$ remains approximately constant up to $\phi_c\simeq0.04$, a threshold that is almost unchanged across the three values of $\mu^2$. After a short plateau, $\beta_{\rm crit}$ becomes more negative, down to $\beta_{\rm crit}\simeq-8$, with only small variations at the third decimal place depending on the value of $\mu^2$.

For the intermediate-pressure, $p_c=5\times10^{-4}~\mathrm{\Msun/km^{3}}$, for pure NS ($\phi_c=0$) scalarization occurs for $\beta \lesssim -5.658.$ for all three values of $\mu^2$. This critical value no longer corresponds to the maximum $\beta_{\rm crit}$ across the domain of $\phi_c$. As $\phi_c$ increases, $\beta_{\rm crit}$ progressively shifts toward less negative values, reaching a maximum that now depends appreciably on the boson mass. Specifically, for $ \mu^2\in\{0.5,\,1.0,\,1.5\},$ we find $\beta_{\rm crit,~max}\in\{-5.013,\,-4.772,\,-4.651\}$, respectively, around $\phi_c\simeq 0.1$, and then $\beta_{\rm crit}$ decreases down to $\beta_{\rm crit}\in\{-8.154,\,-8.034,\,-7.993\}$ at $\phi_c=0.2$, for $\mu^2\in\{0.5,\,1.0,\,1.5\}$, respectively.

Finally, for the high-pressure, $p_c=8\times10^{-4}~\mathrm{\Msun/km^{3}}$, the behavior of $\beta_{\rm crit}(\phi_c)$ is qualitatively similar to that found for the intermediate-pressure case. At $\phi_c=0$, the onset of scalarization occurs at $\beta_{\rm crit}=-7.309$ for all three values of $\mu^2$. As $\phi_c$ increases, the critical value of $\beta$ shifts toward less negative values, reaching $\beta_{\rm crit, max}\in\{-5.577,\,-5.134,\,-4.933\}$ at approximately $\phi_c\simeq0.12$. Compared with the intermediate-pressure case, the value of $\phi_c$ at which the maximum is reached shifts toward larger bosonic amplitudes. Beyond this maximum, $\beta_{\rm crit}$ decreases down to $\beta_{\rm crit}\in\{-8.074,\,-7.953,\,-7.953\}$ at $\phi_c=0.2$, again for $\mu^2\in\{0.5,\,1.0,\,1.5\}$. 

Overall, Fig.~\ref{fig:onset} shows that the onset of scalarization is a genuinely multiparametric feature that depends strongly on $p_c$, $\phi_c$, $\mu^2$, and $\beta$. Nevertheless, the threshold curves $\beta_{\rm crit}(\phi_c)$ tend to become more monotonic and less sensitive to $\mu^2$ as $p_c$ decreases.

%%%%%%%%%%%%%%%%%%%%%%%%%%%%%%%%%%%%%%%%%%%%%%%%%%%%%%%%%%%%
%%%%%%%%%%%%%%%%%%%%%%%%%%%%%%%%%%%%%%%%%%%%%%%%%%%%%%%%%%%%
%%%%%%%%%%%%%%%%%%%%%%%%%%%%%%%%%%%%%%%%%%%%%%%%%%%%%%%%%%%%
\section{Conclusions}
\label{sec:conclusions}
%%%%%%%%%%%%%%%%%%%%%%%%%%%%%%%%%%%%%%%%%%%%%%%%%%%%%%%%%%%%
%%%%%%%%%%%%%%%%%%%%%%%%%%%%%%%%%%%%%%%%%%%%%%%%%%%%%%%%%%%%
%%%%%%%%%%%%%%%%%%%%%%%%%%%%%%%%%%%%%%%%%%%%%%%%%%%%%%%%%%%%
We have numerically constructed static, spherical FBS models in the context of the DEF scalar-tensor theory of gravity with a massless scalar field $\varphi$ and coupling function $a(\varphi)$. Fermionic matter is described by a polytropic perfect fluid, whereas a canonical, massive complex scalar field describes bosonic matter. The equilibrium problem combines the hydrostatic condition characteristic of NSs, the bosonic eigenfrequency condition as in BSs, and the asymptotic condition of the DEF scalar field. As consistency checks, our numerical implementation successfully recovers previously reported scalarized NSs and BSs. 

A scalarized FBS cannot be understood as a simple superposition of its pure NS and BS limits. Changes in the relative matter content, boson mass, central bosonic amplitude, and DEF coupling can produce substantial and highly nonlinear modifications to the degree of scalarization, internal matter distribution, radii, mass, and compactness. Therefore, no single observable or parameter variation is sufficient to characterize these configurations. For this reason, we have systematically explored the solution space through several complementary approaches and projections, in particular two-dimensional mass-radius surfaces. Each representation isolates a different aspect of the same multidimensional family and, taken together, provides both a global and a detailed picture of these new solutions. Accordingly, we organize the remainder of the conclusions into separate subsections.

%%%%%%%%%%%%%%%%%%%%%%%%%%%%%%%%%%%%%%%%%%%%%%
\subsection{Role of the bosonic field}
%%%%%%%%%%%%%%%%%%%%%%%%%%%%%%%%%%%%%%%%%%%%%%
The bosonic sector can either suppress or enhance scalarization. In the low-pressure regime, increasing the central bosonic amplitude, $\phi_c$, generally reduces $\varphi_c$ and induces a faster decay of the DEF field inside scalarized FBSs. Increasing the boson mass contributes to confining bosonic distributions and makes the DEF-field profiles less extended. In this regime, the fermionic matter is typically confined inside a bosonic halo.

At sufficiently high pressures, the situation differs. Larger $\mu^2$ can contribute to produce less confined DEF scalar-field profiles. However, the effect of $\phi_c$ is not systematic. In this regime, the typical configuration is a bosonic core confined within fermionic matter.

In both regimes, increasing $\phi_c$ can favor the displacement of the maximum of $\varphi(r)$ away from the origin, giving rise to shell-like DEF-field profiles.

%%%%%%%%%%%%%%%%%%%%%%%%%%%%%%%%%%%%%%%%%%%%%%
\subsection{Sequences of solutions}
%%%%%%%%%%%%%%%%%%%%%%%%%%%%%%%%%%%%%%%%%%%%%%
Mass-radius relations of FBSs are multiparametric. We observe that higher $\mu^2$ reduces the allowed radii for scalarized models, but increases the maximum allowed value of $\varphi_c$. Also, scalarization enhances the allowed maximum mass, which is even larger as $\phi_c$ decreases.

Pure BSs scalarize over a comparatively narrow interval of $\phi_c$, while introducing fermionic matter changes the extent of this interval. Moreover, increasing the boson mass can enhance the maximum value of $|\varphi_c|$.

Regarding compactness, non-scalarized solutions can be more compact when fermionic central pressures are low, whereas scalarized configurations are typically more compact at intermediate and high pressures. In our particular intermediate-pressure family, the highest compactness is reached slightly above $\simeq 0.33$ by a scalarized FBS with $\phi_c \simeq 0.03$ and $\mu^2=1.5$.

%%%%%%%%%%%%%%%%%%%%%%%%%%%%%%%%%%%%%%%%%%%%%%
\subsection{Onset of scalarization}
%%%%%%%%%%%%%%%%%%%%%%%%%%%%%%%%%%%%%%%%%%%%%%
The critical coupling parameter for the onset of scalarization, $\beta_{\rm crit}$, depends on the fermionic central pressure $p_c$, the bosonic central amplitude $\phi_c$, and the boson mass $\mu^2$.
At low central pressures, the scalarization domain is comparatively insensitive to $\mu^2$, and its threshold $\beta_{\rm crit}(\phi_c)$ decreases monotonically with $\phi_c$. At intermediate and high pressures, however, the threshold develops a non-monotonic dependence on $\phi_c$. In these regimes, most FBSs can scalarize at $\beta$ less negative than their corresponding pure NS for small and intermediate $\phi_c$.

The trend of Fig.~\ref{fig:onset} suggests that, in massless DEF gravity with the coupling function under consideration, with a polytropic EOS and a free bosonic field, the less negative value of $\beta$ at which a FBS can scalarize should be slightly below $-4.3$, depending on $p_c$. This value seems almost independent of $\mu^2$, and requires low $p_c$ and $\phi_c \lesssim 0.04$, so it makes sense that it approaches the threshold for scalarization of pure NSs, $-4.35$~\cite{Damour:1993hw,Harada:1998ge}.

%%%%%%%%%%%%%%%%%%%%%%%%%%%%%%%%%%%%%%%%%%%%%%%%%%
\subsection{Prospects for future work}
%%%%%%%%%%%%%%%%%%%%%%%%%%%%%%%%%%%%%%%%%%%%%%%%%%
In order to isolate the interplay among fermionic, bosonic, and gravitational scalar sectors, we have adopted a deliberately minimal model: a polytropic fluid, a free bosonic field, a spherically symmetric spacetime, and a massless DEF scalar field. These assumptions made it possible to map a large multidimensional solution space, but each of them should eventually be relaxed.

One of the most immediate questions concerns the stability of scalarized and non-scalarized solutions. Because mixed stars possess two independently conserved matter quantities, the usual one-parameter turning-point criterion cannot be applied without qualification. A complete analysis should therefore examine stability across the multidimensional equilibrium space, extending the methods employed for FBSs in GR~\cite{Valdez-Alvarado:2012rct,DiGiovanni:2020frc,Nyhan:2022pda,NdVCoCMjCOaOn2026}.

A complementary step is a linear perturbation analysis of the coupled fluid, bosonic, metric, and DEF-scalar sectors. Scalar-led perturbations should clarify whether the onset of scalarization is associated with a tachyonic instability~\cite{Harada:1997mr}, while the complete spectrum would determine how bosonic and fermionic oscillation modes mix in DEF gravity.

Another generalization is to endow the gravitational scalar with a nonzero mass, which would replace the long-range scalar tail by a faster decay, contributing to suppress scalar radiation on sufficiently large scales~\cite{PhysRevD.92.124016,Ramazanoglu:2016kul,Yazadjiev:2016pcb,Doneva:2016xmf,Sperhake:2017itk,Staykov:2018hhc,Rosca-Mead:2019seq,Rosca-Mead:2020bzt,Huang:2021tpu,Kuan:2023hrh,Asakawa:2023obq,Degollado:2024oyo,Lam:2024wpq}.

Moreover, rotating scalarized FBSs are expected to possess an even richer structure because their components may not respond identically to angular momentum. Thus, rotation should modify the maximum mass, compactness, scalarization threshold, and relative spatial distribution of the two matter sectors, while introducing additional multipolar structure and potentially new families of solutions. Progress in different fronts can be found in \cite{Doneva:2013qva, Doneva:2016xmf,Doneva:2018ouu,Staykov:2023ose,Mourelle:2024qgo,Cipriani:2025wem,Fakhry:2026cmi}.

Once stable and more realistic families have been identified, it will be valuable to investigate whether their observables satisfy approximate universal relations. Such relations have been extensively studied for compact fluid stars in alternative theories of gravity
\cite{Yagi:2013awa,Yagi:2013sva,Haskell:2013vha,Yagi:2016bkt,Rezzolla:2017aly,Doneva:2017jop,Jiang:2019pzw,Torres-Forne:2019zwz,Pappas:2018csu,Sun:2020qkj,Adam:2020aza,Godzieba:2021vnz,Pretel:2024pem,Yip:2025uzd}.
Similar universality properties have been identified for BSs
\cite{Adam:2022nlq,Vaglio:2022flq,Adam:2023qxj,Adam:2024zqr,Adam:2025ktm} and other exotic objects~\cite{Kumar:2026ddm,Panotopoulos:2026xwj}. It remains to investigate whether scalarized FBSs preserve any of these relations, obey new multicomponent generalizations, or instead introduce violations through their matter components and scalar charge, as happens in \cite{Kuan:2026duo}. Calculating their tidal deformabilities, higher-order multipole moments, moments of inertia, and oscillation spectra would address this question and provide observables that can distinguish them from ordinary NSs, pure BSs, and BHs \cite{Sennett:2017etc,Cardoso:2017cfl,Diedrichs:2023trk, Vaglio:2022flq}.

Finally, a long-term prospect is the fully nonlinear simulation of isolated and binary scalarized FBSs. Such simulations would reveal whether scalarization can be dynamically triggered or suppressed during inspiral and merger, how the two matter components are redistributed, and whether GW signals carry identifiable signatures of the mixed internal structure. Similar studies have already demonstrated the relevance of dynamical scalarization and strong-field effects for other compact objects in scalar-tensor gravity
\cite{Barausse:2012da,Shibata:2013pra,Palenzuela:2013hsa,Mendes:2021fon,Ma:2023sok,Kuan:2023hrh,Lam:2024wpq}.
Extending this program to mixed FBSs would connect the equilibrium phenomenology developed here with observable GW and multimessenger signatures.

%%%%%%%%%%%%%%%%%%%%%%%%%%%%%%%%%%%%%%%%%%%%%%%%%%%%%%%%%%%%
%%%%%%%%%%%%%%%%%%%%%%%%%%%%%%%%%%%%%%%%%%%%%%%%%%%%%%%%%%%%
%%%%%%%%%%%%%%%%%%%%%%%%%%%%%%%%%%%%%%%%%%%%%%%%%%%%%%%%%%%%
\begin{acknowledgments}
%%%%%%%%%%%%%%%%%%%%%%%%%%%%%%%%%%%%%%%%%%%%%%%%%%%%%%%%%%%%
%%%%%%%%%%%%%%%%%%%%%%%%%%%%%%%%%%%%%%%%%%%%%%%%%%%%%%%%%%%%
%%%%%%%%%%%%%%%%%%%%%%%%%%%%%%%%%%%%%%%%%%%%%%%%%%%%%%%%%%%%
JCM is funded by the UNAM-DGAPA Postdoctoral Fellowship Program, and by the UNAM-PAPIIT project No. IN108526. SF is supported by the Generalitat Valenciana and the European Social Fund through the APOSTD 2025 fellowship (CIAPOS/2024/461), and partially by MICIU/AEI and FEDER/EU (grant PID2024-159689NB-C21) and the EU Horizon Europe programme (NewFunFiCO, grant No. 101086251). Numerical simulations were carried out at the UNAM-\textsc{LAMOD} facility.
\end{acknowledgments}

%%%%%%%%%%%%%%%%%%%%%%%%%%%%%%%%%%%%%%%%%%%%%%%%%%%%%%%%%%%%
%%%%%%%%%%%%%%%%%%%%%%%%%%%%%%%%%%%%%%%%%%%%%%%%%%%%%%%%%%%%
%%%%%%%%%%%%%%%%%%%%%%%%%%%%%%%%%%%%%%%%%%%%%%%%%%%%%%%%%%%%
\bibliographystyle{apsrev4-2}
\bibliography{biblio}% Produces the bibliography via BibTeX.

@article{Reitze:2019iox,
    author = "Reitze, David and others",
    title = "{Cosmic Explorer: The U.S. Contribution to Gravitational-Wave Astronomy beyond LIGO}",
    eprint = "1907.04833",
    archivePrefix = "arXiv",
    primaryClass = "astro-ph.IM",
    reportNumber = "LIGO-P1900316",
    journal = "Bull. Am. Astron. Soc.",
    volume = "51",
    number = "7",
    pages = "035",
    year = "2019"
}

@article{daRocha:2021sqd,
    author = "da Rocha, Roldao",
    title = "{Gravitational decoupling of generalized Horndeski hybrid stars}",
    eprint = "2111.11995",
    archivePrefix = "arXiv",
    primaryClass = "gr-qc",
    doi = "10.1140/epjc/s10052-021-09971-3",
    journal = "Eur. Phys. J. C",
    volume = "82",
    number = "1",
    pages = "34",
    year = "2022"
}

@article{Novak:1997hw,
    author = "Novak, Jerome",
    title = "{Spherical neutron star collapse in tensor - scalar theory of gravity}",
    eprint = "gr-qc/9707041",
    archivePrefix = "arXiv",
    doi = "10.1103/PhysRevD.57.4789",
    journal = "Phys. Rev. D",
    volume = "57",
    pages = "4789--4801",
    year = "1998"
}

@article{NdVCoCMjCOaOn2026,
    author = "N{\'u}{\~n}ez, Daniel and Velasco-Castillo, Omar and Castelo Mourelle, Jorge and Cruz-Osorio, Alejandro and Ortiz, N{\'e}stor",
    title = "{Stability of Fermion-Boson Stars}",
    eprint = "",
    archivePrefix = "",
    primaryClass = "",
    doi = "",
    journal = "In preparation",
    volume = "",
    number = "",
    pages = "",
    year = "2026"
}

@article{CMjOnFs2026a,
    author = "Castelo Mourelle, Jorge and Fakhry, Saeed and Ortiz, N{\'e}stor",
    title = "{Fermion-Boson Stars can screen Scalar-Tensor Gravity}",
    eprint = "",
    archivePrefix = "",
    primaryClass = "",
    doi = "",
    journal = "Under review",
    volume = "",
    number = "",
    pages = "",
    year = "2026"
}

@article{Shibata:2013pra,
    author = "Shibata, Masaru and Taniguchi, Keisuke and Okawa, Hirotada and Buonanno, Alessandra",
    title = "{Coalescence of binary neutron stars in a scalar-tensor theory of gravity}",
    eprint = "1310.0627",
    archivePrefix = "arXiv",
    primaryClass = "gr-qc",
    doi = "10.1103/PhysRevD.89.084005",
    journal = "Phys. Rev. D",
    volume = "89",
    number = "8",
    pages = "084005",
    year = "2014"
}

@article{Sotani:2014tua,
    author = "Sotani, Hajime",
    title = "{Scalar gravitational waves from relativistic stars in scalar-tensor gravity}",
    eprint = "1402.5699",
    archivePrefix = "arXiv",
    primaryClass = "astro-ph.HE",
    reportNumber = "YITP-14-15",
    doi = "10.1103/PhysRevD.89.064031",
    journal = "Phys. Rev. D",
    volume = "89",
    number = "6",
    pages = "064031",
    year = "2014"
}

@article{Sotani:2005qx,
    author = "Sotani, Hajime and Kokkotas, Kostas D.",
    title = "{Stellar oscillations in scalar-tensor theory of gravity}",
    eprint = "gr-qc/0506060",
    archivePrefix = "arXiv",
    doi = "10.1103/PhysRevD.71.124038",
    journal = "Phys. Rev. D",
    volume = "71",
    pages = "124038",
    year = "2005"
}

@article{Palenzuela:2015ima,
    author = "Palenzuela, Carlos and Liebling, Steven L.",
    title = "{Constraining scalar-tensor theories of gravity from the most massive neutron stars}",
    eprint = "1510.03471",
    archivePrefix = "arXiv",
    primaryClass = "gr-qc",
    doi = "10.1103/PhysRevD.93.044009",
    journal = "Phys. Rev. D",
    volume = "93",
    number = "4",
    pages = "044009",
    year = "2016"
}

@article{Novak:1998rk,
    author = "Novak, Jerome",
    title = "{Neutron star transition to strong scalar field state in tensor scalar gravity}",
    eprint = "gr-qc/9806022",
    archivePrefix = "arXiv",
    doi = "10.1103/PhysRevD.58.064019",
    journal = "Phys. Rev. D",
    volume = "58",
    pages = "064019",
    year = "1998"
}

@article{Harada:1997mr,
    author = "Harada, Tomohiro",
    title = "{Stability analysis of spherically symmetric star in scalar - tensor theories of gravity}",
    eprint = "gr-qc/9706014",
    archivePrefix = "arXiv",
    reportNumber = "KUNS-1439",
    doi = "10.1143/PTP.98.359",
    journal = "Prog. Theor. Phys.",
    volume = "98",
    pages = "359--379",
    year = "1997"
}

@article{Harada:1998ge,
    author = "Harada, Tomohiro",
    title = "{Neutron stars in scalar tensor theories of gravity and catastrophe theory}",
    eprint = "gr-qc/9801049",
    archivePrefix = "arXiv",
    reportNumber = "KUNS-1476",
    doi = "10.1103/PhysRevD.57.4802",
    journal = "Phys. Rev. D",
    volume = "57",
    pages = "4802--4811",
    year = "1998"
}

@article{Mendes:2021fon,
    author = "Mendes, Raissa F. P. and Ortiz, N{\'e}stor and Stergioulas, Nikolaos",
    title = "{Nonlinear dynamics of oscillating neutron stars in scalar-tensor gravity}",
    eprint = "2107.07036",
    archivePrefix = "arXiv",
    primaryClass = "gr-qc",
    doi = "10.1103/PhysRevD.104.104036",
    journal = "Phys. Rev. D",
    volume = "104",
    number = "10",
    pages = "104036",
    year = "2021"
}

@article{Palenzuela:2013hsa,
    author = "Palenzuela, Carlos and Barausse, Enrico and Ponce, Marcelo and Lehner, Luis",
    title = "{Dynamical scalarization of neutron stars in scalar-tensor gravity theories}",
    eprint = "1310.4481",
    archivePrefix = "arXiv",
    primaryClass = "gr-qc",
    doi = "10.1103/PhysRevD.89.044024",
    journal = "Phys. Rev. D",
    volume = "89",
    number = "4",
    pages = "044024",
    year = "2014"
}

@article{Barausse:2012da,
    author = "Barausse, Enrico and Palenzuela, Carlos and Ponce, Marcelo and Lehner, Luis",
    title = "{Neutron-star mergers in scalar-tensor theories of gravity}",
    eprint = "1212.5053",
    archivePrefix = "arXiv",
    primaryClass = "gr-qc",
    doi = "10.1103/PhysRevD.87.081506",
    journal = "Phys. Rev. D",
    volume = "87",
    pages = "081506",
    year = "2013"
}

@article{Ma:2023sok,
    author = "Ma, Sizheng and Varma, Vijay and Stein, Leo C. and Foucart, Francois and Duez, Matthew D. and Kidder, Lawrence E. and Pfeiffer, Harald P. and Scheel, Mark A.",
    title = "{Numerical simulations of black hole-neutron star mergers in scalar-tensor gravity}",
    eprint = "2304.11836",
    archivePrefix = "arXiv",
    primaryClass = "gr-qc",
    doi = "10.1103/PhysRevD.107.124051",
    journal = "Phys. Rev. D",
    volume = "107",
    number = "12",
    pages = "124051",
    year = "2023"
}

@article{Brihaye:2020klz,
    author = "Brihaye, Y. and Ducobu, L. and Hartmann, B.",
    title = "{Boson and neutron stars with increased density}",
    eprint = "2004.08292",
    archivePrefix = "arXiv",
    primaryClass = "gr-qc",
    doi = "10.1016/j.physletb.2020.135906",
    journal = "Phys. Lett. B",
    volume = "811",
    pages = "135906",
    year = "2020"
}

@article{Brihaye:2019puo,
    author = "Brihaye, Yves and Hartmann, Betti",
    title = "{Spontaneous scalarization of boson stars}",
    eprint = "1903.10471",
    archivePrefix = "arXiv",
    primaryClass = "gr-qc",
    doi = "10.1007/JHEP09(2019)049",
    journal = "JHEP",
    volume = "09",
    pages = "049",
    year = "2019"
}

@article{Alcubierre:2010ea,
    author = "Alcubierre, Miguel and Degollado, Juan C. and Nunez, Dario and Ruiz, Milton and Salgado, Marcelo",
    title = "{Dynamic transition to spontaneous scalarization in boson stars}",
    eprint = "1003.4767",
    archivePrefix = "arXiv",
    primaryClass = "gr-qc",
    doi = "10.1103/PhysRevD.81.124018",
    journal = "Phys. Rev. D",
    volume = "81",
    pages = "124018",
    year = "2010"
}

@article{Ruiz:2012jt,
    author = "Ruiz, Milton and Degollado, Juan Carlos and Alcubierre, Miguel and Nunez, Dario and Salgado, Marcelo",
    title = "{Induced scalarization in boson stars and scalar gravitational radiation}",
    eprint = "1207.6142",
    archivePrefix = "arXiv",
    primaryClass = "gr-qc",
    doi = "10.1103/PhysRevD.86.104044",
    journal = "Phys. Rev. D",
    volume = "86",
    pages = "104044",
    year = "2012"
}

@article{Whinnett:1999sc,
    author = "Whinnett, A. W.",
    title = "{Spontaneous scalarization and boson stars}",
    eprint = "gr-qc/9911052",
    archivePrefix = "arXiv",
    doi = "10.1103/PhysRevD.61.124014",
    journal = "Phys. Rev. D",
    volume = "61",
    pages = "124014",
    year = "2000"
}

@article{Whinnett:1999ma,
    author = "Whinnett, A. W. and Torres, Diego F.",
    title = "{Charged scalar tensor boson stars: Equilibrium, stability and evolution}",
    eprint = "gr-qc/9905017",
    archivePrefix = "arXiv",
    doi = "10.1103/PhysRevD.60.104050",
    journal = "Phys. Rev. D",
    volume = "60",
    pages = "104050",
    year = "1999"
}

@article{Asakawa:2023obq,
    author = "Asakawa, Naomichi and Sekiguchi, Yuichiro",
    title = "{Constraining self-interactions of a massive scalar field using scalar gravitational waves from stellar core collapse}",
    eprint = "2308.15052",
    archivePrefix = "arXiv",
    primaryClass = "gr-qc",
    doi = "10.1103/PhysRevD.108.044060",
    journal = "Phys. Rev. D",
    volume = "108",
    number = "4",
    pages = "044060",
    year = "2023"
}

@article{Rosca-Mead:2019seq,
    author = "Rosca-Mead, Roxana and Moore, Christopher J. and Agathos, Michalis and Sperhake, Ulrich",
    title = "{Inverse-chirp signals and spontaneous scalarisation with self-interacting potentials in stellar collapse}",
    eprint = "1903.09704",
    archivePrefix = "arXiv",
    primaryClass = "gr-qc",
    doi = "10.1088/1361-6382/ab256f",
    journal = "Class. Quant. Grav.",
    volume = "36",
    number = "13",
    pages = "134003",
    year = "2019"
}

@article{Huang:2021tpu,
    author = "Huang, Da and Geng, Chao-Qiang and Kuan, Hao-Jui",
    title = "{Scalar gravitational wave signals from core collapse in massive scalar{\textendash}tensor gravity with triple-scalar interactions}",
    eprint = "2106.13065",
    archivePrefix = "arXiv",
    primaryClass = "gr-qc",
    doi = "10.1088/1361-6382/ac35ab",
    journal = "Class. Quant. Grav.",
    volume = "38",
    number = "24",
    pages = "245006",
    year = "2021"
}

@article{PhysRevD.92.124016,
  title = {Spontaneous-scalarization-induced dark matter and variation of the gravitational constant},
  author = {Chen, Pisin and Suyama, Teruaki and Yokoyama, Jun'ichi},
  journal = {Phys. Rev. D},
  volume = {92},
  issue = {12},
  pages = {124016},
  numpages = {15},
  year = {2015},
  month = {Dec},
  publisher = {American Physical Society},
  doi = {10.1103/PhysRevD.92.124016},
  url = {https://link.aps.org/doi/10.1103/PhysRevD.92.124016}
}

@article{Ramazanoglu:2016kul,
    author = "Ramazano\u{g}lu, Fethi M. and Pretorius, Frans",
    title = "{Spontaneous Scalarization with Massive Fields}",
    eprint = "1601.07475",
    archivePrefix = "arXiv",
    primaryClass = "gr-qc",
    doi = "10.1103/PhysRevD.93.064005",
    journal = "Phys. Rev. D",
    volume = "93",
    number = "6",
    pages = "064005",
    year = "2016"
}

@article{Yazadjiev:2016pcb,
    author = "Yazadjiev, Stoytcho S. and Doneva, Daniela D. and Popchev, Dimitar",
    title = "{Slowly rotating neutron stars in scalar-tensor theories with a massive scalar field}",
    eprint = "1602.04766",
    archivePrefix = "arXiv",
    primaryClass = "gr-qc",
    doi = "10.1103/PhysRevD.93.084038",
    journal = "Phys. Rev. D",
    volume = "93",
    number = "8",
    pages = "084038",
    year = "2016"
}

@article{Doneva:2016xmf,
    author = "Doneva, Daniela D. and Yazadjiev, Stoytcho S.",
    title = "{Rapidly rotating neutron stars with a massive scalar field\textemdash{}structure and universal relations}",
    eprint = "1607.03299",
    archivePrefix = "arXiv",
    primaryClass = "gr-qc",
    doi = "10.1088/1475-7516/2016/11/019",
    journal = "JCAP",
    volume = "11",
    pages = "019",
    year = "2016"
}

@article{Sperhake:2017itk,
    author = "Sperhake, Ulrich and Moore, Christopher J. and Rosca, Roxana and Agathos, Michalis and Gerosa, Davide and Ott, Christian D.",
    title = "{Long-lived inverse chirp signals from core collapse in massive scalar-tensor gravity}",
    eprint = "1708.03651",
    archivePrefix = "arXiv",
    primaryClass = "gr-qc",
    doi = "10.1103/PhysRevLett.119.201103",
    journal = "Phys. Rev. Lett.",
    volume = "119",
    number = "20",
    pages = "201103",
    year = "2017"
}

@article{Staykov:2018hhc,
    author = "Staykov, Kalin V. and Popchev, Dimitar and Doneva, Daniela D. and Yazadjiev, Stoytcho S.",
    title = "{Static and slowly rotating neutron stars in scalar\textendash{}tensor theory with self-interacting massive scalar field}",
    eprint = "1805.07818",
    archivePrefix = "arXiv",
    primaryClass = "gr-qc",
    doi = "10.1140/epjc/s10052-018-6064-x",
    journal = "Eur. Phys. J. C",
    volume = "78",
    number = "7",
    pages = "586",
    year = "2018"
}

@article{Rosca-Mead:2020bzt,
    author = "Rosca-Mead, Roxana and Moore, Christopher J. and Sperhake, Ulrich and Agathos, Michalis and Gerosa, Davide",
    title = "{Structure of neutron stars in massive scalar-tensor gravity}",
    eprint = "2007.14429",
    archivePrefix = "arXiv",
    primaryClass = "gr-qc",
    doi = "10.3390/sym12091384",
    journal = "Symmetry",
    volume = "12",
    number = "9",
    pages = "1384",
    year = "2020"
}

@article{Kuan:2023hrh,
    author = "Kuan, Hao-Jui and Van Aelst, Karim and Lam, Alan Tsz-Lok and Shibata, Masaru",
    title = "{Binary neutron star mergers in massive scalar-tensor theory: Quasiequilibrium states and dynamical enhancement of the scalarization}",
    eprint = "2309.01709",
    archivePrefix = "arXiv",
    primaryClass = "gr-qc",
    doi = "10.1103/PhysRevD.108.064057",
    journal = "Phys. Rev. D",
    volume = "108",
    number = "6",
    pages = "064057",
    year = "2023"
}

@article{Lam:2024wpq,
    author = "Lam, Alan Tsz-Lok and Kuan, Hao-Jui and Shibata, Masaru and Van Aelst, Karim and Kiuchi, Kenta",
    title = "{Binary neutron star mergers in massive scalar-tensor theory: Properties of postmerger remnants}",
    eprint = "2406.05211",
    archivePrefix = "arXiv",
    primaryClass = "gr-qc",
    doi = "10.1103/PhysRevD.110.104018",
    journal = "Phys. Rev. D",
    volume = "110",
    number = "10",
    pages = "104018",
    year = "2024"
}

@article{Antoniadis:2013pzd,
    author = "Antoniadis, John and others",
    title = "{A Massive Pulsar in a Compact Relativistic Binary}",
    eprint = "1304.6875",
    archivePrefix = "arXiv",
    primaryClass = "astro-ph.HE",
    doi = "10.1126/science.1233232",
    journal = "Science",
    volume = "340",
    pages = "6131",
    year = "2013"
}

@article{Bertotti:2003rm,
    author = "Bertotti, B. and Iess, L. and Tortora, P.",
    title = "{A test of general relativity using radio links with the Cassini spacecraft}",
    doi = "10.1038/nature01997",
    journal = "Nature",
    volume = "425",
    pages = "374--376",
    year = "2003"
}

@article{Tuna:2022qqr,
    author = {Tuna, Semih and {\"U}nl{\"u}t{\"u}rk, K{\i}van{\c{c}} {\.I}. and Ramazano{\u{g}}lu, Fethi M.},
    title = "{Constraining scalar-tensor theories using neutron star mass and radius measurements}",
    eprint = "2204.02138",
    archivePrefix = "arXiv",
    primaryClass = "gr-qc",
    doi = "10.1103/PhysRevD.105.124070",
    journal = "Phys. Rev. D",
    volume = "105",
    number = "12",
    pages = "124070",
    year = "2022"
}

@article{Niu:2021nic,
    author = "Niu, Rui and Zhang, Xing and Wang, Bo and Zhao, Wen",
    title = "{Constraining Scalar-tensor Theories Using Neutron Star{\textendash}Black Hole Gravitational Wave Events}",
    eprint = "2105.13644",
    archivePrefix = "arXiv",
    primaryClass = "gr-qc",
    doi = "10.3847/1538-4357/ac1d4f",
    journal = "Astrophys. J.",
    volume = "921",
    number = "2",
    pages = "149",
    year = "2021"
}

@article{Shao:2017gwu,
    author = "Shao, Lijing and Sennett, Noah and Buonanno, Alessandra and Kramer, Michael and Wex, Norbert",
    title = "{Constraining nonperturbative strong-field effects in scalar-tensor gravity by combining pulsar timing and laser-interferometer gravitational-wave detectors}",
    eprint = "1704.07561",
    archivePrefix = "arXiv",
    primaryClass = "gr-qc",
    reportNumber = "LIGO-P1700073",
    doi = "10.1103/PhysRevX.7.041025",
    journal = "Phys. Rev. X",
    volume = "7",
    number = "4",
    pages = "041025",
    year = "2017"
}

@article{Freire:2012mg,
    author = "Freire, Paulo C. C. and Wex, Norbert and Esposito-Farese, Gilles and Verbiest, Joris P. W. and Bailes, Matthew and Jacoby, Bryan A. and Kramer, Michael and Stairs, Ingrid H. and Antoniadis, John and Janssen, Gemma H.",
    title = "{The relativistic pulsar-white dwarf binary PSR J1738+0333 II. The most stringent test of scalar-tensor gravity}",
    eprint = "1205.1450",
    archivePrefix = "arXiv",
    primaryClass = "astro-ph.GA",
    doi = "10.1111/j.1365-2966.2012.21253.x",
    journal = "Mon. Not. Roy. Astron. Soc.",
    volume = "423",
    pages = "3328",
    year = "2012"
}

@article{Berti:2015itd,
    author = "Berti, Emanuele and others",
    title = "{Testing General Relativity with Present and Future Astrophysical Observations}",
    eprint = "1501.07274",
    archivePrefix = "arXiv",
    primaryClass = "gr-qc",
    doi = "10.1088/0264-9381/32/24/243001",
    journal = "Class. Quant. Grav.",
    volume = "32",
    pages = "243001",
    year = "2015"
}

@article{Doneva:2022ewd,
    author = "Doneva, Daniela D. and Ramazano{\u{g}}lu, Fethi M. and Silva, Hector O. and Sotiriou, Thomas P. and Yazadjiev, Stoytcho S.",
    title = "{Spontaneous scalarization}",
    eprint = "2211.01766",
    archivePrefix = "arXiv",
    primaryClass = "gr-qc",
    doi = "10.1103/RevModPhys.96.015004",
    journal = "Rev. Mod. Phys.",
    volume = "96",
    number = "1",
    pages = "015004",
    year = "2024"
}

@book{Fujii_Maeda_2003,
place={Cambridge},
series={Cambridge Monographs on Mathematical Physics},
title={The Scalar-Tensor Theory of Gravitation},
publisher={Cambridge University Press},
author={Fujii, Yasunori and Maeda, Kei-ichi},
year={2003},
collection={Cambridge Monographs on Mathematical Physics},
doi="https://doi.org/10.1017/CBO9780511535093"
}

@article{deSousa:1995ye,
    author = "de Sousa, Claudio M. G. and Tomazelli, J. L.",
    title = "{A Model for stars of interacting bosons and fermions}",
    eprint = "gr-qc/9507043",
    archivePrefix = "arXiv",
    reportNumber = "IC-95-187",
    doi = "10.1103/PhysRevD.58.123003",
    journal = "Phys. Rev. D",
    volume = "58",
    pages = "123003",
    year = "1998"
}

@article{Pisano:1995yk,
    author = "Pisano, F. and Tomazelli, J. L.",
    title = "{Stars of WIMPs}",
    eprint = "gr-qc/9509022",
    archivePrefix = "arXiv",
    doi = "10.1142/S0217732396000667",
    journal = "Mod. Phys. Lett. A",
    volume = "11",
    pages = "647--652",
    year = "1996"
}

@article{Zhou:2025kku,
    author = "Zhou, Xiao-Ding and Chen, Tian-Shun and Wu, Si-Man and Zhang, Kilar",
    title = "{Stability Boundary of Neutron-Dark Matter Mixed Stars}",
    journal = "",
    volume = "",
    eprint = "2512.01641",
    archivePrefix = "arXiv",
    primaryClass = "astro-ph.HE",
    doi = "10.1088/1674-1137/ae86dd",
    month = "12",
    year = "2025"
}

@article{Nyhan:2022pda,
    author = "Nyhan, Joseph E. and Kain, Ben",
    title = "{Dynamical evolution of fermion-boson stars with realistic equations of state}",
    eprint = "2206.07715",
    archivePrefix = "arXiv",
    primaryClass = "gr-qc",
    doi = "10.1103/PhysRevD.105.123016",
    journal = "Phys. Rev. D",
    volume = "105",
    number = "12",
    pages = "123016",
    year = "2022"
}

@article{DiGiovanni:2020frc,
    author = "Di Giovanni, Fabrizio and Fakhry, Saeed and Sanchis-Gual, Nicolas and Degollado, Juan Carlos and Font, Jos{\'e} A.",
    title = "{Dynamical formation and stability of fermion-boson stars}",
    eprint = "2006.08583",
    archivePrefix = "arXiv",
    primaryClass = "gr-qc",
    doi = "10.1103/PhysRevD.102.084063",
    journal = "Phys. Rev. D",
    volume = "102",
    number = "8",
    pages = "084063",
    year = "2020"
}

@article{Q:2025ycf,
    author = "Q., Roberto D. Alba and Chagoya, Javier and Roque, Armando A.",
    title = "{Compact stars in Einstein-scalar-Gauss-Bonnet gravity: Regular and divergent scalar field configurations}",
    eprint = "2508.13273",
    archivePrefix = "arXiv",
    primaryClass = "gr-qc",
    doi = "10.1103/ts7l-tr7m",
    journal = "Phys. Rev. D",
    volume = "112",
    number = "8",
    pages = "084029",
    year = "2025"
}

@article{Roque:2021lvr,
    author = "Roque, Armando A. and Ure{\~n}a-L{\'o}pez, L. Arturo",
    title = "{Horndeski fermion{\textendash}boson stars}",
    eprint = "2109.14747",
    archivePrefix = "arXiv",
    primaryClass = "gr-qc",
    doi = "10.1088/1361-6382/ac4614",
    journal = "Class. Quant. Grav.",
    volume = "39",
    number = "4",
    pages = "044001",
    year = "2022"
}

@article{Valdez-Alvarado:2012rct,
    author = "Valdez-Alvarado, Susana and Palenzuela, Carlos and Alic, Daniela and Ure{\~n}a-L{\'o}pez, L. Arturo",
    title = "{Dynamical evolution of fermion-boson stars}",
    eprint = "1210.2299",
    archivePrefix = "arXiv",
    primaryClass = "gr-qc",
    doi = "10.1103/PhysRevD.87.084040",
    journal = "Phys. Rev. D",
    volume = "87",
    number = "8",
    pages = "084040",
    year = "2013"
}

@article{Hui:2021tkt,
    author = "Hui, Lam",
    title = "{Wave Dark Matter}",
    eprint = "2101.11735",
    archivePrefix = "arXiv",
    primaryClass = "astro-ph.CO",
    doi = "10.1146/annurev-astro-120920-010024",
    journal = "Ann. Rev. Astron. Astrophys.",
    volume = "59",
    pages = "247--289",
    year = "2021"
}

@article{Visinelli:2021uve,
    author = "Visinelli, Luca",
    title = "{Boson stars and oscillatons: A review}",
    eprint = "2109.05481",
    archivePrefix = "arXiv",
    primaryClass = "gr-qc",
    doi = "10.1142/S0218271821300068",
    journal = "Int. J. Mod. Phys. D",
    volume = "30",
    number = "15",
    pages = "2130006",
    year = "2021"
}

@article{Jetzer:1991jr,
    author = "Jetzer, Philippe",
    title = "{Boson stars}",
    reportNumber = "ZU-TH-25-91",
    doi = "10.1016/0370-1573(92)90123-H",
    journal = "Phys. Rept.",
    volume = "220",
    pages = "163--227",
    year = "1992"
}

@article{Jetzer:1990xa,
    author = "Jetzer, P.",
    title = "{Stability of Combined Boson - Fermion Stars}",
    reportNumber = "CERN-TH-5685-90",
    doi = "10.1016/0370-2693(90)90952-3",
    journal = "Phys. Lett. B",
    volume = "243",
    pages = "36--40",
    year = "1990"
}

@article{Liebling:2012fv,
    author = "Liebling, Steven L. and Palenzuela, Carlos",
    title = "{Dynamical boson stars}",
    eprint = "1202.5809",
    archivePrefix = "arXiv",
    primaryClass = "gr-qc",
    doi = "10.1007/s41114-023-00043-4",
    journal = "Living Rev. Rel.",
    volume = "26",
    number = "1",
    pages = "1",
    year = "2023"
}

@article{Chatziioannou:2024jsr,
    author = "Chatziioannou, Katerina and Cromartie, H. Thankful and Gandolfi, Stefano and Tews, Ingo and Radice, David and Steiner, Andrew W. and Watts, Anna L.",
    title = "{Neutron stars and the dense matter equation of state}",
    eprint = "2407.11153",
    archivePrefix = "arXiv",
    primaryClass = "nucl-th",
    reportNumber = "LA-UR-23-22545",
    doi = "10.1103/ymsq-cfcw",
    journal = "Rev. Mod. Phys.",
    volume = "97",
    number = "4",
    pages = "045007",
    year = "2025"
}

@article{Baiotti:2016qnr,
    author = "Baiotti, Luca and Rezzolla, Luciano",
    title = "{Binary neutron star mergers: a review of Einstein{\textquoteright}s richest laboratory}",
    eprint = "1607.03540",
    archivePrefix = "arXiv",
    primaryClass = "gr-qc",
    doi = "10.1088/1361-6633/aa67bb",
    journal = "Rept. Prog. Phys.",
    volume = "80",
    number = "9",
    pages = "096901",
    year = "2017"
}

@article{Damour:1992we,
    author = "Damour, Thibault and Esposito-Farese, Gilles",
    title = "{Tensor multiscalar theories of gravitation}",
    reportNumber = "IHES-P-91-93, CPT-91-PE-2542",
    doi = "10.1088/0264-9381/9/9/015",
    journal = "Class. Quant. Grav.",
    volume = "9",
    pages = "2093--2176",
    year = "1992"
}

@article{Damour:1993hw,
    author = "Damour, Thibault and Esposito-Farese, Gilles",
    title = "{Nonperturbative strong field effects in tensor - scalar theories of gravitation}",
    reportNumber = "IHES-P-93-1, CPT-93-PE-2868",
    doi = "10.1103/PhysRevLett.70.2220",
    journal = "Phys. Rev. Lett.",
    volume = "70",
    pages = "2220--2223",
    year = "1993"
}

@article{Fakhry:2026cmi,
    author = "Fakhry, Saeed and Mourelle, Jorge Castelo and Sanchis-Gual, Nicolas and Doneva, Daniela and Yazadjiev, Stoytcho and Font, Jos{\'e} A.",
    title = "{Rotating fermion-boson stars in R-squared gravity}",
    eprint = "2607.04744",
    archivePrefix = "arXiv",
    primaryClass = "gr-qc",
    doi = "10.1103/y3l6-5xsv",
    journal = "Phys. Rev. D",
    volume = "114",
    number = "6",
    pages = "064037",
    year = "2026"
}

@article{Henriques:1989ez,
    author = "Henriques, A. B. and Liddle, Andrew R. and Moorhouse, R. G.",
    title = "{Combined Boson - Fermion Stars: Configurations and Stability}",
    reportNumber = "GUTPA-89-10-1",
    doi = "10.1016/0550-3213(90)90514-E",
    journal = "Nucl. Phys. B",
    volume = "337",
    pages = "737--761",
    year = "1990"
}

@article{GWTC1_PRX,
  doi = {10.1103/PhysRevX.9.031040},
  url = {https://doi.org/10.1103/PhysRevX.9.031040},
  year = {2019},
  month = sep,
  publisher = {American Physical Society},
  volume = {9},
  number = {3},
  author = {B. P. Abbott and others},
  title = {{GWTC}-1: A Gravitational-Wave Transient Catalog of Compact Binary Mergers Observed by {LIGO} and Virgo during the First and Second Observing Runs},
  eprint = {1811.12907},
  archivePrefix = {arXiv},
  primaryClass = {gr-qc},
  journal = {Physical Review X}
}

@article{LIGOScientific:2020ibl,
    author = "Abbott, R. and others",
    collaboration = "LIGO Scientific, Virgo",
    title = "{GWTC-2: Compact Binary Coalescences Observed by LIGO and Virgo During the First Half of the Third Observing Run}",
    eprint = "2010.14527",
    archivePrefix = "arXiv",
    primaryClass = "gr-qc",
    reportNumber = "P2000061",
    doi = "10.1103/PhysRevX.11.021053",
    journal = "Phys. Rev. X",
    volume = "11",
    pages = "021053",
    year = "2021"
}

@article{Clesse:2020ghq,
    author = "Clesse, Sebastien and Garcia-Bellido, Juan",
    title = "{GW190425, GW190521 and GW190814: Three candidate mergers of primordial black holes from the QCD epoch}",
    eprint = "2007.06481",
    archivePrefix = "arXiv",
    primaryClass = "astro-ph.CO",
    reportNumber = "IFT-UAM/CSIC-20-108",
    doi = "10.1016/j.dark.2022.101111",
    journal = "Phys. Dark Univ.",
    volume = "38",
    pages = "101111",
    year = "2022"
}

@article{Hild:2010id,
    author = "Hild, S. and others",
    title = "{Sensitivity Studies for Third-Generation Gravitational Wave Observatories}",
    eprint = "1012.0908",
    archivePrefix = "arXiv",
    primaryClass = "gr-qc",
    doi = "10.1088/0264-9381/28/9/094013",
    journal = "Class. Quant. Grav.",
    volume = "28",
    pages = "094013",
    year = "2011"
}

@article{Punturo:2010zza,
    author = "Punturo, M. and others",
    editor = "Marka, Zsuzsa and Marka, Szabolcs",
    title = "{The third generation of gravitational wave observatories and their science reach}",
    doi = "10.1088/0264-9381/27/8/084007",
    journal = "Class. Quant. Grav.",
    volume = "27",
    pages = "084007",
    year = "2010"
}

@article{Ackley:2020atn,
    author = "Ackley, K. and others",
    title = "{Neutron Star Extreme Matter Observatory: A kilohertz-band gravitational-wave detector in the global network}",
    eprint = "2007.03128",
    archivePrefix = "arXiv",
    primaryClass = "astro-ph.HE",
    doi = "10.1017/pasa.2020.39",
    journal = "Publ. Astron. Soc. Austral.",
    volume = "37",
    pages = "e047",
    year = "2020"
}

@article{LIGOScientific:2016wof,
    author = "Abbott, Benjamin P and others",
    collaboration = "LIGO Scientific",
    title = "{Exploring the Sensitivity of Next Generation Gravitational Wave Detectors}",
    eprint = "1607.08697",
    archivePrefix = "arXiv",
    primaryClass = "astro-ph.IM",
    reportNumber = "LIGO-P1600143",
    doi = "10.1088/1361-6382/aa51f4",
    journal = "Class. Quant. Grav.",
    volume = "34",
    number = "4",
    pages = "044001",
    year = "2017"
}

@article{Das:2021yny,
    author = "Das, H. C. and Kumar, Ankit and Patra, S. K.",
    title = "{Dark matter admixed neutron star as a possible compact component in the GW190814 merger event}",
    eprint = "2109.01853",
    archivePrefix = "arXiv",
    primaryClass = "astro-ph.HE",
    doi = "10.1103/PhysRevD.104.063028",
    journal = "Phys. Rev. D",
    volume = "104",
    number = "6",
    pages = "063028",
    year = "2021"
}

@article{Vattis:2020iuz,
    author = "Vattis, Kyriakos and Goldstein, Isabelle S. and Koushiappas, Savvas M.",
    title = "{Could the 2.6  $M_\odot$ object in GW190814 be a primordial black hole?}",
    eprint = "2006.15675",
    archivePrefix = "arXiv",
    primaryClass = "astro-ph.HE",
    doi = "10.1103/PhysRevD.102.061301",
    journal = "Phys. Rev. D",
    volume = "102",
    number = "6",
    pages = "061301",
    year = "2020"
}

@article{LIGOScientific:2020zkf,
    author = "Abbott, R. and others",
    collaboration = "LIGO Scientific, Virgo",
    title = "{GW190814: Gravitational Waves from the Coalescence of a 23 Solar Mass Black Hole with a 2.6 Solar Mass Compact Object}",
    eprint = "2006.12611",
    archivePrefix = "arXiv",
    primaryClass = "astro-ph.HE",
    reportNumber = "LIGO-P190814",
    doi = "10.3847/2041-8213/ab960f",
    journal = "Astrophys. J. Lett.",
    volume = "896",
    number = "2",
    pages = "L44",
    year = "2020"
}

@article{Degollado:2024oyo,
    author = "Degollado, Juan Carlos and Ortiz, N{\'e}stor and Salgado, Marcelo",
    title = "{Dynamical transition to spontaneous scalarization in neutron stars: The massive scalar field scenario}",
    eprint = "2407.08124",
    archivePrefix = "arXiv",
    primaryClass = "gr-qc",
    doi = "10.1103/PhysRevD.110.084011",
    journal = "Phys. Rev. D",
    volume = "110",
    number = "8",
    pages = "084011",
    year = "2024"
}

@article{KAGRA:2021vkt,
    author = "Abbott, R. and others",
    collaboration = "KAGRA, VIRGO, LIGO Scientific",
    title = "{GWTC-3: Compact Binary Coalescences Observed by LIGO and Virgo during the Second Part of the Third Observing Run}",
    eprint = "2111.03606",
    archivePrefix = "arXiv",
    primaryClass = "gr-qc",
    reportNumber = "LIGO-P2000318",
    doi = "10.1103/PhysRevX.13.041039",
    journal = "Phys. Rev. X",
    volume = "13",
    number = "4",
    pages = "041039",
    year = "2023"
}

@article{Evstafyeva:2023kfg,
    author = "Evstafyeva, Tamara and Rosca-Mead, Roxana and Sperhake, Ulrich and Brugmann, Bernd",
    title = "{Boson stars in massless and massive scalar-tensor gravity}",
    eprint = "2310.05200",
    archivePrefix = "arXiv",
    primaryClass = "gr-qc",
    doi = "10.1103/PhysRevD.108.104064",
    journal = "Phys. Rev. D",
    volume = "108",
    number = "10",
    pages = "104064",
    year = "2023"
}

@article{Mendes:2016fby,
    author = "Mendes, Raissa F. P. and Ortiz, N{\'e}stor",
    title = "{Highly compact neutron stars in scalar-tensor theories of gravity: Spontaneous scalarization versus gravitational collapse}",
    eprint = "1604.04175",
    archivePrefix = "arXiv",
    primaryClass = "gr-qc",
    doi = "10.1103/PhysRevD.93.124035",
    journal = "Phys. Rev. D",
    volume = "93",
    number = "12",
    pages = "124035",
    year = "2016"
}

@article{LIGOScientific:2021usb,
    author = "Abbott, R. and others",
    collaboration = "LIGO Scientific, VIRGO",
    title = "{GWTC-2.1: Deep extended catalog of compact binary coalescences observed by LIGO and Virgo during the first half of the third observing run}",
    eprint = "2108.01045",
    archivePrefix = "arXiv",
    primaryClass = "gr-qc",
    reportNumber = "LIGO-P2100063",
    doi = "10.1103/PhysRevD.109.022001",
    journal = "Phys. Rev. D",
    volume = "109",
    number = "2",
    pages = "022001",
    year = "2024"
}

@article{GWTC4,
    author = "Abac, A. G. and others",
    collaboration = "LIGO Scientific, VIRGO, KAGRA",
    title = "{GWTC-4.0: Updating the Gravitational-Wave Transient Catalog with Observations from the First Part of the Fourth LIGO-Virgo-KAGRA Observing Run}",
    eprint = "2508.18082",
    archivePrefix = "arXiv",
    primaryClass = "gr-qc",
    reportNumber = "LIGO-P2400386",
    doi = "10.3847/2041-8213/ae2c74",
    journal = "Astrophys. J. Lett.",
    volume = "1004",
    number = "2",
    pages = "L22",
    year = "2026"
}

@article{GW230529,
    author = "Abac, A. G. and others",
    collaboration = "LIGO Scientific, KAGRA, VIRGO",
    title = "{Observation of Gravitational Waves from the Coalescence of a 2.5{\textendash}4.5 M $_{⊙}$ Compact Object and a Neutron Star}",
    eprint = "2404.04248",
    archivePrefix = "arXiv",
    primaryClass = "astro-ph.HE",
    reportNumber = "LIGO-P2300352",
    doi = "10.3847/2041-8213/ad5beb",
    journal = "Astrophys. J. Lett.",
    volume = "970",
    number = "2",
    pages = "L34",
    year = "2024"
}

@article{DiGiovanni:2021ejn,
    author = "Di Giovanni, Fabrizio and Sanchis-Gual, Nicolas and Cerd\'a-Dur\'an, Pablo and Font, Jos\'e Antonio",
    title = "{Can fermion-boson stars reconcile multimessenger observations of compact stars?}",
    eprint = "2110.11997",
    archivePrefix = "arXiv",
    primaryClass = "gr-qc",
    doi = "10.1103/PhysRevD.105.063005",
    journal = "Phys. Rev. D",
    volume = "105",
    number = "6",
    pages = "063005",
    year = "2022"
}

@article{Henriques:1989ar,
    author = "Henriques, A. B. and Liddle, Andrew R. and Moorhouse, R. G.",
    title = "{COMBINED BOSON - FERMION STARS}",
    reportNumber = "GU-TPA-89/9-1",
    doi = "10.1016/0370-2693(89)90623-0",
    journal = "Phys. Lett. B",
    volume = "233",
    pages = "99",
    year = "1989"
}

@article{Adam:2022nlq,
    author = "Adam, Christoph and Castelo, Jorge and Garc\'\i{}a Mart\'\i{}n-Caro, Alberto and Huidobro, Miguel and V\'azquez, Ricardo and Wereszczynski, Andrzej",
    title = "{Universal relations for rotating boson stars}",
    eprint = "2203.16558",
    archivePrefix = "arXiv",
    primaryClass = "gr-qc",
    doi = "10.1103/PhysRevD.106.123022",
    journal = "Phys. Rev. D",
    volume = "106",
    number = "12",
    pages = "123022",
    year = "2022"
}

@article{Rezzolla:2017aly,
    author = "Rezzolla, Luciano and Most, Elias R. and Weih, Lukas R.",
    title = "{Using gravitational-wave observations and quasi-universal relations to constrain the maximum mass of neutron stars}",
    eprint = "1711.00314",
    archivePrefix = "arXiv",
    primaryClass = "astro-ph.HE",
    doi = "10.3847/2041-8213/aaa401",
    journal = "Astrophys. J. Lett.",
    volume = "852",
    number = "2",
    pages = "L25",
    year = "2018"
}

@book{press1992numerical,
  title={Numerical recipes in C},
  author={Press, William H and Teukolsky, Saul A and Vetterling, William T and Flannery, Brian P},
  volume={2},
  number={1},
  year={1992},
  publisher={Cambridge university press Cambridge}
}

@article{Haskell:2013vha,
    author = "Haskell, Brynmor and Ciolfi, Riccardo and Pannarale, Francesco and Rezzolla, Luciano",
    title = "{On the universality of I-Love-Q relations in magnetized neutron stars}",
    eprint = "1309.3885",
    archivePrefix = "arXiv",
    primaryClass = "astro-ph.SR",
    doi = "10.1093/mnrasl/slt161",
    journal = "Mon. Not. Roy. Astron. Soc.",
    volume = "438",
    number = "1",
    pages = "L71--L75",
    year = "2014"
}

@article{Pretel:2024pem,
    author = "Pretel, Juan M. Z. and Zhang, Chen",
    title = "{Universal relations for anisotropic interacting quark stars}",
    eprint = "2401.12519",
    archivePrefix = "arXiv",
    primaryClass = "nucl-th",
    doi = "10.1088/1475-7516/2024/10/032",
    journal = "JCAP",
    volume = "10",
    pages = "032",
    year = "2024"
}

@article{Yip:2025uzd,
    author = "Yip, Chun-Ming and Lau, Shu Yan and Yagi, Kent",
    title = "{Universal relations for elastic hybrid stars and quark stars}",
    eprint = "2510.15149",
    archivePrefix = "arXiv",
    primaryClass = "astro-ph.HE",
    doi = "10.1103/cflw-5fps",
    journal = "Phys. Rev. D",
    volume = "114",
    number = "4",
    pages = "043024",
    year = "2026"
}

@mastersthesis{Lai:2004fw,
    author = "Lai, Chi-Wai",
    title = "{A Numerical study of boson stars}",
    eprint = "gr-qc/0410040",
    archivePrefix = "arXiv",
    reportNumber = "UMI-NQ99499",
    type = "Other thesis",
    month = "9",
    year = "2004"
}

@article{Doneva:2018ouu,
    author = "Doneva, Daniela D. and Yazadjiev, Stoytcho S. and Stergioulas, Nikolaos and Kokkotas, Kostas D.",
    title = "{Differentially rotating neutron stars in scalar-tensor theories of gravity}",
    eprint = "1807.05449",
    archivePrefix = "arXiv",
    primaryClass = "gr-qc",
    doi = "10.1103/PhysRevD.98.104039",
    journal = "Phys. Rev. D",
    volume = "98",
    number = "10",
    pages = "104039",
    year = "2018"
}

@article{Doneva:2013qva,
    author = "Doneva, Daniela D. and Yazadjiev, Stoytcho S. and Stergioulas, Nikolaos and Kokkotas, Kostas D.",
    title = "{Rapidly rotating neutron stars in scalar-tensor theories of gravity}",
    eprint = "1309.0605",
    archivePrefix = "arXiv",
    primaryClass = "gr-qc",
    doi = "10.1103/PhysRevD.88.084060",
    journal = "Phys. Rev. D",
    volume = "88",
    number = "8",
    pages = "084060",
    year = "2013"
}

@article{Staykov:2023ose,
    author = "Staykov, Kalin V. and Doneva, Daniela D. and Heisenberg, Lavinia and Stergioulas, Nikolaos and Yazadjiev, Stoytcho S.",
    title = "{Differentially rotating scalarized neutron stars with realistic postmerger profiles}",
    eprint = "2303.07769",
    archivePrefix = "arXiv",
    primaryClass = "gr-qc",
    reportNumber = "ET-0053A-23",
    doi = "10.1103/PhysRevD.108.024058",
    journal = "Phys. Rev. D",
    volume = "108",
    number = "2",
    pages = "024058",
    year = "2023"
}

@article{Cardoso:2017cfl,
    author = "Cardoso, Vitor and Franzin, Edgardo and Maselli, Andrea and Pani, Paolo and Raposo, Guilherme",
    title = "{Testing strong-field gravity with tidal Love numbers}",
    eprint = "1701.01116",
    archivePrefix = "arXiv",
    primaryClass = "gr-qc",
    doi = "10.1103/PhysRevD.95.084014",
    journal = "Phys. Rev. D",
    volume = "95",
    number = "8",
    pages = "084014",
    year = "2017",
    note = "[Addendum: Phys.Rev.D 95, 089901 (2017)]"
}

@article{Sennett:2017etc,
    author = "Sennett, Noah and Hinderer, Tanja and Steinhoff, Jan and Buonanno, Alessandra and Ossokine, Serguei",
    title = "{Distinguishing Boson Stars from Black Holes and Neutron Stars from Tidal Interactions in Inspiraling Binary Systems}",
    eprint = "1704.08651",
    archivePrefix = "arXiv",
    primaryClass = "gr-qc",
    doi = "10.1103/PhysRevD.96.024002",
    journal = "Phys. Rev. D",
    volume = "96",
    number = "2",
    pages = "024002",
    year = "2017"
}

@article{Doneva:2017jop,
    author = "Doneva, Daniela D. and Pappas, George",
    title = "{Universal Relations and Alternative Gravity Theories}",
    eprint = "1709.08046",
    archivePrefix = "arXiv",
    primaryClass = "gr-qc",
    doi = "10.1007/978-3-319-97616-7_13",
    journal = "Astrophys. Space Sci. Libr.",
    volume = "457",
    pages = "737--806",
    year = "2018"
}

@article{Pappas:2018csu,
    author = "Pappas, George and Doneva, Daniela D. and Sotiriou, Thomas P. and Yazadjiev, Stoytcho S. and Kokkotas, Kostas D.",
    title = "{Multipole moments and universal relations for scalarized neutron stars}",
    eprint = "1812.01117",
    archivePrefix = "arXiv",
    primaryClass = "gr-qc",
    doi = "10.1103/PhysRevD.99.104014",
    journal = "Phys. Rev. D",
    volume = "99",
    number = "10",
    pages = "104014",
    year = "2019"
}

@article{Bustillo:2020syj,
    author = "Bustillo, Juan Calder\'on and Sanchis-Gual, Nicolas and Torres-Forn\'e, Alejandro and Font, Jos\'e A. and Vajpeyi, Avi and Smith, Rory and Herdeiro, Carlos and Radu, Eugen and Leong, Samson H. W.",
    title = "{GW190521 as a Merger of Proca Stars: A Potential New Vector Boson of $8.7\times 10^{-13}$  eV}",
    eprint = "2009.05376",
    archivePrefix = "arXiv",
    primaryClass = "gr-qc",
    reportNumber = "LIGO DCC:P-2000353",
    doi = "10.1103/PhysRevLett.126.081101",
    journal = "Phys. Rev. Lett.",
    volume = "126",
    number = "8",
    pages = "081101",
    year = "2021"
}

@article{Yagi:2013awa,
    author = "Yagi, Kent and Yunes, Nicolas",
    title = "{I-Love-Q Relations in Neutron Stars and their Applications to Astrophysics, Gravitational Waves and Fundamental Physics}",
    eprint = "1303.1528",
    archivePrefix = "arXiv",
    primaryClass = "gr-qc",
    doi = "10.1103/PhysRevD.88.023009",
    journal = "Phys. Rev. D",
    volume = "88",
    number = "2",
    pages = "023009",
    year = "2013"
}

@article{Sun:2020qkj,
    author = "Sun, Wenjie and Wen, Dehua and Wang, Jue",
    title = "{New quasi-universal relations for static and rapid rotating neutron stars}",
    eprint = "2008.02958",
    archivePrefix = "arXiv",
    primaryClass = "gr-qc",
    doi = "10.1103/PhysRevD.102.023039",
    journal = "Phys. Rev. D",
    volume = "102",
    number = "2",
    pages = "023039",
    year = "2020"
}

@article{Torres-Forne:2019zwz,
    author = {Torres-Forn\'e, Alejandro and Cerd\'a-Dur\'an, Pablo and Obergaulinger, Martin and M\"uller, Bernhard and Font, Jos\'e A.},
    title = "{Universal Relations for Gravitational-Wave Asteroseismology of Protoneutron Stars}",
    eprint = "1902.10048",
    archivePrefix = "arXiv",
    primaryClass = "gr-qc",
    doi = "10.1103/PhysRevLett.127.239901",
    journal = "Phys. Rev. Lett.",
    volume = "123",
    number = "5",
    pages = "051102",
    year = "2019",
    note = "[Erratum: Phys.Rev.Lett. 127, 239901 (2021)]"
}

@article{Yagi:2013sva,
    author = "Yagi, Kent",
    title = "{Multipole Love Relations}",
    eprint = "1311.0872",
    archivePrefix = "arXiv",
    primaryClass = "gr-qc",
    doi = "10.1103/PhysRevD.89.043011",
    journal = "Phys. Rev. D",
    volume = "89",
    number = "4",
    pages = "043011",
    year = "2014",
    note = "[Erratum: Phys.Rev.D 96, 129904 (2017), Erratum: Phys.Rev.D 97, 129901 (2018)]"
}

@article{Godzieba:2021vnz,
    author = "Godzieba, Daniel A. and Radice, David",
    title = "{High-Order Multipole and Binary Love Number Universal Relations}",
    eprint = "2109.01159",
    archivePrefix = "arXiv",
    primaryClass = "astro-ph.HE",
    doi = "10.3390/universe7100368",
    journal = "Universe",
    volume = "7",
    number = "10",
    pages = "368",
    year = "2021"
}

@article{Jiang:2019pzw,
    author = "Jiang, Rongrong and Wen, Dehua and Chen, Houyuan",
    title = "{Universal behavior of a compact star based upon the gravitational binding energy}",
    eprint = "1911.10935",
    archivePrefix = "arXiv",
    primaryClass = "nucl-th",
    doi = "10.1103/PhysRevD.100.123010",
    journal = "Phys. Rev. D",
    volume = "100",
    number = "12",
    pages = "123010",
    year = "2019"
}

@article{Vaglio:2022flq,
    author = "Vaglio, Massimo and Pacilio, Costantino and Maselli, Andrea and Pani, Paolo",
    title = "{Multipolar structure of rotating boson stars}",
    eprint = "2203.07442",
    archivePrefix = "arXiv",
    primaryClass = "gr-qc",
    doi = "10.1103/PhysRevD.105.124020",
    journal = "Phys. Rev. D",
    volume = "105",
    number = "12",
    pages = "124020",
    year = "2022"
}

@article{Adam:2020aza,
    author = "Adam, Christoph and Garc\'\i{}a Mart\'\i{}n-Caro, Alberto and Huidobro, Miguel and V\'azquez, Ricardo and Wereszczynski, Andrzej",
    title = "{Quasiuniversal relations for generalized Skyrme stars}",
    eprint = "2011.08573",
    archivePrefix = "arXiv",
    primaryClass = "hep-th",
    doi = "10.1103/PhysRevD.103.023022",
    journal = "Phys. Rev. D",
    volume = "103",
    number = "2",
    pages = "023022",
    year = "2021"
}

@article{Adam:2023qxj,
    author = "Adam, Christoph and Castelo, Jorge and Garc{\'\i}a Mart{\'\i}n-Caro, Alberto and Huidobro, Miguel and Wereszczynski, Andrzej",
    title = "{Effective no-hair relations for spinning boson stars}",
    eprint = "2305.06181",
    archivePrefix = "arXiv",
    primaryClass = "gr-qc",
    doi = "10.1103/PhysRevD.108.043015",
    journal = "Phys. Rev. D",
    volume = "108",
    number = "4",
    pages = "043015",
    year = "2023"
}

@article{Adam:2024zqr,
    author = "Adam, Christoph and Mourelle, Jorge Castelo and dos Santos Costa Filho, Etevaldo and Herdeiro, Carlos A. R. and Wereszczynski, Andrzej",
    title = "{Universal relations for rotating scalar and vector boson stars}",
    eprint = "2406.07613",
    archivePrefix = "arXiv",
    primaryClass = "gr-qc",
    doi = "10.1103/PhysRevD.110.084017",
    journal = "Phys. Rev. D",
    volume = "110",
    number = "8",
    pages = "084017",
    year = "2024"
}

@article{Panotopoulos:2026xwj,
    author = "Panotopoulos, Grigoris and Rinc{\'o}n, {\'A}ngel and Lopes, Ilidio",
    title = "{Slowly rotating condensate dark stars beyond the mean-field approximation}",
    eprint = "2607.06898",
    archivePrefix = "arXiv",
    primaryClass = "gr-qc",
    doi = "10.1140/epjc/s10052-026-16183-0",
    journal = "Eur. Phys. J. C",
    volume = "86",
    number = "8",
    pages = "992",
    year = "2026"
}

@article{Kumar:2026ddm,
    author = "Kumar, Ankit and Sotani, Hajime",
    title = "{Slowly rotating two-fluid neutron stars: Coupled frame-dragging, inertia splitting, and universal relations}",
    eprint = "2603.12613",
    archivePrefix = "arXiv",
    primaryClass = "astro-ph.HE",
    doi = "10.1103/559f-kl69",
    journal = "Phys. Rev. D",
    volume = "113",
    number = "6",
    pages = "063055",
    year = "2026"
}

@article{Adam:2025ktm,
    author = "Adam, Christoph and Mourelle, Jorge Castelo and Garc{\'\i}a Mart{\'\i}n-Caro, Alberto and Wereszczynski, Andrzej",
    title = "{Spinning boson stars in nonlinear sigma models and universal relations}",
    eprint = "2502.20923",
    archivePrefix = "arXiv",
    primaryClass = "gr-qc",
    doi = "10.1103/z5v2-f3ms",
    journal = "Phys. Rev. D",
    volume = "112",
    number = "2",
    pages = "023004",
    year = "2025"
}

@article{Yagi:2016bkt,
    author = "Yagi, Kent and Yunes, Nicol\'as",
    title = "{Approximate Universal Relations for Neutron Stars and Quark Stars}",
    eprint = "1608.02582",
    archivePrefix = "arXiv",
    primaryClass = "gr-qc",
    doi = "10.1016/j.physrep.2017.03.002",
    journal = "Phys. Rept.",
    volume = "681",
    pages = "1--72",
    year = "2017"
}

@article{Diedrichs:2023trk,
    author = {Diedrichs, Robin Fynn and Becker, Niklas and Jockel, C{\'e}dric and Christian, Jan-Erik and Sagunski, Laura and Schaffner-Bielich, J{\"u}rgen},
    title = "{Tidal deformability of fermion-boson stars: Neutron stars admixed with ultralight dark matter}",
    eprint = "2303.04089",
    archivePrefix = "arXiv",
    primaryClass = "gr-qc",
    doi = "10.1103/PhysRevD.108.064009",
    journal = "Phys. Rev. D",
    volume = "108",
    number = "6",
    pages = "064009",
    year = "2023"
}

@article{bustillo2023searching,
    author = "Calderon Bustillo, Juan and Sanchis-Gual, Nicolas and Leong, Samson H. W. and Chandra, Koustav and Torres-Forne, Alejandro and Font, Jose A. and Herdeiro, Carlos and Radu, Eugen and Wong, Isaac C. F. and Li, Tjonnie G. F.",
    title = "{Searching for vector boson-star mergers within LIGO-Virgo intermediate-mass black-hole merger candidates}",
    eprint = "2206.02551",
    archivePrefix = "arXiv",
    primaryClass = "gr-qc",
    doi = "10.1103/PhysRevD.108.123020",
    journal = "Phys. Rev. D",
    volume = "108",
    number = "12",
    pages = "123020",
    year = "2023"
}

@article{Valdez-Alvarado:2024eze,
    author = "Valdez-Alvarado, Susana and Cruz-Osorio, Alejandro and D{\'a}vila, J. M. and Ure{\~n}a-L{\'o}pez, L. Arturo and Becerril, Ricardo",
    title = "{Observations of compact stars and fermion-boson stars with a quartic self-interaction}",
    eprint = "2402.12328",
    archivePrefix = "arXiv",
    primaryClass = "gr-qc",
    doi = "10.1088/1361-6382/adc361",
    journal = "Class. Quant. Grav.",
    volume = "42",
    number = "8",
    pages = "085001",
    year = "2025"
}

@article{lee2021could,
    author = "Lee, Billy K. K. and Chu, Ming-chung and Lin, Lap-Ming",
    title = "{Could the GW190814 Secondary Component Be a Bosonic Dark Matter Admixed Compact Star?}",
    eprint = "2110.05538",
    archivePrefix = "arXiv",
    primaryClass = "astro-ph.HE",
    doi = "10.3847/1538-4357/ac2735",
    journal = "Astrophys. J.",
    volume = "922",
    number = "2",
    pages = "242",
    year = "2021"
}

@article{Mourelle:2024qgo,
    author = "Mourelle, Jorge Castelo and Adam, Christoph and Calder{\'o}n Bustillo, Juan and Sanchis-Gual, Nicolas",
    title = "{Rotating fermion-boson stars}",
    eprint = "2403.13052",
    archivePrefix = "arXiv",
    primaryClass = "gr-qc",
    doi = "10.1103/PhysRevD.110.123019",
    journal = "Phys. Rev. D",
    volume = "110",
    number = "12",
    pages = "123019",
    year = "2024"
}

@article{Konstantinou:2024ynd,
    author = "Konstantinou, Andreas",
    title = "{The Effect of a Dark Matter Core on the Structure of a Rotating Neutron Star}",
    eprint = "2405.01487",
    archivePrefix = "arXiv",
    primaryClass = "astro-ph.HE",
    doi = "10.3847/1538-4357/ad4701",
    journal = "Astrophys. J.",
    volume = "968",
    number = "2",
    pages = "83",
    year = "2024"
}

@article{Cipriani:2025wem,
    author = "Cipriani, Lorenzo and Sagun, Violetta and Staykov, Kalin V. and Doneva, Daniela D. and Yazadjiev, Stoytcho S.",
    title = "{Differentially rotating neutron stars with dark matter cores}",
    eprint = "2512.05898",
    archivePrefix = "arXiv",
    primaryClass = "astro-ph.HE",
    doi = "10.1103/4rs2-973y",
    journal = "Phys. Rev. D",
    volume = "114",
    number = "2",
    pages = "023039",
    year = "2026"
}

@misc{Kuan:2026duo,
    author = "Kuan, Hao-Jui and Lam, Alan Tsz-Lok and Noronha-Hostler, Jacquelyn and Yunes, Nicol{\'a}s",
    title = "{Donutization Inside Neutron Stars: Shell-Localized Scalar Fields}",
    eprint = "2605.26228",
    archivePrefix = "arXiv",
    primaryClass = "gr-qc",
    year = "2026"
}
%%%%%%%%%%%%%%%%%%%%%%%%%%%%%%%%%%%%%%%%%%%%%%%%%%%%%%%%%%%%
%%%%%%%%%%%%%%%%%%%%%%%%%%%%%%%%%%%%%%%%%%%%%%%%%%%%%%%%%%%%
%%%%%%%%%%%%%%%%%%%%%%%%%%%%%%%%%%%%%%%%%%%%%%%%%%%%%%%%%%%%

%%%%%%%%%%%%%%%%%%%%%%%%%%%%%%%%%%%%%%%%%%%%%%%%%%%%%%%%%%%%
%%%%%%%%%%%%%%%%%%%%%%%%%%%%%%%%%%%%%%%%%%%%%%%%%%%%%%%%%%%%
%%%%%%%%%%%%%%%%%%%%%%%%%%%%%%%%%%%%%%%%%%%%%%%%%%%%%%%%%%%%
\appendix
%%%%%%%%%%%%%%%%%%%%%%%%%%%%%%%%%%%%%%%%%%%%%%%%%%%%%%%%%%%%
%%%%%%%%%%%%%%%%%%%%%%%%%%%%%%%%%%%%%%%%%%%%%%%%%%%%%%%%%%%%
%%%%%%%%%%%%%%%%%%%%%%%%%%%%%%%%%%%%%%%%%%%%%%%%%%%%%%%%%%%%

%%%%%%%%%%%%%%%%%%%%%%%%%%%%%%%%%%%%%%%%%%%%%%%%%%%%%%%%%%%%
%%%%%%%%%%%%%%%%%%%%%%%%%%%%%%%%%%%%%%%%%%%%%%%%%%%%%%%%%%%%
\section{Validation in limiting cases}
\label{app:validation}
%%%%%%%%%%%%%%%%%%%%%%%%%%%%%%%%%%%%%%%%%%%%%%%%%%%%%%%%%%%%
%%%%%%%%%%%%%%%%%%%%%%%%%%%%%%%%%%%%%%%%%%%%%%%%%%%%%%%%%%%%
We have verified that our numerical framework recovers previously reported solutions in the two limiting cases obtained by removing either the bosonic or the fermionic component.

In the pure NS limit, for $\beta=-6$ and the same polytropic EOS used in Ref.~\cite{Mendes:2016fby}, our code reproduces the behavior of the baryonic mass $M_b$ and of the central DEF scalar field $\lvert\varphi_c\rvert$ reported in Fig.~4 of Ref.~\cite{Mendes:2016fby} for their model M2. For a more stringent quantitative comparison, Table~\ref{tab:validation_M2} displays properties of the representative M2 solution reported in Table~I of Ref.~\cite{Mendes:2016fby}, together with the corresponding values obtained by our code. The relative differences remain below $0.6\%$ for all the reported quantities.
\begin{table}[h!]
\centering
\caption{
Comparison between solution M2 of Table~I of
Ref.~\cite{Mendes:2016fby} and the corresponding pure
fermionic solution obtained with our code using
$r_{\rm cut}=500$. The relative difference is defined as
$\Delta_X=100|X_{\rm ours}-X_{\rm ref}|/|X_{\rm ref}|$.
}
\label{tab:validation_M2}
\renewcommand{\arraystretch}{1.12}
\setlength{\tabcolsep}{6pt}
\begin{tabular}{lccc}
\hline\hline
Quantity & Reference~\cite{Mendes:2016fby} & This work & $\Delta_X$ [\%] \\
\hline
$\rho_{0c}/\rho_{\rm ref}$
    & $8.0747$ & $8.07560$ & $0.011$ \\
$M_b\,[M_\odot]$
    & $2.3594$ & $2.36032$ & $0.039$ \\
$M_{\rm ADM}\,[M_\odot]$
    & $1.9459$ & $1.94484$ & $0.054$ \\
$M/R_s$
    & $0.209$ & $0.20868$ & $0.154$ \\
$|\varphi_c-\varphi_0|$
    & $0.284$ & $0.283769$ & $0.081$ \\
$|Q_\varphi|\,[M_\odot]$
    & $1.23$ & $1.23698$ & $0.567$ \\
\hline\hline
\end{tabular}
\end{table}

In the pure BS limit, we compare against results of Ref.~\cite{Evstafyeva:2023kfg} with $\beta=-7$. Our code reproduces the solutions shown in the top-left panel of Fig.~6 of Ref.~\cite{Evstafyeva:2023kfg}, including the radial behavior of both the bosonic field and the DEF scalar field (see middle part of Fig.~\ref{fig:all_def}).

It is important to note that our numerical scheme uses the central bosonic amplitude $\phi_c$ as an input parameter, rather than fixing the frequency $\omega$. Nevertheless, the resulting frequencies reproduce the values reported in Ref.~\cite{Evstafyeva:2023kfg} with relative differences of only
\begin{equation}
\Delta\omega_{\rm M1}=0.001323\%,
\qquad
\Delta\omega_{\rm M2}=0.000487\%.
\end{equation}
The corresponding values of $\phi_c$ and of the central gravitational scalar field are also consistent with those quoted in the reference.

%%%%%%%%%%%%%%%%%%%%%%%%%%%%%%%%%%%%%%%%%%%%%%%%%%%%%%%%%%%%
%%%%%%%%%%%%%%%%%%%%%%%%%%%%%%%%%%%%%%%%%%%%%%%%%%%%%%%%%%%%
\section{Global quantities}
\label{app:magnitudes}
%%%%%%%%%%%%%%%%%%%%%%%%%%%%%%%%%%%%%%%%%%%%%%%%%%%%%%%%%%%%
%%%%%%%%%%%%%%%%%%%%%%%%%%%%%%%%%%%%%%%%%%%%%%%%%%%%%%%%%%%%

Here, we summarize the global quantities used to characterize equilibrium configurations, which are computed in the Einstein frame using the areal radial coordinate \(\tilde r\), although observables and magnitudes are reported as functions of the Jordan-frame radial coordinate given in Eq.~\eqref{eq:jordan_radius}, $r=a[\varphi(\tilde r)]\tilde r$.

%%%%%%%%%%%%%%%%%%%%%%%%%%%%%%%%%%%%%%%%%%%%%%%%%%%%%%%%%%%%
%%%%%%%%%%%%%%%%%%%%%%%%%%%%%%%%%%%%%%%%%%%%%%%%%%%%%%%%%%%%
\subsection{Total gravitational mass}
\label{app:total_gravitational_mass}
%%%%%%%%%%%%%%%%%%%%%%%%%%%%%%%%%%%%%%%%%%%%%%%%%%%%%%%%%%%%
%%%%%%%%%%%%%%%%%%%%%%%%%%%%%%%%%%%%%%%%%%%%%%%%%%%%%%%%%%%%
For a given stellar model in the Einstein frame, the Misner-Sharp mass function, $\tilde m(\tilde r)$, is obtained from the radial metric function $B(\tilde r)$ according to
\begin{equation}
\tilde m(\tilde r)
=
\frac{\tilde r}{2}
\left[
1-\frac{1}{B(\tilde r)}
\right].
\label{eq:einstein_mass_profile}
\end{equation}
The corresponding Jordan-frame quasilocal mass is
\begin{equation}
m_{\rm J}(r)
=
\frac{r}{2}
\left[
1-\frac{1}{B_{\rm J}(r)}
\right],
\label{eq:jordan_mass_profile}
\end{equation}
where, according to Eq.~\eqref{eq:jordan_metric},
$B_{\rm J}(r)\,
=
a^{2}(\varphidef)B(\tilde r)\,$.
In our spherically symmetric spacetime, the total gravitational mass of a stellar model is given by the ADM mass,
\begin{equation}
M_{\rm ADM}
=
\lim_{\tilde r\rightarrow\infty}
\tilde m(\tilde r),
\label{eq:ADM_mass_definition}
\end{equation}
which we approximate at the outer boundary of our numerical domain, $\tilde r_{\max}$, as
\begin{equation}
M_{\rm ADM}
\simeq
\tilde m(\tilde r_{\max}).
\label{eq:numerical_ADM_mass}
\end{equation}

For the coupling function used in this work, Eq.~\eqref{eq:def_factor}, with \(\alpha_{0}=0\), we demand solutions to satisfy $\varphi(\tilde r) \to 0$ as $\tilde r \to \infty$. Therefore, $a(\varphidef) = 1+\mathcal{O}(\tilde r^{-2})$
in the asymptotic region, and thus the Einstein- and Jordan-frame ADM masses coincide.
The finite-radius quantities \(\tilde m(\tilde r_{\max})\) and \(m_{\rm J}[r(\tilde r_{\max})]\), however, need not be numerically identical. Our code retains the Jordan-frame quasilocal mass at the cutoff for diagnostics, while the reported total mass is given by Eq.~\eqref{eq:numerical_ADM_mass}.

%%%%%%%%%%%%%%%%%%%%%%%%%%%%%%%%%%%%%%%%%%%%%%%%%%%%%%%%%%%%
%%%%%%%%%%%%%%%%%%%%%%%%%%%%%%%%%%%%%%%%%%%%%%%%%%%%%%%%%%%%
\subsection{Fermionic and bosonic radii}
\label{app:radii}
%%%%%%%%%%%%%%%%%%%%%%%%%%%%%%%%%%%%%%%%%%%%%%%%%%%%%%%%%%%%
%%%%%%%%%%%%%%%%%%%%%%%%%%%%%%%%%%%%%%%%%%%%%%%%%%%%%%%%%%%%

For configurations containing a fermionic component, the surface is defined by the first zero of the fluid pressure,
\begin{equation}
p(\tilde r_{\rm F})=0.
\label{eq:NS_surface}
\end{equation}
In our numerical implementation, the surface position is obtained by interpolating the zero of the fluid pseudoenthalpy between the last interior point and the first vacuum point. The Jordan-frame fermionic radius is then
\begin{equation}
\RF
=
a[\varphidef(\tilde r_{\rm F})]\,\tilde r_{\rm F}.
\label{eq:jordan_NS_radius}
\end{equation}

The bosonic field has no sharp surface.
Given the total bosonic Noether charge of a solution, $N_{\rm B}(\tilde r_{\max})$ (see Appendix~\ref{sec:app_bosonic_rest_mass} below), the Einstein-frame radius containing \(99\%\) of this charge, $\tilde r_{\rm B}^{99}$, satisfies
\begin{equation}
N_{\rm B}(\tilde r_{\rm B}^{99})
=
0.99N_{\rm B}(\tilde r_{\max}).
\label{eq:RBS99_charge_definition}
\end{equation}
We thus define the Jordan-frame bosonic radius containing \(99\%\) of the total Noether charge as
\begin{equation}
\RB
=
a[\varphidef(\tilde r_{\rm B}^{99})]\,
\tilde r_{\rm B}^{99}.
\label{eq:jordan_BS_radius}
\end{equation}
Since the bosonic rest mass is proportional to the Noether charge, this is equivalently the radius containing \(99\%\) of the bosonic rest mass.

For diagnostic purposes, our code also evaluates the radius
\(R_{\rm F}^{99}\) containing \(99\%\) of the fermionic rest mass and the radius \(R_{\rm M}^{99}\) containing \(99\%\) of the integrated positive gravitational-mass density.

%%%%%%%%%%%%%%%%%%%%%%%%%%%%%%%%%%%%%%%%%%%%%%%%%%%%%%%%%%%%
%%%%%%%%%%%%%%%%%%%%%%%%%%%%%%%%%%%%%%%%%%%%%%%%%%%%%%%%%%%%
\subsection{Scalar charge}
\label{app:scalar_charge}
%%%%%%%%%%%%%%%%%%%%%%%%%%%%%%%%%%%%%%%%%%%%%%%%%%%%%%%%%%%%
%%%%%%%%%%%%%%%%%%%%%%%%%%%%%%%%%%%%%%%%%%%%%%%%%%%%%%%%%%%%
For all the stellar models in this work, the asymptotic behavior of the massless DEF scalar field is
\begin{equation}
\varphidef(\tilde r)
=
\varphidef_{\infty}
+
\frac{Q_{\varphi}}{\tilde r}
+
\mathcal{O}(\tilde r^{-2}),
\label{eq:DEF_asymptotic_charge}
\end{equation}
where $\varphidef_{\infty}=0$ and $Q_\varphi$ is the scalar charge, which is therefore
\begin{equation}
Q_{\varphi}
=
-\lim_{\tilde r\rightarrow\infty}
\tilde r^{2}\varphidef'(\tilde r).
\label{eq:scalar_charge}
\end{equation}
On the finite numerical domain, our code evaluates
\begin{equation}
Q_{\varphi}
\simeq
-\tilde r_{\max}^{2}
\varphidef'(\tilde r_{\max}),
\label{eq:numerical_scalar_charge}
\end{equation}
which gives the scalar charge in internal length units. Its value in kilometers is obtained by multiplying by the geometrized solar-mass length,
\begin{equation}
Q_{\varphi}[{\rm km}]
=
\frac{GM_{\odot}}{c^{2}}
Q_{\varphi}[{\rm code}].
\label{eq:scalar_charge_conversion}
\end{equation}

%%%%%%%%%%%%%%%%%%%%%%%%%%%%%%%%%%%%%%%%%%%%%%%%%%%%%%%%%%%%
%%%%%%%%%%%%%%%%%%%%%%%%%%%%%%%%%%%%%%%%%%%%%%%%%%%%%%%%%%%%
\subsection{Fermionic particle number and rest mass}
\label{sec:app_fermionic_rest_mass}
%%%%%%%%%%%%%%%%%%%%%%%%%%%%%%%%%%%%%%%%%%%%%%%%%%%%%%%%%%%%
%%%%%%%%%%%%%%%%%%%%%%%%%%%%%%%%%%%%%%%%%%%%%%%%%%%%%%%%%%%%

The fluid rest-mass density \(\rho_{0}\), and the fermion number density \(n_{\rm F}\), are physical Jordan-frame matter variables related through
\begin{equation}
\rho_{0}
=
m_{\rm F}n_{\rm F},
\label{eq:rest_mass_baryon_density}
\end{equation}
where $m_{\rm F}$ is the average rest-mass per fermion.
The conserved fermionic particle number enclosed within an Einstein-frame coordinate radius \(\tilde r\) is
\begin{equation}
N_{\rm F}(\tilde r)
=
4\pi
\int_{0}^{\tilde r}
\dd\tilde r'\,
\tilde r'^{\,2}
\sqrt{B(\tilde r')}
a^{3}[\varphidef(\tilde r')]
n_{\rm F}(\tilde r').
\label{eq:baryon_number}
\end{equation}
The factor \(a^{3}(\varphidef)\) follows from the conformal transformation of the conserved matter current and of the spatial volume element.
The total fermionic particle number is
\begin{equation}
N_{\rm F}
=
N_{\rm F}(\tilde r_{\rm F}),
\label{eq:total_baryon_number}
\end{equation}
because the number density vanishes outside the fluid surface of radius $\tilde r_{\rm F}$. The corresponding fermionic rest mass is
\begin{equation}
M_{\rm 0,F}
=
m_{\rm F}N_{\rm F}.
\label{eq:total_baryonic_mass}
\end{equation}
Equivalently, in solar-mass units,
\begin{equation}
\frac{M_{\rm 0,F}}{M_{\odot}}
=
\frac{m_{\rm F}N_{\rm F}}{M_{\odot}},
\qquad
N_{\rm F}
=
\frac{M_{\rm 0,F}M_{\odot}}{m_{\rm F}},
\label{eq:baryonic_mass_solar}
\end{equation}
where \(M_{\rm 0,F}\) on the right-hand side of the second equality is understood as its numerical value in units of \(M_{\odot}\).

In the geometrized variables used by the code, the fermionic rest mass is evaluated as
\begin{equation}
M_{\rm 0,F}
=
4\pi
\int_{0}^{\tilde r_{\max}}
\dd\tilde r\,
\tilde r^{2}
\sqrt{B(\tilde r)}
a^{3}[\varphidef(\tilde r)]
\rho_{0}(\tilde r),
\label{eq:baryonic_mass_code}
\end{equation}
where the integrand vanishes outside the fermionic surface.

%%%%%%%%%%%%%%%%%%%%%%%%%%%%%%%%%%%%%%%%%%%%%%%%%%%%%%%%%%%%
%%%%%%%%%%%%%%%%%%%%%%%%%%%%%%%%%%%%%%%%%%%%%%%%%%%%%%%%%%%%
\subsection{Bosonic Noether charge and rest mass}
\label{sec:app_bosonic_rest_mass}
%%%%%%%%%%%%%%%%%%%%%%%%%%%%%%%%%%%%%%%%%%%%%%%%%%%%%%%%%%%%
%%%%%%%%%%%%%%%%%%%%%%%%%%%%%%%%%%%%%%%%%%%%%%%%%%%%%%%%%%%%

For the complex bosonic field, the global \(U(1)\) symmetry gives rise to the conserved current
\begin{equation}
j^{\mu}
=
-i\,a^{2}(\varphidef)
\left(
\Phi^{*}\tilde\nabla^{\mu}\Phi
-
\Phi\tilde\nabla^{\mu}\Phi^{*}
\right),
\label{eq:bosonic_noether_current}
\end{equation}
up to the overall normalization convention adopted for the bosonic action. For the harmonic ansatz in Eq.~\eqref{eq:boson_ansatz}, the enclosed Noether charge computed by our code is
\begin{equation}
N_{\rm B}(\tilde r)
=
8\pi\omega
\int_{0}^{\tilde r}
\dd\tilde r'\,
\tilde r'^{\,2}
\sqrt{
\frac{B(\tilde r')}{A(\tilde r')}
}
a^{2}[\varphidef(\tilde r')]
\phib^{2}(\tilde r').
\label{eq:boson_number}
\end{equation}

Here, \(\omega\) is the frequency normalized with respect to the asymptotic time coordinate, for which \(A(\infty)=1\). The factors of \(\hbar\) and \(c\) are absorbed into the geometrized field normalization used by the code.

The total bosonic Noether charge is evaluated at the outer numerical boundary,
\begin{equation}
N_{\rm B}
=
N_{\rm B}(\tilde r_{\max}).
\label{eq:total_boson_number}
\end{equation}
In the code normalization, the bosonic rest mass is
\begin{equation}
M_{\rm 0,B}
=
\mu N_{\rm B},
\label{eq:bosonic_rest_mass_code}
\end{equation}
where \(\mu\) is the boson mass parameter appearing in the Klein-Gordon potential. When expressed as a physical particle number, the conversion is
\begin{equation}
\mathcal{N}_{\rm B}
=
\frac{
M_{\rm 0,B}M_{\odot}
}{
m_{\rm B}
},
\label{eq:physical_boson_number}
\end{equation}
where \(M_{\rm 0,B}\) is expressed in solar-mass units and \(m_{\rm B}\) is the physical mass of one boson.

%%%%%%%%%%%%%%%%%%%%%%%%%%%%%%%%%%%%%%%%%%%%%%%%%%%%%%%
\section{Maximum of DEF scalar field}
\label{app:varphimax}
%%%%%%%%%%%%%%%%%%%%%%%%%%%%%%%%%%%%%%%%%%%%%%%%%%%%%%%
As discussed throughout this paper, we generally use the central value $|\varphi_c|$ as a measure of the degree of scalarization, even though the maximum value of the DEF scalar field does not necessarily coincide with its central value. The main reason for this choice is that $\varphi_c$ is the value of the DEF field at the origin, which is fixed by the outer boundary condition, and therefore provides a natural \textit{a posteriori} parametrization of equilibrium solutions. 

\begin{widetext}

\begin{figure*}[htbp]
    \centering
    \includegraphics[width=0.32\columnwidth]{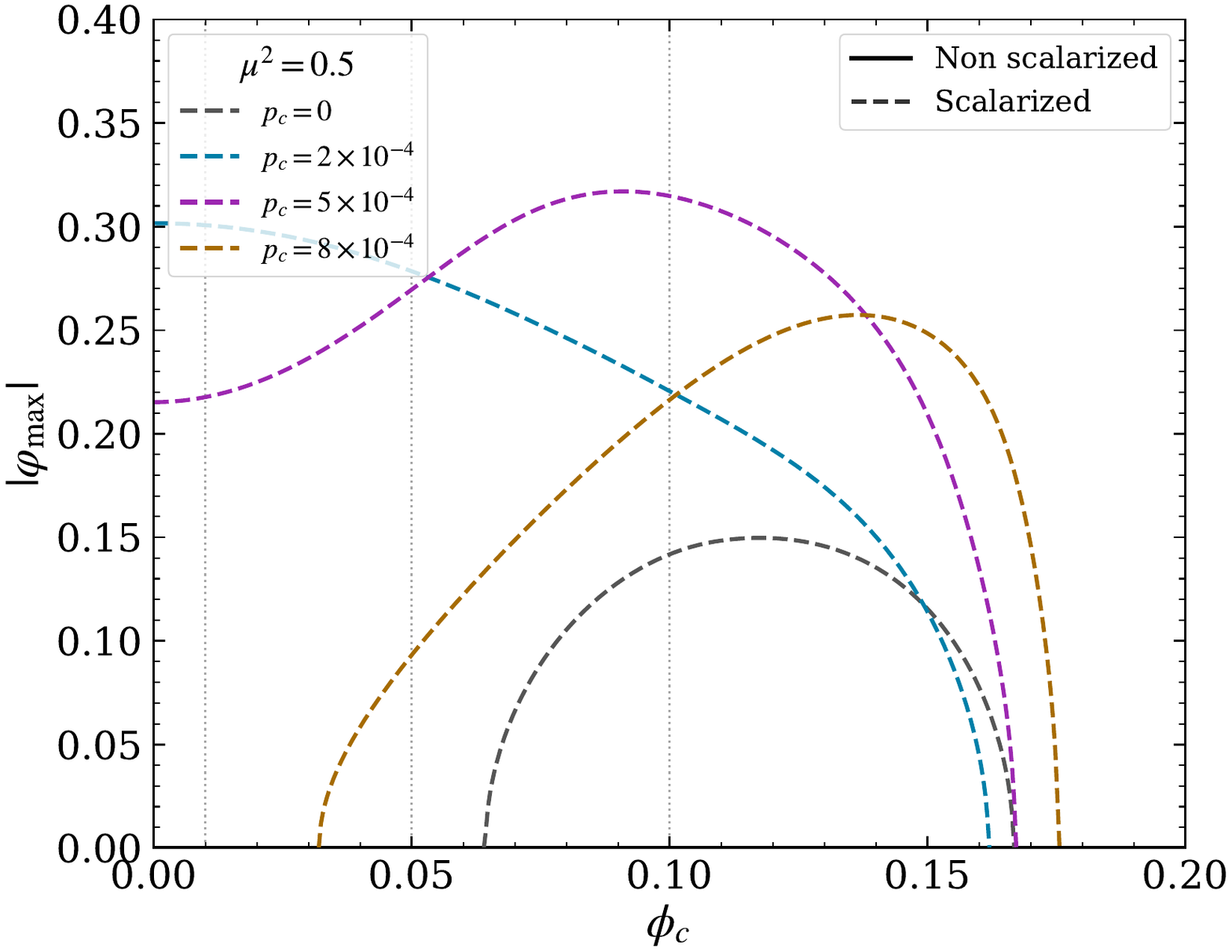}
     \includegraphics[width=0.32\columnwidth]{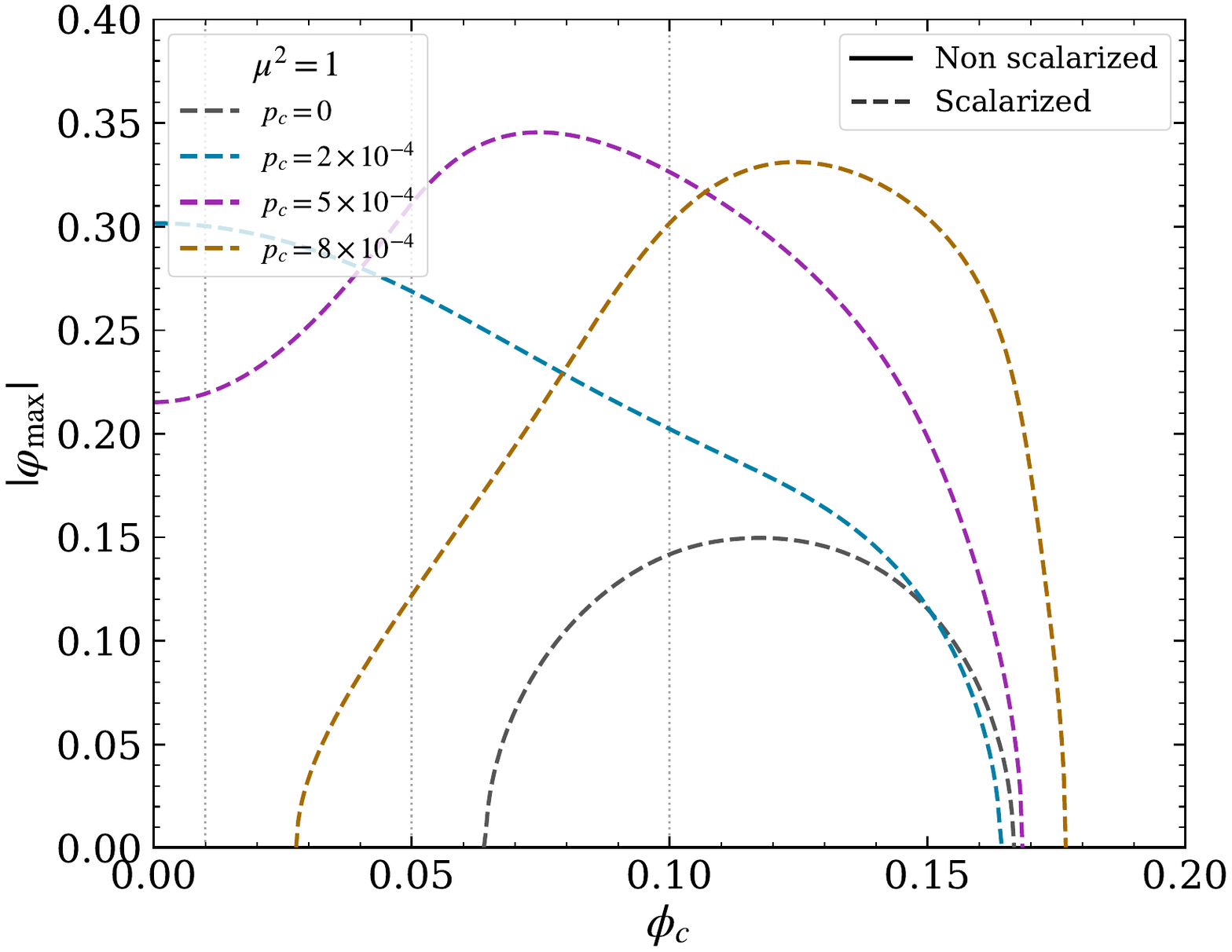}
      \includegraphics[width=0.32\columnwidth]{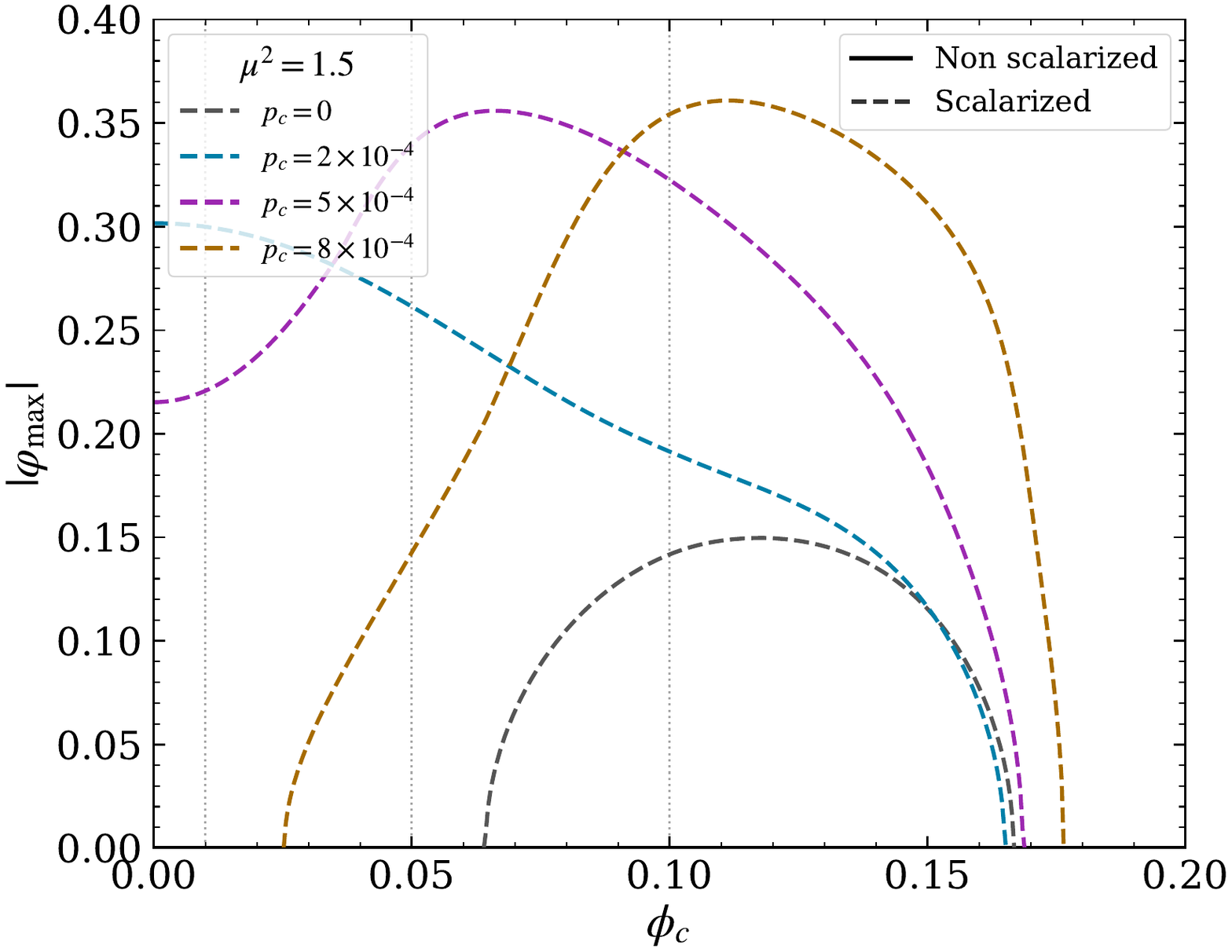}
       
    \caption{Maximum value of the DEF scalar field, $|\varphi_{\text{max}}|$, as a function of the central bosonic field amplitude, $\phi_c$, for four fixed central pressures, $p_c$. The left, central, and right panels correspond to $\mu^2=0.5$, $1.0$, and $1.5$, respectively. Dashed colored lines identify different $p_c$ families and show how the maximum DEF field varies across the scanned $\phi_c$ range. Vertical gray lines correspond to $\phi_c = 0.01$, 0.05, and 0.1.}
    \label{fig_varphimax_phi}
\end{figure*}
\end{widetext}

For completeness, in Fig.~\ref{fig_varphimax_phi} we present the counterpart of Fig.~\ref{fig_varphi_phi} using $|\varphi_{\rm max}|$ instead. The resulting diagrams show the same qualitative trends and the same interplay among the relevant parameters, namely $\mu^2$, $p_c$, and $\phi_c$. This confirms that the conclusions drawn from the analysis based on $|\varphi_c|$ would remain virtually identical if the scalarization strength were characterized instead by $|\varphi_{\rm max}|$.

\end{document}